\documentclass[10pt]{iopart}
\usepackage{epsfig}
\usepackage{graphicx}
\newcommand{\writingdate}{October 23rd, 2002}

\newcommand{\BN}{\vspace*{3mm}\begin{quote}\hrule\sl $~$\\ NOTE: }
\newcommand{\EN}{\\ \hrule\end{quote}\vspace*{3mm}}
\newcommand{\ndelta}{{\overline{\delta}}}
\newcommand{\room}{\rule[-0.3cm]{0cm}{0.8cm}}

\newcommand{\vsp}{\vspace*{3mm}}
\newcommand{\be}{\begin{equation}}
\newcommand{\ee}{\end{equation}}
\newcommand{\bd}{\begin{displaymath}}
\newcommand{\ed}{\end{displaymath}}
\newcommand{\bdm}{\begin{displaymath}}
\newcommand{\edm}{\end{displaymath}}
\newcommand{\bea}{\begin{eqnarray}}
\newcommand{\eea}{\end{eqnarray}}

\newcommand{\extr}{~{\rm extr}}

\newcommand{\bra}{\langle}
\newcommand{\ket}{\rangle}
\newcommand{\bigbra}{\left\langle\room}
\newcommand{\bigket}{\right\rangle\room}
\newcommand{\bras}{\langle\!\langle}
\newcommand{\kets}{\rangle\!\rangle}
\newcommand{\bigbras}{\left\langle\!\!\!\left\langle\room}
\newcommand{\bigkets}{\room\right\rangle\!\!\!\right\rangle}
\newcommand{\order}{{\cal O}}
\newcommand{\minus}{\!-\!}
\newcommand{\plus}{\!+\!}

\newcommand{\bk}{\mbox{\boldmath $k$}}

\newcommand{\bq}{\mbox{\boldmath $q$}}

\newcommand{\bx}{\mbox{\boldmath $x$}}
\newcommand{\by}{\mbox{\boldmath $y$}}

\newcommand{\bC}{\mbox{\boldmath $C$}}

\newcommand{\bE}{\mbox{\boldmath $E$}}

\newcommand
{\bnul}{\mbox{\boldmath $0$}}

\newcommand{\hq}{\hat{q}}

\newcommand{\hbq}{\hat{\mbox{\boldmath $q$}}}

\newcommand{\btheta}{\mbox{\boldmath $\theta$}}

\newcommand{\unity}{{\bf 1}\hspace{-2mm}{\bf I}}

\begin{document}

\title[Frustrated coupled oscillators with random fields --  \writingdate]
{Partially and fully frustrated coupled oscillators\\ with random
pinning fields}
\author{A C C Coolen \dag~ and C P\'{e}rez-Vicente \ddag}

\address{\dag
Department of Mathematics, King's College London, The Strand,
London WC2R 2LS, UK}
\address{\ddag
Departament de Fisica Fonamental, Facultat de Fisica, Universitat
de Barcelona, 08028 Barcelona, Spain }

 \ead{\tt
tcoolen@mth.kcl.ac.uk, conrad@ffn.ub.es}

\begin{abstract}
We have studied two specific models of frustrated and disordered
coupled  Kuramoto oscillators, all driven with the same natural
frequency, in the presence of random external pinning fields. Our
models are structurally similar, but differ in their degree of
bond frustration and in their finite size ground state properties
(one has random ferro- and anti-ferromagnetic interactions; the
other has random chiral interactions). We have calculated the
equilibrium properties of both models in the thermodynamic limit
using the replica method, with emphasis on the role played by
symmetries of the pinning field distribution, leading to explicit
predictions for observables, transitions, and phase diagrams. For
absent pinning fields our two models are found to behave
identically, but pinning fields (provided with appropriate
statistical properties) break this symmetry. Simulation data lend
satisfactory support to our theoretical predictions.
\end{abstract}

\section{Introduction}

The dynamics and the analysis of models of interacting oscillators
have received an increasing amount of attention during the last
few decades \cite{Winfr}. These models are excellent for studying
synchronization phenomena and other features related to the
temporal activity of populations of microscopic elements with
limit cycle behaviour. They have been studied in many  fields,
including physics, chemistry and biology \cite{Kuram1}. One of the
most successful models  was proposed by Kuramoto \cite{Kuram2},
who assumed that each member of the population can be modeled as
an oscillator moving in a globally attracting limit cycle of
constant amplitude. The interactions between the oscillators are
sufficiently weak to ensure that perturbations will not move them
away from their individual limit cycles. Only one degree of
freedom, the phase $\theta_i$, is then required to describe the
dynamics of each oscillator $i$. The latter evolves according to
$\frac{ d}{dt}\theta_i = \omega_i + \sum_j K_{ij} f(\theta_i -
\theta_j)$. Here $\omega_i$ is the natural frequency of oscillator
$i$ (drawn randomly from some distribution), $K_{ij}$ is the
coupling strength between oscillators $i$ and $j$, and $f(x)$ a
non-linear function. Kuramoto initially considered mean field
coupling strengths $K_{ij} = K/N$, where $N$ is the total number
of oscillators, and $f(x)=\sin(x)$. This particular model has been
investigated intensively. Generally, for some critical (positive)
value of the coupling $K$ a phase transition occurs from a state
where all the oscillators run incoherently to a state where a
certain degree of synchronization emerges spontaneously. The
details  of this transition depend on the the distribution of
natural frequencies \cite{Strog,Bonil1,Bonil2,Aceb} and the nature
of the interactions (viz. short range interactions
\cite{Daid92,Stro88}, random site disorder \cite{Boni93}, or even
more complex types \cite{Arenas94}). Kuramoto-type models  have in
addition been studied extensively in the presence of noise
\cite{Strog,Saka88,Reimann99}.

The equations of the Kuramoto model have also been derived in the
context of neural networks \cite{Schu,Abb} and Josephson junctions
\cite{Wies,PB222a,PB222b}. There, however, one usually obtains an
extra term $A_{ij}$ in the argument of the nonlinear function:
\be
\frac{ d}{dt}\theta_i = \omega_i + \sum_j K_{ij} f(\theta_i -
\theta_j + A_{ij}) \label{eq:Kuramoto}
 \ee
The simplest case, $A_{ij} = \alpha$ for all $(i,j)$, was studied
in \cite{Sakaku}. Here the degree of spontaneous synchronization
was found to be maximal when $\alpha=0$, decreasing monotonically
to zero as $\alpha \rightarrow \frac{\pi}{2}$. A similar, but more
complicated case is the subject of \cite{Raeetal}. However, the
problem is far from solved for the case of quenched bond disorder.
Here
 only simulation studies \cite{Daido,Still} have been published
so far;  these emphasize the crucial role played by the disorder
in determining the long time properties of the model, as measured
by correlation functions. The situation is even more complex when
random pinning fields are added, leading to further competition.

The simplest models with randomness in both interactions and
external fields are those where all oscillators have the same
natural frequency, $\omega_i=\omega$; here one can switch to a new
 basis and transform away the $\omega_i$. Moreover, if the matrix
$\{A_{ij}\}$ fulfills certain symmetry criteria, then
(\ref{eq:Kuramoto})  can be written as a gradient descent process,
enabling (upon adding Gaussian white noise) an equilibrium
statistical mechanical analysis. Low dimensional versions of the
resulting system  have also been studied; they are equivalent to
frustrated XY models, used to describe Josephson junction arrays.
Here the sum over interacting pairs in the Hamiltonian is
restricted to neighbours which form a plaquette, with $\theta_i$
denoting the orientation of the spin at site $i$, and $A_{ij}$
denoting the bond angle, such that the plaquette sum
$\sum_{(i,j)\in ~{\rm plaq}} A_{ij} = 2\pi f$, where $f$ is a
measure of the frustration. For $f=1/2$ one has the so-called
fully frustrated model. Both the critical behaviour and the
symmetries of these low dimensional systems are known to depend
highly discontinuously on $f$ \cite{Choi}. Random external fields
trigger local fluctuations of $f$, giving rise to a complex
phenomenology which is still subject of study. It would be
interesting to investigate whether such features remain in
mean-field models. To our knowledge, this has not yet been
studied; the effect of random fields has so far only been analyzed
in systems without quenched disorder \cite{Arenas95}, or for
$m$-vector spin glass models \cite{Elder82,Cragg,Elder83} (but for
local uni-axial an-isotropic fields of a different nature than the
ones discussed here).

Our paper is structured as follows. We first define two mean-field
models with disorder in both bonds and pinning fields, and their
appropriate macroscopic observables. The first model is of a
conventional type, with individual pairs of oscillators trying to
either fully synchronize or fully anti-synchronize, depending on
the sign of their interaction. The second model is less
conventional in that neuron pairs prefer phase differences
$A_{ij}$ of either $\pi/2$ or $-\pi/2$. For both models the
presence of random external pinning forces increases energetic
conflicts and frustration further. In section three we solve both
models in equilibrium, using the replica method; we make the
replica-symmetric ansatz and calculate the conditions for replicon
instabilities. In sections four and five we study the effects of
global symmetries. Finally, in section six we present results (in
the form of phase diagrams and the temperature and field strength
dependence of observables) for a number of specific choices for
the pinning field distributions, and validate our predictions via
numerical simulations.

\section{Model definitions}

We study systems of $N$ coupled Kuramoto oscillators (or XY spins)
with external pinning fields, described by Hamiltonians of the
form
\be
H=-\frac{2K}{\sqrt{N}}\sum_{i<j}\cos(\theta_i\minus\theta_j\plus
A_{ij}) -h\sum_i \cos(\theta_i\minus\phi_i) \label{eq:Hamiltonian}
\ee where $\theta_i\in[0,2\pi]$ denotes the phase of the $i$-th
oscillator. The natural dynamics leading to a Boltzmann state with
Hamiltonian (\ref{eq:Hamiltonian}) is the Langevin equation
$\frac{d}{dt}\theta_i=-\partial H/\partial \theta_i+\xi_i(t)$,
with Gaussian random forces $\xi_i(t)$ which obey
$\bra\xi_i(t)\ket=0$ and
$\bra\xi_i(t)\xi_j(t^\prime)\ket=2T\delta_{ij}\delta[t\minus
t^\prime]$. Working out this expression gives
\begin{eqnarray}
\frac{d}{dt}\theta_i&=& -\frac{2K}{\sqrt{N}} \sum_{k\neq i}\left[
\room \cos(\tilde{A}_{ik})\sin(\theta_i\minus\theta_k)
+\sin(\tilde{A}_{ik})\cos(\theta_i\minus\theta_k) \right]
\nonumber \\ &&
 -h\sin(\theta_i\minus\phi_i) +\xi_i(t)
\label{eq:forces}
\end{eqnarray}
with $\tilde{A}_{ik}=A_{ik}$ for $i<j$, and
$\tilde{A}_{ik}=-A_{ki}$ for $i>j$. We introduce disorder by
drawing the pinning angles $\phi_i$ independently at random from a
distribution $p(\phi)$, and the relative angles $A_{ij}$
independently at random from a distribution $P(A)$, with
$\int\!dA~P(A)\cos(A)=0$ (to ensure that the bond disorder will
retain significance for $N\to\infty$). This system will generally
exhibit competition between alignment of the spins to the pinning
fields and the realization of prescribed relative angles
 between pairs, and have a high degree
of frustration. Each pair $(i,j)$ of spins try to achieve
$\theta_i\minus\theta_j\plus \tilde{A}_{ij}=0$, so a
bond-frustrated loop $\{i_1,\ldots,i_L\}$ is one where
$\tilde{A}_{i_1 i_2}\plus\tilde{A}_{i_2 i_3}\plus \ldots \plus
\tilde{A}_{i_{L-1} i_L}\plus \tilde{A}_{i_L i_1}\neq 0$ (mod
$2\pi$). We simplify the bond average by choosing the $\{A_{ij}\}$
to be binary: $P(A)=\frac{1}{2}\delta[A\minus A^\star]
+\frac{1}{2}\delta[A\minus A^\star\minus\pi]$. Without loss of
generality we may take $-\frac{\pi}{2}<A^\star\leq\frac{\pi}{2}$.
However, only for $A^\star\in\{0,\frac{1}{2}\pi\}$ will it be
possible to calculate the disorder-averaged free energy for the
system (\ref{eq:Hamiltonian}) in terms of a standard replica mean
field theory; these two cases are the subject of our paper.

\begin{figure}[t]
\vspace*{-11mm}
\newcommand{\here}{\makebox(0,0)}
\newcommand{\blob}{\here{\circle*{5}}}
\newcommand{\point}{\here{\circle*{2}}}
\newcommand{\ua}{\mbox{\huge\boldmath $\uparrow$}}
\newcommand{\da}{\mbox{\huge\boldmath $\downarrow$}}
\newcommand{\la}{\mbox{\huge\boldmath $\leftarrow$}}
\newcommand{\ra}{\mbox{\huge\boldmath $\rightarrow$}}
\setlength{\unitlength}{0.20mm}

\begin{picture}(500,200)(-170,0)
\put(-180,60){Model I:} \put(-180,30){$A_{ij}\in\{0,\pi\}$}
 \put(-5,20){\blob}\put(55,20){\blob}\put(-5,20){\line(1,0){60}}
 \put(-17,11){\ua}\put(43,11){\ua}
 \put(-11,-16){\em 1}\put(49,-16){\em 2}
\put(-5,80){\blob}\put(55,80){\blob}
\multiput(-7,80)(5,0){12}{\point}
\put(-17,71){\ua}\put(43,71){\da} \put(-11,105){\em
1}\put(49,105){\em 2}
 \put(138,70){\ua}
 \put(125,30){\blob}\put(175,30){\blob}\put(150,79){\blob}
 \put(123,30){\line(1,2){25}}\put(173,30){\line(-1,2){25}} \put(125,30){\line(1,0){50}}
 \put(113,21){\ua}\put(163,21){\ua}
 \put(111,-6){\em 3}\put(173,-6){\em 2}\put(144,104){\em 1}
\put(238,70){\ua} \put(225,30){\blob}\put(275,30){\blob}
\put(250,79){\blob} \put(223,30){\line(1,2){25}}
\put(273,30){\line(-1,2){25}} \multiput(223,30)(5,0){10}{\point}
 \put(211,-6){\em 3}\put(273,-6){\em 2}\put(244,104){\em 1}
 \put(243,37){\Large ?}
\put(338,70){\ua} \put(325,30){\blob} \put(375,30){\blob}
\put(350,79){\blob} \multiput(323,30)(5,0){10}{\point}
 \put(323,30){\line(1,2){25}}
\multiput(348,79)(2.5,-5){10}{\point}
 \put(311,-6){\em 3}\put(373,-6){\em 2}\put(344,104){\em 1}
  \put(313,21){\ua}\put(363,21){\da}
\put(438,70){\ua} \put(425,30){\blob} \put(475,30){\blob}
\put(450,79){\blob} \multiput(423,30)(5,0){10}{\point}
\multiput(423,30)(2.5,5){10}{\point}
\multiput(473,30)(-2.5,5){10}{\point}
 \put(411,-6){\em 3}\put(473,-6){\em 2}\put(444,104){\em 1}
 \put(443,37){\Large ?}
\end{picture}
\vspace*{-5mm}

\begin{picture}(500,200)(-170,0)
\put(-180,60){Model II:}
\put(-180,30){$A_{ij}\in\{-\frac{\pi}{2},\frac{\pi}{2}\}$}
 \put(-5,20){\blob}\put(55,20){\blob}\put(-5,20){\line(1,0){60}}
 \put(-17,11){\ua}\put(31,9){\ra}
 \put(-11,-16){\em 1}\put(49,-16){\em 2}
\put(-5,80){\blob}\put(55,80){\blob}
\multiput(-7,80)(5,0){12}{\point}
\put(-17,71){\ua}\put(29,70){\la} \put(-11,105){\em
1}\put(49,105){\em 2}
 \put(138,70){\ua}
 \put(125,30){\blob}\put(175,30){\blob}\put(150,79){\blob}
 \put(123,30){\line(1,2){25}}\put(173,30){\line(-1,2){25}} \put(125,30){\line(1,0){50}}
 \put(143,37){\Large ?}
 \put(111,-6){\em 3}\put(173,-6){\em 2}\put(144,104){\em 1}
\put(238,70){\ua} \put(225,30){\blob}\put(275,30){\blob}
\put(250,79){\blob} \put(223,30){\line(1,2){25}}
\put(273,30){\line(-1,2){25}} \multiput(223,30)(5,0){10}{\point}
 \put(211,-6){\em 3}\put(273,-6){\em 2}\put(244,104){\em 1}
 \put(243,37){\Large ?}
\put(338,70){\ua} \put(325,30){\blob} \put(375,30){\blob}
\put(350,79){\blob} \multiput(323,30)(5,0){10}{\point}
 \put(323,30){\line(1,2){25}}
\multiput(348,79)(2.5,-5){10}{\point}
 \put(311,-6){\em 3}\put(373,-6){\em 2}\put(344,104){\em 1}
 \put(343,37){\Large ?}
\put(438,70){\ua} \put(425,30){\blob} \put(475,30){\blob}
\put(450,79){\blob} \multiput(423,30)(5,0){10}{\point}
\multiput(423,30)(2.5,5){10}{\point}
\multiput(473,30)(-2.5,5){10}{\point}
 \put(411,-6){\em 3}\put(473,-6){\em 2}\put(444,104){\em 1}
 \put(443,37){\Large ?}
\end{picture}

\vspace*{5mm}
 \caption{The low energy states of the
 two models (\ref{eq:model1}) (upper row) and (\ref{eq:model2}) (lower row),
 for $N\in\{2,3\}$ and
 in the absence of pinning fields. Arrows indicate the orientations of the
 vectors ${\bf
 S}_i=(\cos(\theta_i),\sin(\theta_i))$.
 Without pinning fields  only relative angles are important, so we may
  put $\theta_1=\frac{1}{2}\pi$.
 The bonds
$J_{ij}\in\{-1,1\}$ are drawn as solid ($1$) and dotted ($-1$)
line segments, respectively, which connect the vectors ${\bf
 S}_i$ and ${\bf S}_j$.
 Frustrated clusters, where no state exists which minimizes all terms in the
 Hamiltonian simultaneously,  are indicated with `?'.
In model I the triplet $(1,2,3)$ is bond-frustrated if an odd
number of the bonds $\{J_{12},J_{13},J_{23}\}$ equals $-1$. In
model II the triplet $(1,2,3)$ is {\em always} bond-frustrated.
 }
 \label{fig:0}
\end{figure}
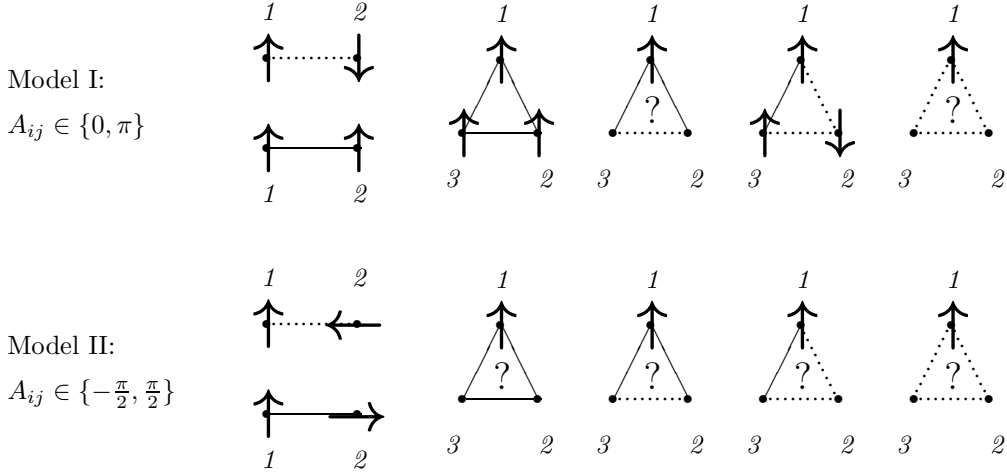

Our first (conventional) model corresponds to
$P(A_{ij})=\frac{1}{2}\delta[A_{ij}]
+\frac{1}{2}\delta[A_{ij}\minus\pi]$:
\be
{\rm Model~I:}~~~~~ H=-\frac{2K}{\sqrt{N}} \sum_{i<j}
J_{ij}\cos(\theta_i\minus\theta_j) -h\sum_i
\cos(\theta_i\minus\phi_i) \label{eq:model1} \ee with
$J_{ij}=\cos(A_{ij})\in\{-1,1\}$. In this model energy
minimization will translate for any pair $(i,j)$ of oscillators
into the objectives
\begin{eqnarray*}
J_{ij}=1: &~~~~& \theta_i\minus \theta_j \to 0~~~~~~({\rm
mod}~2\pi)\\  J_{ij}=-1: &~~~~& \theta_i\minus \theta_j \to
\pi~~~~~~({\rm mod}~2\pi)
\end{eqnarray*}
(i.e. ferro- and anti-ferromagnetic interactions, respectively).
The Hamiltonian (\ref{eq:model1}) defines a Gibbs measure
corresponding to the Langevin forces $f_i= -\partial
H/\partial\theta_i= - (2K/\sqrt{N}) \sum_{k\neq i}
J_{ik}\sin(\theta_i\minus\theta_k) - h\sin(\theta_i\minus\phi_i)$,
provided we define $J_{ij}=J_{ji}$.
 A triplet of spins $(i,j,k)$ is now bond-frustrated if an odd number of
the relative angles involved equals $\minus 1$, and unfrustrated
otherwise (see figure \ref{fig:0}). Our second (less conventional)
model is obtained for
$P(A_{ij})=\frac{1}{2}\delta[A_{ij}\minus\frac{\pi}{2}]+\frac{1}{2}\delta[A_{ij}\plus\frac{\pi}{2}]$:
\be
{\rm Model~II:}~~~~~ H=-\frac{2K}{\sqrt{N}}
\sum_{i<j}J_{ij}\sin(\theta_i\minus\theta_j)-h\sum_i\cos(\theta_i\minus\phi_i)
\label{eq:model2} \ee with $J_{ij}=-\sin(A_{ij})\in\{-1,1\}$. Here
energy minimization will translate for any pair $(i,j)$ of
oscillators into the objectives
\begin{eqnarray*}
J_{ij}=1: &~~~~& \theta_i\minus \theta_j \to ~\pi/2~~~~~~~({\rm
mod}~2\pi)\\
 J_{ij}=-1: &~~~~&
\theta_i\minus \theta_j \to -\pi/2~~~~~~({\rm mod}~2\pi)
\end{eqnarray*}
The Hamiltonian (\ref{eq:model1}) defines a Gibbs measure
corresponding to the Langevin forces $f_i=- \partial
H/\partial\theta_i=- (2K/\sqrt{N}) \sum_{k\neq
i}J_{ij}\cos(\theta_i\minus\theta_k)
-h\sin(\theta_i\minus\phi_i)$, provided we define $J_{ij}=\minus
J_{ji}$. Now {\em every} triplet $(i,j,k)$ is bond-frustrated,
since with $A_{ij}\in\{-\frac{\pi}{2},\frac{\pi}{2}\}$ one always
has $\tilde{A}_{ij}+\tilde{A}_{jk}+\tilde{A}_{ik}=\ell \pi/2$,
with $\ell$ integer and odd (see figure \ref{fig:0}).

It is not a priori clear whether and how the solutions of models
(\ref{eq:model1}) and (\ref{eq:model2}) are related. For instance,
finding their ground states for $N=3$ reduces to minimizing
$H_{\rm
I}(\theta_1,\theta_2)=J_1\cos(\theta_1)+J_2\cos(\theta_2)+J_3\cos(\theta_1\minus\theta_2)$
and $H_{\rm
II}(\theta_1,\theta_2)=J_1\sin(\theta_1)+J_2\sin(\theta_2)+J_3\sin(\theta_1\minus\theta_2)$,
respectively (since rotation invariance allows us to put
$\theta_3=0$), with $J_i\in\{-1,1\}$ (chosen randomly). Solving
this simple problem reveals that the ground state energy $E_{\rm
I}$ of $H_{\rm I}$ is dependent on the bonds $\{J_i\}$, $E_{\rm
I}\in\{\minus 1,\minus 3\}$. Averaging over the bond realizations
gives $\overline{E_{\rm I}}=-2$. In contrast, for the ground state
energy of model II one finds $E_{\rm II}=-\frac{3}{2}\sqrt{3}$,
independent of the bond realization.

Pinning fields break the global rotation invariance of
(\ref{eq:Hamiltonian}) which is present  for $h=0$, and
simplifications of our equations will thus have to be based on
symmetries of the pinning field distribution. Any non-zero value
of the average pinning direction $\bra\phi\ket_\phi$ can be
transformed away, via $\theta_i\to \theta_i+\bra \phi\ket_\phi$,
without affecting the interaction terms in our Hamiltonian. Thus,
without loss of generality we may choose $\bra \phi\ket_\phi=0$.
In the remainder of this paper we restrict ourselves, for
simplicity, to reflection-symmetric pinning field distributions,
where $p(\phi)=p(\minus \phi)$ for all $\phi\in[\minus\pi,\pi]$.
However, the free energies of our models
(\ref{eq:model1},\ref{eq:model2}) are both invariant\footnote{The
transformation (\ref{eq:gauge}) implies $\phi_i\to
\phi_i+\lambda+\tau_i\frac{\pi}{2}$, where $\tau_i\in\{-1,1\}$.
The variables $\lambda$ and $\{\tau_i\}$ can be gauged away by
putting $\theta_i\to \theta_i+\lambda+\tau_i\frac{\pi}{2}$ in
combination with $J_{ij}\to
J_{ij}\cos(\frac{\pi}{2}(\tau_i-\tau_j))$.}
 under the
transformation
\be
p(\phi)~\rightarrow~\frac{1}{2}(1\plus\eta)p(\phi-\lambda-\frac{\pi}{2})+
\frac{1}{2}(1\minus\eta)p(\phi-\lambda+\frac{\pi}{2})
\label{eq:gauge} \ee for any $\eta\in[\minus 1,1]$ and
$\lambda\in[0,2\pi]$, so we solve implicitly for a larger family
of models. For high temperatures we expect the symmetries of
$p(\phi)$ to be inherited by the solutions of our models. We
define a complex (single-site) disorder-averaged susceptibility
$\chi$, upon adding small perturbations to the external pinning
angles $\phi_i$:
\be
\chi=\frac{1}{iNh}\sum_i \overline{\left[
e^{-i\phi_i}\frac{\partial}{\partial \phi_i}\bra
e^{i\theta_i}\ket\right]} \label{eq:chi} \ee Here $\bra\ldots\ket$
and $\overline{[\ldots]}$  denote equilibrium and disorder
averages, respectively. For perfectly linear single-site response,
i.e. $\bra e^{i\theta_i}\ket=x ~e^{i\phi_i}$, definition
(\ref{eq:chi}) would give $\chi=x/h$. Evaluating (\ref{eq:chi})
for systems with Hamiltonians of the form (\ref{eq:Hamiltonian})
gives
\be
\chi=\frac{\beta}{iN}\sum_i\overline{\left[\room \bra
\sin(\theta_i\minus\phi_i) e^{i(\theta_i-\phi_i)}\ket-
\bra\sin(\theta_i\minus\phi_i)\ket\bra e^{i(\theta_i-\phi_i)}\ket
\right]} \label{eq:chi_workedout} \ee

\section{Equilibrium analysis}

\subsection{Solution via replica theory}

We calculate the disorder-averaged free energy per spin
$\overline{f}=-\lim_{N\to\infty}(\beta N)^{-1}\overline{\log Z}$
for the Hamiltonian (\ref{eq:Hamiltonian}), using the identity
$\overline{\log Z}=\lim_{n\to 0}n^{-1}\log \overline{Z^n}$ and the
standard manipulations of conventional replica theory
\cite{Mezardetal}, giving \bd \overline{f} =
-\lim_{N\to\infty}\lim_{n\to 0}\frac{1}{\beta Nn}\log
\int\!\!\!\cdots\!\!\int\!\left[ \prod_{\alpha=1}^n
d\btheta^\alpha \right] \overline{e^{-\beta\sum_\alpha
H(\btheta^\alpha)}} \ed The disorder average, with
$P(A)=\frac{1}{2}\delta[A\minus A^\star]+\frac{1}{2}\delta[A\minus
A^\star\minus \pi]$,
 gives:
\bd
\overline{e^{-\beta \sum_{\alpha} H(\btheta^\alpha)}}
=
e^{\beta h\sum_{i\alpha}\cos(\theta^\alpha_i\minus\phi_i)}
\prod_{i<j}\cosh[\frac{2\beta K}{\sqrt{N}}\sum_\alpha
\cos(\theta^\alpha_i\minus\theta^\alpha_j\plus A^\star)] \ed
Expansion for large $N$, however, only leads to an expression in
terms of the usual single-site replica order parameters
 if $\sin(2A^\star)=0$, i.e. for
$A^\star\in\{0,\frac{\pi}{2}\}$ (\ref{eq:model1},\ref{eq:model2}).
Here \bd
 \hspace*{-25mm}
 \overline{e^{-\beta \sum_{\alpha} H(\btheta^\alpha)}}
 = e^{\beta h\sum_{i\alpha}\cos(\theta^\alpha_i\minus\phi_i) +2\beta^2 K^2
 N\sin^2(A^\star) \sum_{\alpha\beta} \left[
 q^{ss}_{\alpha\beta}(\btheta) q^{cc}_{\alpha\beta}(\btheta) -
 q^{sc}_{\alpha\beta}(\btheta) q^{cs}_{\alpha\beta}(\btheta)\right]
 + \order(N^{0})}
\ed
\bd
 \hspace*{-5mm}\times~e^{\beta^2
 K^2 N\cos^2(A^\star) \sum_{\alpha\beta} \left[
q^{cc}_{\alpha\beta}(\btheta)q^{cc}_{\alpha\beta}(\btheta) +
q^{ss}_{\alpha\beta}(\btheta) q^{ss}_{\alpha\beta}(\btheta) +
q^{cs}_{\alpha\beta}(\btheta) q^{cs}_{\alpha\beta}(\btheta) +
q^{sc}_{\alpha\beta}(\btheta) q^{sc}_{\alpha\beta}(\btheta)
\right]} \ed with \vspace*{-3mm}
 \begin{eqnarray*}
 q^{cc}_{\alpha\beta}(\btheta)=
\frac{1}{N}\sum_i\cos(\theta_i^\alpha)\cos(\theta_i^\beta),&&~~~~~~
q^{ss}_{\alpha\beta}(\btheta)=
\frac{1}{N}\sum_i\sin(\theta_i^\alpha)\sin(\theta_i^\beta) \\
q^{sc}_{\alpha\beta}(\btheta)=
\frac{1}{N}\sum_i\sin(\theta_i^\alpha)\cos(\theta_i^\beta),&&~~~~~~
q^{cs}_{\alpha\beta}(\btheta)=
\frac{1}{N}\sum_i\cos(\theta_i^\alpha)\sin(\theta_i^\beta)
\end{eqnarray*} We proceed as usual, isolating
observables via appropriate $\delta$-distributions (which
introduces conjugate order parameters), leading to site
factorisation. The result is
\be
\overline{f} = -\lim_{n\to 0}\frac{1}{n}\extr \left\{
\Phi[\{\hbq^{\star\star},\bq^{\star\star}\}]+\Psi[\{\hbq^{\star\star}\}]
\right\} \label{eq:f_rsb} \ee
\begin{eqnarray}
\Phi[\ldots]&=&
\frac{i}{\beta}\sum_{\star\star}\sum_{\alpha\beta}\hq^{\star\star}_{\alpha\beta}
q^{\star\star}_{\alpha\beta} +\beta K^2
\cos^2(A^\star)\sum_{\star\star}\sum_{\alpha\beta}
[q^{\star\star}_{\alpha\beta}]^2 \nonumber
\\
&&
 +2\beta K^2 \sin^2(A^\star)
\sum_{\alpha\beta}\left[q^{ss}_{\alpha\beta} q^{cc}_{\alpha\beta}
- q^{sc}_{\alpha\beta} q^{cs}_{\alpha\beta}\right] \label{eq:Phi}
\\
 \beta\Psi[\ldots]
&=&
 \bra \log\int\!d\btheta~M(\btheta|\{\hbq^{\star\star}\})\ket_\phi
\label{eq:Psi}
\end{eqnarray}
with $(\star\star)\in\{(cc),(ss),(sc),(cs)\}$, and with
\begin{eqnarray}
\hspace*{-5mm} M(\btheta|\{\hbq^{\star\star}\}) &=&
 e^{\beta
h\sum_{\alpha}\cos(\theta^\alpha\minus\phi) -i\sum_{\alpha\beta}[
\hq^{cc}_{\alpha\beta}\cos(\theta_\alpha)\cos(\theta_\beta)
+\hq^{cs}_{\alpha\beta}\cos(\theta_\alpha)\sin(\theta_\beta)]}
\nonumber
\\
&& \times
 e^{-i\sum_{\alpha\beta}[
\hq^{ss}_{\alpha\beta}\sin(\theta_\alpha)\sin(\theta_\beta)
+\hq^{sc}_{\alpha\beta}\sin(\theta_\alpha)\cos(\theta_\beta)]}
\end{eqnarray}
 One always has
$q^{cs}_{\lambda\rho}=q^{sc}_{\rho\lambda}$, which allows us to
simplify the saddle point equations. For both models we find upon
varying the conjugate order parameters
$\{\hat{q}_{\alpha\beta}^{\star\star}\}$:
\be
q^{cc}_{\lambda\rho}
=
\bigbra
\frac{\int\!d\btheta~\cos(\theta_\lambda)\cos(\theta_\rho)
M(\btheta|\{\hbq^{\star\star}\})}
{\int\!d\btheta~
M(\btheta|\{\hbq^{\star\star}\})
}
\bigket_{\!\!\phi}
\label{eq:fullsaddle_3}
\ee
\be
q^{ss}_{\lambda\rho}
=
\bigbra
\frac{\int\!d\btheta~\sin(\theta_\lambda)\sin(\theta_\rho)
M(\btheta|\{\hbq^{\star\star}\})}
{\int\!d\btheta~
M(\btheta|\{\hbq^{\star\star}\})}
\bigket_{\!\!\phi}
\label{eq:fullsaddle_4}
\ee
\be
q^{sc}_{\lambda\rho}
=
\bigbra \frac{\int\!d\btheta~\sin(\theta_\lambda)\cos(\theta_\rho)
M(\btheta|\{\hbq^{\star\star}\})}
{\int\!d\btheta~M(\btheta|\{\hbq^{\star\star}\})}
\bigket_{\!\!\phi} \label{eq:fullsaddle_5} \ee The models I
($A^\star=0$) and II ($A^\star=\frac{\pi}{2}$) differ only in the
equations resulting from variation of the order parameters
 $\{q_{\alpha\beta}^{\star\star}\}$:
\begin{eqnarray}
{\rm Model~I:}&~~~~&
 \begin{array}{lll}
 \hq^{cc}_{\alpha\beta}= 2i\beta^2 K^2 q^{cc}_{\alpha\beta},&~~~&
 \hq^{ss}_{\alpha\beta}= 2i\beta^2 K^2 q^{ss}_{\alpha\beta}\\[1mm]
 \hq^{cs}_{\alpha\beta} =2i\beta^2 K^2 q^{cs}_{\alpha\beta},&~~~&
 \hq^{sc}_{\alpha\beta} =2i\beta^2 K^2 q^{sc}_{\alpha\beta}
 \end{array}
 \label{eq:hateqns_I} \\[1mm]
 {\rm Model~I\!I:}&~~~~&
 \begin{array}{lll}
 \hq^{cc}_{\alpha\beta}= 2i\beta^2 K^2 q^{ss}_{\alpha\beta},&&
 \hq^{ss}_{\alpha\beta}= 2i\beta^2 K^2 q^{cc}_{\alpha\beta}
 \\[1mm]
 \hq^{cs}_{\alpha\beta} =-2i\beta^2 K^2 q^{sc}_{\alpha\beta},&&
 \hq^{sc}_{\alpha\beta}=-2i\beta^2 K^2 q^{cs}_{\alpha\beta}
 \end{array}
\label{eq:hateqns_II}
\end{eqnarray}

\subsection{Replica-symmetric solutions}

We now make the replica-symmetric (or ergodic) ansatz (RS) for the
saddle-points:
\be
q_{\alpha\beta}^{\star\star}= (Q_{\star\star}\minus
q_{\star\star})\delta_{\alpha\beta} +q_{\star\star} \ee
\be
\hq^{\star\star}_{\alpha\beta}=\frac{1}{2}i(2\beta K)^2
 (R_{\star\star}\minus r_{\star\star})\delta_{\alpha\beta}
 +\frac{1}{2}i(2\beta K)^2 r_{\star\star}
\label{eq:rs_ansatz} \ee Insertion into the saddle-point equations
tells us that \bd
R_{cc}=\cos^2(A^\star)Q_{cc}\plus\sin^2(A^\star)Q_{ss}~~~~~~~
r_{cc}=\cos^2(A^\star)q_{cc}\plus\sin^2(A^\star)q_{ss} \ed \bd
R_{ss}=\cos^2(A^\star)Q_{ss}\plus\sin^2(A^\star)Q_{cc}~~~~~~~
r_{ss}=\cos^2(A^\star)q_{ss}\plus\sin^2(A^\star)q_{cc} \ed \bd
R_{cs}=\cos(2A^\star)Q_{cs}~~~~~~~~~~~~~~~~~~
~~~~~~~r_{cs}=\cos(2A^\star)q_{cs} \ed
 The identity $Q_{cc}\plus Q_{ss}=1$ allows us to reduce the problem
to solving five coupled non-linear equations. We will compactify
our notation by using the short-hands
\bd
 \bras
 \ldots\kets=\!\int\!\!DxDyDuDv \ldots~~~~~~~ \bra
 f(\theta)\ket_\star=\frac{\int\!d\theta~f(\theta)M(\theta|x,y,u,v,\phi)}
 {\int\!d\theta~M(\theta|x,y,u,v,\phi)}
\ed We can write the five remaining saddle-point equations as
\be
Q_{cc}=
\bra\bras ~\bra \cos^2(\theta)\ket_\star ~\kets\ket_{\phi}
~~~~~~~~
Q_{cs}=
\bra\bras ~\bra \cos(\theta)\sin(\theta)\ket_\star~ \kets\ket_{\phi}
\label{eq:saddle_Qs}
\ee
\be
q_{cc}= \bra\bras~\bra\cos(\theta)\ket_\star^2~\kets\ket_{\phi}
~~~~~~~~~~~ q_{ss}=
\bra\bras~\bra\sin(\theta)\ket_\star^2~\kets\ket_{\phi} \ee
\be
q_{cs}= \bra\bras~\bra\cos(\theta)\ket_\star
\bra\sin(\theta)\ket_\star~ \kets\ket_{\phi} \label{eq:saddle_qs}
\ee
Tracing back the physical meaning of these observables gives:
\begin{eqnarray}
\hspace*{-15mm}
 Q_{cc}&\!=\!&\lim_{N\to\infty}\frac{1}{N}\sum_i \overline{\bra
 \cos^2(\theta_i)\ket},~~~~~~
 Q_{cs}=\lim_{N\to\infty}\frac{1}{N}\sum_i \overline{\bra
 \cos(\theta_i)\sin(\theta_i)\ket} \label{eq:meaning_1}
\\
\hspace*{-15mm}
 q_{cc}&\!=\!&\lim_{N\to\infty}\frac{1}{N}\sum_i \overline{\bra
 \cos(\theta_i)\ket^2},~~~~~~
 q_{ss}=\lim_{N\to\infty}\frac{1}{N}\sum_i \overline{\bra
 \sin(\theta_i)\ket^2} \label{eq:meaning_2}
\\
\hspace*{-15mm}
 q_{cs}&\!=\!&\lim_{N\to\infty}\frac{1}{N}\sum_i \overline{\bra
 \cos(\theta_i)\ket \bra \sin(\theta_i)\ket} \label{eq:meaning_3}
\end{eqnarray}
Our two models differ only in the form taken by the effective
single spin measure $M(\theta|x,y,u,v,\phi)$ in the RS
saddle-point equations:
\begin{eqnarray}
 \hspace*{-15mm}
 M_{\rm
 I}(\theta|x,y,u,v,\phi)&=& e^{\beta h\cos(\theta\minus\phi) +(\beta
 K)^2\left[ [2Q_{cc}-1+q_{ss}-q_{cc}]\cos(2\theta)
 +2(Q_{cs}-q_{cs})\sin(2\theta) \right]}
\nonumber
\\
&& \hspace*{-25mm}\times~e^{\beta K \cos(\theta)\sqrt{2}[x
\sqrt{2}\sqrt{q_{cc}}
 +(u-iv)\sqrt{q_{cs}}]+\beta K
\sin(\theta)\sqrt{2}[y\sqrt{2}\sqrt{q_{ss}} +(u+iv)\sqrt{q_{cs}}]}
\label{eq:RSMeasureI}
\\
 \hspace*{-15mm}
 M_{\rm I\!I}(\theta|x,y,u,v,\phi)&=&
 e^{\beta h\cos(\theta\minus\phi) -(\beta K)^2\left[
 [2Q_{cc}-1+q_{ss}-q_{cc}]\cos(2\theta)
 +2(Q_{cs}-q_{cs})\sin(2\theta) \right]}
 \nonumber
\\
&& \hspace*{-25mm}\times~e^{\beta K \cos(\theta)\sqrt{2}[x
\sqrt{2}\sqrt{q_{ss}}
 +(iu+v)\sqrt{q_{cs}}]+\beta K
\sin(\theta)\sqrt{2}[y\sqrt{2}\sqrt{q_{cc}}
+(iu-v)\sqrt{q_{cs}}]}
\label{eq:RSMeasureII}
\end{eqnarray}
(apart, in both cases, from a constant pre-factor $e^{(\beta
K)^2[1-q_{cc}-q_{ss}]}$). To compactify future notation we define
the following short-hands
\be
\gamma_{cc}=\bra
\cos^2(\theta)\ket_\star\minus\bra\cos(\theta)\ket_\star^2
~~~~~~~~~~~~~ \gamma_{ss}=\bra
\sin^2(\theta)\ket_\star\minus\bra\sin(\theta)\ket_\star^2
\label{eq:gammas1} \ee
\be
\gamma_{cs}=\bra
\cos(\theta)\sin(\theta)\ket_\star\minus\bra\cos(\theta)\ket_\star\bra
\sin(\theta)\ket_\star \label{eq:gammas2} \ee Note that in
replica-symmetric states one has
 $\bra\bras\gamma_{cc}\kets\ket_\phi=Q_{cc}\minus q_{cc}$,
 $\bra\bras\gamma_{ss}\kets\ket_\phi=1\minus Q_{cc}\minus q_{ss}$,
$\bra\bras\gamma_{cs}\kets\ket_\phi=Q_{cs}\minus q_{cs}$, and
$|\gamma_{cs}|\leq \frac{1}{2}(\gamma_{cc}+\gamma_{ss})$. \vsp

Similarly we can work out the RS free energy per oscillator,  by
insertion of the RS ansatz into (\ref{eq:f_rsb}), followed by
standard manipulations. This results in
\be
\hspace*{-10mm}
 \overline{f}={\rm
 extr}_{\{Q_{\star\star},q_{\star\star}\}}f[\{Q_{\star\star},q_{\star\star}\}]
 \label{eq:fRS}
 \ee
\be
\hspace*{-10mm}
 f[\ldots]=\beta K^2
U -\frac{1}{\beta}\int\!Dx Dy Du Dv\bigbra
\log\int\!d\theta~M(\theta|x,y,u,v,\phi)\bigket_{\!\!\phi}
 \ee with the effective measures
$M(\theta|\ldots)$ defined in
(\ref{eq:RSMeasureI},\ref{eq:RSMeasureII}), and with
\begin{eqnarray}
U_{\rm I}&=& 2Q_{cc}(Q_{cc}\minus 1)+2Q_{cs}^2 -
q_{cc}(q_{cc}\minus 1) -q_{ss}(q_{ss}\minus 1)-2q_{cs}^2
 \label{eq:UI}
\\
U_{\rm I\!I}&=&-2Q_{cc}(Q_{cc}\minus 1)-1-2Q_{cs}^2 + q_{cc}
+q_{ss}-2q_{cc}q_{ss}+2q_{cs}^2
 \label{eq:UII}
\end{eqnarray}
The nature of the extremum in (\ref{eq:fRS}) follows from the
high-temperature state. Expanding for $\beta\to 0$ gives
$f[\ldots]=\minus \frac{1}{\beta}\log(2\pi)\minus \frac{1}{4}\beta
h^2\plus 2\beta
 K^2 {\rm extr}\{\tilde{U}\}\plus \order(\beta^2)$, with
\begin{eqnarray*}
\tilde{U}_{\rm I}&=&
 Q_{cc}^2-Q_{cc}+Q_{cs}^2 - \frac{1}{2}(q^2_{cc} \plus q^2_{ss})-q_{cs}^2
\\
 \tilde{U}_{\rm I\!I}&=&
-Q^2_{cc}+Q_{cc} -\frac{1}{2}-Q_{cs}^2 -q_{cc}q_{ss}+q_{cs}^2
\end{eqnarray*}
For $\beta\to 0$ we must find the paramagnetic solution
$Q_{cc}=\frac{1}{2}$ and $Q_{cs}=q_{cc}=q_{ss}=q_{cs}=0$, so  the
nature of the desired extremum in (\ref{eq:fRS}) is
\begin{eqnarray}
{\rm I:} &~~~{\rm min~w.r.t.}~\{Q_{cc},Q_{cs}\}, &~~{\rm
max~w.r.t.}~\{q_{cc},q_{ss},q_{cs}\} \label{eq:extremumI}
\\
{\rm I\!I:} &~~~{\rm min~w.r.t.}~\{q_{cs},q_{cc}\minus q_{ss}\},
&~~{\rm max~w.r.t.}~\{Q_{cc},Q_{cs},q_{cc}\plus q_{ss}\}
\label{eq:extremumII}
\end{eqnarray}
Finally,  for RS solutions we can also work out expression
(\ref{eq:chi_workedout}) for $\chi$ in the limit $N\to\infty$:
\begin{eqnarray*}
\frac{\chi_{\rm RS}}{\beta}&=&  \bra \bras~ \bra
\sin^2(\theta\minus\phi)\ket_\star\minus
\bra\sin(\theta\minus\phi)\ket_\star^2~\kets\ket_\phi  \nonumber
\\
&& \hspace*{-6mm}  -i~\bra \bras ~\bra
\sin(\theta\minus\phi)\cos(\theta\minus\phi)\ket_\star\minus
\bra\sin(\theta\minus\phi)\ket_\star\bra\cos(\theta\minus\phi)\ket_\star~\kets\ket_\phi
\end{eqnarray*} Equivalently, with the conventions
(\ref{eq:gammas1},\ref{eq:gammas2}) this can be written as
\begin{eqnarray}
\frac{\chi_{\rm RS}}{\beta}&=&\frac{1}{2}(1\minus q_{cc}\minus
q_{ss}) -\bra\bras~
\frac{1}{2}\cos(2\phi)(\gamma_{cc}\minus\gamma_{ss})
+\sin(2\phi)\gamma_{cs}~\kets\ket_{\phi} \nonumber\\ &&
-i~\bra\bras~
\cos(2\phi)\gamma_{cs}-\frac{1}{2}\sin(2\phi)(\gamma_{cc}\minus
\gamma_{ss})~\kets\ket_{\phi}
 \label{eq:chi_rs} \end{eqnarray}

\subsection{The AT lines}

Here we determine the stability of our RS solution against
`replicon' perturbations: $q^{\star\star}_{\alpha\beta}\to
q^{\star\star,\rm
RS}_{\alpha\beta}+\eta^{\star\star}_{\alpha\beta}$ with
$\eta^{cc}_{\alpha\beta}= \eta^{cc}_{\beta\alpha}$ and
 $\eta^{ss}_{\alpha\beta}=\eta^{ss}_{\beta\alpha}$
(for all $\alpha,\beta$), $\eta^{\star\star}_{\alpha\alpha}=0$ for
all $\alpha$, $\sum_{\alpha}\eta^{\star\star}_{\alpha\beta}=
\sum_{\alpha}\eta^{\star\star}_{\beta\alpha}= 0$ for all $\alpha$,
and with all $|\eta^{\star\star}_{\alpha\beta}|\ll 1$. Details of
the derivation of the equations signaling the this type of
instability (the AT \cite{AT} lines), are given in \ref{app:AT}.
Our calculation is more complicated than that in e.g.
\cite{Elder82,Elder83}, because, due to the pinning fields, we
cannot generally use rotational symmetry; replicon fluctuations
can now have a more complicated structure. We found two types of
AT instabilities (the physical RSB transition associated with
these is the one occurring at the highest temperature). The first
is the same for our two models:
\be
{\rm Models~I~\&~I\!I}:~~~~~~
(T/2K)^{2}=\bra\bras\gamma_{cc}\gamma_{ss}\minus\gamma_{cs}^2\kets\ket_{\phi}
\label{eq:ATmodelIandII} \ee The second AT instability is found to
be model dependent:
\begin{eqnarray}
{\rm Model~I:} &~~~~~~~& {\rm Det}[\bE-(T/2K)^2\unity]=0
 \label{eq:ATmodelI}
\\
{\rm Model~I\!I:} &~~~~~~~& {\rm Det}[\bE-(T/2K)^2\bC]=0
\label{eq:ATmodelII}
\end{eqnarray}
with
\be
\bC= \left(\begin{array}{ccc}1 & 0 & 0\\ 0 & -1 & 0 \\ 0 & 0 & -1
\end{array}
\right) \label{eq:matrixC} \ee
\be
\hspace*{-22mm} \bE=\left(\!\begin{array}{ccc} \bra\bras
\frac{1}{2}(\gamma_{cc}^2\plus \gamma_{ss}^2)\plus
\gamma_{cs}^2\kets\ket_\phi &
\bra\bras\frac{1}{2}(\gamma_{cc}^2\minus
\gamma_{ss}^2)\kets\ket_\phi &
\bra\bras\gamma_{cs}(\gamma_{cc}\plus \gamma_{ss})\kets\ket_\phi
\\[1mm]
\bra\bras \frac{1}{2}(\gamma_{cc}^2\minus
\gamma_{ss}^2)\kets\ket_\phi &
\bra\bras\frac{1}{2}(\gamma_{cc}^2\plus \gamma_{ss}^2)\minus
\gamma_{cs}^2\kets\ket_\phi &
\bra\bras\gamma_{cs}(\gamma_{cc}\minus \gamma_{ss})\kets\ket_\phi
\\[1mm]
\bra\bras \gamma_{cs}(\gamma_{cc}\plus\gamma_{ss})\kets\ket_\phi &
\bra\bras \gamma_{cs}(\gamma_{cc}\minus \gamma_{ss})\kets\ket_\phi
& \bra\bras \gamma_{cc}\gamma_{ss}\plus\gamma_{cs}^2\kets\ket_\phi
\end{array}\!
\right) \label{eq:matrixE}
 \ee

\section{States with global reflection symmetry}

We inspect the effect of global reflection $\theta_i\to
-\theta_i$, on the order parameters (within the RS  ansatz), with
the help of the identifications
(\ref{eq:meaning_1},\ref{eq:meaning_2},\ref{eq:meaning_3}):
\be
\theta_i^\prime=- \theta_i ~~{\rm for~ all}~~i: ~~~~~
\begin{array}{ccccc}
 Q_{cc}^\prime=Q_{cc}, & Q_{cs}^\prime=-Q_{cs}, &\\
 q_{cc}^\prime=q_{cc}, & q_{cs}^\prime=-q_{cs}, & q_{ss}^\prime=q_{ss}
\end{array}
\label{eq:reflection} \ee Invariance under this transformation
implies $Q_{cs}=q_{cs}=0$, and breaking of the associated symmetry
is signaled by a bifurcation of $Q_{cs}\neq 0$ and/or $q_{cs}\neq
0$.

\subsection{Implications for free energy and order parameters}

For reflection-symmetric states the Gaussian variables $\{u,v\}$
disappear from the problem, and the effective single-spin measures
(\ref{eq:RSMeasureI},\ref{eq:RSMeasureII}) simplify to
\begin{eqnarray}
\hspace*{-20mm}
 M_{\rm I}(\theta|x,y,\phi)&=& e^{\beta
h\cos(\theta\minus\phi) +(\beta
K)^2[2Q_{cc}-1+q_{ss}-q_{cc}]\cos(2\theta) +2\beta K[\cos(\theta)x
\sqrt{q_{cc}} +\sin(\theta)y \sqrt{q_{ss}}]} \nonumber
\\
\label{eq:symRSMeasureI}
\\
\hspace*{-20mm} M_{\rm I\!I}(\theta|x,y,\phi)&=&
 e^{\beta h\cos(\theta\minus\phi)
-(\beta K)^2[2Q_{cc}-1+q_{ss}-q_{cc}]\cos(2\theta) +2\beta
K[\cos(\theta)x\sqrt{q_{ss}} +\sin(\theta)y\sqrt{q_{cc}}]}
\nonumber\\
 \label{eq:symRSMeasureII}
\end{eqnarray}
These expressions show, in combination with the underlying
symmetry $p(\phi)=p(-\phi)$, that for any set of functions
$\{k_\ell\}$ (with $\ell=0,1,2,\ldots$):
\be
\bra \bras ~k_0(-\phi)\prod_{\ell>0}\bra
k_\ell(-\theta)\ket_\star~\kets\ket_{\phi}= \bra \bras
~k_0(\phi)\prod_{\ell>0}\bra
k_\ell(\theta)\ket_\star~\kets\ket_{\phi}
\label{eq:reflection_symmetry} \ee One is left with just three
coupled order parameter equations: \be \hspace*{-10mm} Q_{cc}=
\bra\bras \bra \cos^2(\theta)\ket_\star \kets\ket_{\phi}~~~~~~
 q_{cc}=
\bra\bras\bra\cos(\theta)\ket_\star^2\kets\ket_{\phi}~~~~~~
q_{ss}= \bra\bras\bra\sin(\theta)\ket_\star^2\kets\ket_{\phi}
\label{eq:symm_saddle}
 \ee
Inserting $Q_{cs}=q_{cs}=0$ into (\ref{eq:fRS},\ref{eq:UI},
\ref{eq:UII}) shows that for reflection-symmetric states the free
energy per oscillator equals
$\overline{f}=\extr_{\{Q_{cc},q_{ss},q_{cc}\}}f[Q_{cc},q_{ss},q_{cc}]$,
with
\begin{eqnarray*}
f_{\rm I}[Q_{cc},q_{ss},q_{cc}] &=&
 \beta K^2\left[ q_{cc}\plus
 q_{ss} \plus 2Q_{cc}(Q_{cc}\minus 1) \minus q^2_{cc}\minus
 q^2_{ss} \right] \nonumber\\ &&
 \hspace*{10mm}
 -\frac{1}{\beta}\bigbra\int\!\!DxDy~\log
 \int\!d\theta~ M_{\rm I}(\theta|x,y,\phi)
 \bigket_{\!\!\phi}
\\
f_{\rm I\!I}[Q_{cc},q_{ss},q_{cc}] &=&\beta K^2\left[ q_{cc}\plus
 q_{ss} \minus 2Q_{cc}(Q_{cc}\minus 1) \minus 2q_{ss}q_{cc} \minus
 1\right] \nonumber\\ &&
 \hspace*{10mm}
 -\frac{1}{\beta} \bigbra\int\!\!DxDy~\log
 \int\!d\theta~ M_{\rm I\!I}(\theta|x,y,\phi)
 \bigket_{\!\!\phi}
\end{eqnarray*}
 From
these expressions one can  easily extract the replica symmetric
ground states, using a saddle-point argument as $\beta\to\infty$
in the entropic term:
\begin{eqnarray*}
\hspace*{-10mm} \lim_{\beta\to\infty} f_{\rm I}[\ldots]/ \beta K^2
&=&
 q_{cc}\plus q_{ss} \plus 2Q_{cc}(Q_{cc}\minus 1)
\minus q^2_{cc}\minus q^2_{ss} \minus |2Q_{cc}\minus 1\plus
q_{ss}\minus q_{cc}|
\\
\hspace*{-10mm} \lim_{\beta\to\infty}f_{\rm I\!I}[\ldots]/\beta
K^2 &=& q_{cc}\plus q_{ss} \minus 2Q_{cc}(Q_{cc}\minus 1) \minus
2q_{ss}q_{cc} \minus 1 \minus |2Q_{cc}\minus 1\plus q_{ss}\minus
q_{cc}|
\end{eqnarray*}
According to (\ref{eq:extremumI},\ref{eq:extremumII}), we must for
model I minimize with respect to $Q_{cc}$ and maximize with
respect to $\{q_{cc},q_{ss}\}$, whereas for model II we must
minimize with respect to $q_{cc}\minus q_{ss}$ and maximize with
respect to $\{Q_{cc},q_{cc}\plus q_{ss}\}$. For both models this
leads to
\be
q_{cc}\plus q_{ss}=1,~~~~~~Q_{cc}=\frac{1}{2}[1\plus q_{cc}\minus
q_{ss}] \label{eq:groundstate} \ee giving in both cases
$\extr_{Q_{cc},q_{ss},q_{cc}} \lim_{\beta\to\infty} (\beta
K^2)^{-1} f[Q_{cc},q_{ss},q_{cc}]=0$. Hence there is a finite
ground state energy.
 To find the RS ground states of our two models from the family
$q_{cc}\plus q_{ss}=1$, one would have to inspect the next order
in $T$. In addition we must, of course, expect replica symmetry to
be broken for low temperatures.

\subsection{Implications for AT lines and RS susceptibility}

Reflection symmetry is also found to simplify expressions
(\ref{eq:ATmodelI},\ref{eq:ATmodelII}) for the AT instabilities,
due to  $\bra\bras \gamma_{cc}\gamma_{cs}\kets\ket_\phi= \bra\bras
\gamma_{ss}\gamma_{cs}\kets\ket_\phi=0$. In particular, with
(\ref{eq:reflection_symmetry}) one can simplify the matrix
(\ref{eq:matrixE}) to
\be
\hspace*{-22mm} \bE=\left(\!\begin{array}{ccc} \bra\bras
\frac{1}{2}(\gamma_{cc}^2\plus \gamma_{ss}^2)\plus
\gamma_{cs}^2\kets\ket_\phi &
\bra\bras\frac{1}{2}(\gamma_{cc}^2\minus
\gamma_{ss}^2)\kets\ket_\phi & 0
\\[1mm]
\bra\bras \frac{1}{2}(\gamma_{cc}^2\minus
\gamma_{ss}^2)\kets\ket_\phi &
\bra\bras\frac{1}{2}(\gamma_{cc}^2\plus \gamma_{ss}^2)\minus
\gamma_{cs}^2\kets\ket_\phi & 0
\\[1mm]
0 & 0 & \bra\bras
\gamma_{cc}\gamma_{ss}\plus\gamma_{cs}^2\kets\ket_\phi
\end{array}\!
\right) \label{eq:simplermatrixE}
 \ee
 In combination with
(\ref{eq:ATmodelIandII}), and upon rejecting solutions which are
immediately seen not to give the highest transition temperature,
the AT lines of our two models are found to be the solutions of
the following equations, respectively: \begin{eqnarray}
\hspace*{-25mm} {\rm I:}~~~~
  \left[\frac{T}{2K}\right]^2\!
= \max &\!\!\!& \left\{ \frac{1}{2}\bra\!\bras
 \gamma^2_{cc}\plus\gamma^2_{ss}\kets\!\ket_{\phi}
 \plus\sqrt{
 \bra\!\bras\gamma_{cs}^2\kets\!\ket_{\phi}^2\plus
 \frac{1}{4}
 \bra\!\bras\gamma^2_{cc}\minus\gamma^2_{ss}\kets\!\ket_{\phi}^2}~,
 \bra\!\bras\gamma_{cc}\gamma_{ss}\plus \gamma_{cs}^2
 \kets\!\ket_{\phi} \!\right\}
 \nonumber\\
 &&\label{eq:pinning_ATalmost1}
 \\
 \hspace*{-25mm} {\rm I\!I:} ~~~
 \left[\frac{T}{2K}\right]^2\! = \max &\!\!\!& \left\{
 \bra\!\bras\gamma_{cs}^2\kets\!\ket_{\phi}\plus
\sqrt{ \bra\!\bras\gamma^2_{cc}\kets\!\ket_{\phi}
       \bra\!\bras\gamma^2_{ss}\kets\!\ket_{\phi}} ~,
 \bra\!\bras\gamma_{cc}\gamma_{ss}\minus\gamma_{cs}^2\kets\!\ket_{\phi}
 \!\right\}
 \label{eq:pinning_ATalmost2}
\end{eqnarray}
 For both models the left of the two arguments in
the corresponding extremization problems give the required
maximum. To see this, one first notes that \be \hspace*{-20mm}
 \bra\bras \gamma_{cc}\gamma_{ss}\kets\ket_\phi=
 \frac{1}{2}\bra\bras
 \gamma_{cc}^2+\gamma_{ss}^2-(\gamma_{cc}\minus\gamma_{ss})^2\kets\ket_\phi
 \leq
 \frac{1}{2}\bra\bras
 \gamma_{cc}^2+\gamma_{ss}^2\kets\ket_\phi
 \label{eq:gamma_ineq1}
\ee \be \hspace*{-20mm} \bra\!\bras\gamma^2_{cc}\kets\!\ket_{\phi}
       \bra\!\bras\gamma^2_{ss}\kets\!\ket_{\phi}
=\frac{1}{2}\left\{\bra\!\bras\gamma_{cc}^2\plus\gamma_{ss}^2\kets\!\ket_\phi^2
-\bra\!\bras\gamma_{cc}^2\kets\!\ket_\phi^2
-\bra\!\bras\gamma_{cc}^2\kets\!\ket_\phi^2\right\} \geq
\frac{1}{2}\bra\!\bras\gamma_{cc}^2\plus\gamma_{ss}^2\kets\!\ket_\phi^2
\label{eq:gamma_ineq2}
 \ee
For model I we subtract the right argument of
(\ref{eq:pinning_ATalmost1})
 from the left argument and find
\begin{eqnarray*}
\hspace*{-10mm} {\rm LA}-{\rm RA}&=& \frac{1}{2}\bra\!\bras
 (\gamma_{cc}\minus \gamma_{ss})^2\kets\!\ket_{\phi}
 -
 \bra\!\bras \gamma_{cs}^2
 \kets\!\ket_{\phi}
 +\sqrt{
 \bra\!\bras\gamma_{cs}^2\kets\!\ket_{\phi}^2\plus
 \frac{1}{4}
 \bra\!\bras\gamma^2_{cc}\minus\gamma^2_{ss}\kets\!\ket_{\phi}^2}
 \\
 &\geq & \frac{1}{2}\bra\!\bras
 (\gamma_{cc}\minus \gamma_{ss})^2\kets\!\ket_{\phi}
~\geq ~0
\end{eqnarray*}
Thus the maximum in (\ref{eq:pinning_ATalmost1}) is always
realized by the left argument. Similarly, upon subtracting the
right argument from the left argument in
 (\ref{eq:pinning_ATalmost2}), using (\ref{eq:gamma_ineq1},\ref{eq:gamma_ineq2}), we find
\begin{eqnarray*}
\hspace*{-10mm} {\rm LA}-{\rm RA}&=&
 2\bra\!\bras\gamma_{cs}^2\kets\!\ket_{\phi}\plus
\sqrt{ \bra\!\bras\gamma^2_{cc}\kets\!\ket_{\phi}
       \bra\!\bras\gamma^2_{ss}\kets\!\ket_{\phi}} -
 \bra\!\bras\gamma_{cc}\gamma_{ss}\kets\!\ket_{\phi}
 \\
 &\geq &
\sqrt{ \bra\!\bras\gamma^2_{cc}\kets\!\ket_{\phi}^2
       \bra\!\bras\gamma^2_{ss}\kets\!\ket_{\phi}^2} -
 \frac{1}{2}\bra\!\bras
 \gamma_{cc}^2\plus\gamma_{ss}^2\kets\!\ket_\phi
=
\frac{1}{2}(\sqrt{2}\minus
1)\bra\!\bras\gamma_{cc}^2\plus\gamma_{ss}^2\kets\!\ket_\phi \geq
0
\end{eqnarray*}
Hence also the maximum in (\ref{eq:pinning_ATalmost2}) is realized
by the left argument. We conclude that
\begin{eqnarray}
\hspace*{-15mm}
 {\rm AT~line~I:} &~~~&
\left[\frac{T}{2K}\right]^2
=
\frac{1}{2}\bra\!\bras
 \gamma^2_{cc}\plus\gamma^2_{ss}\kets\!\ket_{\phi}
 \plus\sqrt{
 \bra\!\bras\gamma_{cs}^2\kets\!\ket_{\phi}^2\plus
 \frac{1}{4}
 \bra\!\bras\gamma^2_{cc}\minus\gamma^2_{ss}\kets\!\ket_{\phi}^2}
 \label{eq:pinning_AT_I}
\\
\hspace*{-15mm}
 {\rm AT~line~I\!I:} &~~~&
\left[\frac{T}{2K}\right]^2
=
\bra\!\bras\gamma_{cs}^2\kets\!\ket_{\phi}\plus \sqrt{
\bra\!\bras\gamma^2_{cc}\kets\!\ket_{\phi}
       \bra\!\bras\gamma^2_{ss}\kets\!\ket_{\phi}}
 \label{eq:pinning_AT_II}
\end{eqnarray}
\vsp

 Finally, in reflection symmetric states with $p(\phi)=p(-\phi)$
 $\forall\phi$, where we may use identity
 (\ref{eq:reflection_symmetry}),
   the replica-symmetric
susceptibility (\ref{eq:chi_rs}) is found to be purely real:
\begin{eqnarray}
\frac{\chi_{\rm RS}}{\beta}&=&\frac{1}{2}(1\minus q_{cc}\minus
q_{ss}) -\bra\!\bras
\frac{1}{2}\cos(2\phi)(\gamma_{cc}\minus\gamma_{ss})
\plus\sin(2\phi)\gamma_{cs} \kets\!\ket_{\phi}
\label{eq:pinning_chi}
\end{eqnarray}

\subsection{Reflection symmetry breaking transitions}

 Assuming reflection symmetry breaking (to states
 with $Q_{cs}\neq 0$ and/or $q_{cs}\neq 0$)
to happen via second order transitions, allows us to determine its
occurrence by studying the relevant entries of the RS Hessian of
the free energy. Such transitions occur when \be \hspace*{-5mm}
{\rm Det}~\left|\begin{array}{cc} \room
 \partial^2 f[Q_{cc,q_{ss},q_{cc}}]/\partial Q_{cs}^2  &
 \partial^2 f[Q_{cc,q_{ss},q_{cc}}]/\partial Q_{cs}\partial q_{cs} \\
 \room\partial^2 f[Q_{cc,q_{ss},q_{cc}}]/\partial Q_{cs}\partial
 q_{cs}&
 \partial^2 f[Q_{cc,q_{ss},q_{cc}}]/\partial q_{cs}^2
 \end{array}\right|=0
 \label{eq:reflection_trans}
\ee The relevant matrix elements of the Hessian are given in
\ref{app:fderivatives}. We can deal with both our two models
simultaneously upon defining the variable $\tau\in\{-1,1\}$, where
$\tau=1$ for model I and $\tau=-1$ for model II. This allows us to
write:
\begin{eqnarray}
 \frac{1}{\beta K^2}\frac{\partial^2 f}{\partial Q_{cs}^2} &=&
4\tau-8(\beta K)^2\lambda_1
\\
 \frac{1}{\beta K^2} \frac{\partial^2 f}{\partial q_{cs}^2}
 &=& -4\tau + 8(\beta K)^2\lambda_2
\\
 \frac{1}{\beta K^2}\frac{\partial^2 f}{\partial Q_{cs}\partial
 q_{cs}}
&=&
 4(\beta K)^2\lambda_3
\end{eqnarray}
 They involve
 \begin{eqnarray}
 \hspace*{-5mm}
 \lambda_1
 &=&
\frac{1}{2}\bra\bras
 \bra
 \sin^2(2\theta)
 \ket_\star
 \minus
\bra \sin(2\theta) \ket_\star^2
 \kets\ket_{\phi}
 \label{eq:lambda_1}
\\
 \hspace*{-5mm}
\lambda_2&=& \bra\bras [\bra\sin^2(\theta)\ket_\star\minus
\bra\sin(\theta)\ket_\star^2]
 [\bra\cos^2(\theta)\ket_\star\minus 3\bra\cos(\theta)\ket_\star^2]
 \nonumber
\\&&
 \hspace*{21mm} + [\bra\cos^2(\theta)\ket_\star\minus \bra\cos(\theta)\ket_\star^2]
 [\bra\sin^2(\theta)\ket_\star\minus 3\bra\sin(\theta)\ket_\star^2]
 \nonumber
\\&&
 \hspace*{-20mm}
 +2[\bra\sin(\theta)\cos(\theta)\ket_\star\minus \bra\sin(\theta)\ket_\star\bra\cos(\theta)\ket_\star]
 [\bra\sin(\theta)\cos(\theta)\ket_\star\minus 3\bra\sin(\theta)\ket_\star\bra\cos(\theta)\ket_\star]
 \kets\ket_\phi\nonumber
 \\
 \label{eq:lambda_2}
 \\
  \hspace*{-5mm}
\lambda_3&=& 2\bra\bras
 \bra\sin(2\theta)\cos(\theta)\ket_\star\bra\sin(\theta)\ket_\star
   +\bra\sin(2\theta)\sin(\theta)\ket_\star\bra \cos(\theta)\ket_\star
   \nonumber
   \\&&\hspace*{25mm}
   -2\bra\sin(2\theta)\ket_\star
 \bra\sin(\theta)\ket_\star\bra\cos(\theta)\ket_\star\kets\ket_\phi
 \label{eq:lambda_3}
\end{eqnarray}
Insertion of the second derivatives into
(\ref{eq:reflection_trans}), using the above short-hands
(\ref{eq:lambda_1},\ref{eq:lambda_2},\ref{eq:lambda_3}) reveals
that  reflection symmetry breaking transitions are marked by the
highest temperature for which the following functions $\Sigma^{\rm
ref}_{\rm I,II}(T)$ become negative:
\begin{eqnarray}\hspace*{-10mm}
{\rm Model~I}:&~~~~& \Sigma^{\rm ref}_{\rm I}(T)=
(T/K)^2\!-\sqrt{(\lambda_1\minus \lambda_2)^2\minus
\lambda_3^2}-\lambda_1-\lambda_2
 \label{eq:reflec_trans_I}
 \\
 \hspace*{-10mm}
{\rm Model~II}:&~~~~& \Sigma^{\rm ref}_{\rm II}(T)=
(T/K)^2\!-\sqrt{(\lambda_1\minus \lambda_2)^2\minus
\lambda_3^2}+\lambda_1+\lambda_2
 \label{eq:reflec_trans_II}
\end{eqnarray}
Since the $\{\lambda_i\}$ are bounded, expressions
 (\ref{eq:reflec_trans_I},\ref{eq:reflec_trans_II}) confirm that
reflection symmetry will always be stable for sufficiently high
temperatures.

\section{States with global rotation symmetry}

For uniformly distributed pinning angles, i.e.
$p(\phi)=(2\pi)^{-1}$, we may expect the macroscopic state to have
 global rotation symmetry for sufficiently high temperatures, in
addition to reflection symmetry. We therefore inspect the effect
of global rotations $\theta_i\to \theta_i+\psi$ on the order
parameters (within the RS ansatz), with the help of
(\ref{eq:meaning_1},\ref{eq:meaning_2},\ref{eq:meaning_3}):
\be
\hspace*{-15mm} \theta_i^\prime= \theta_i+\psi ~~{\rm for all}~~i:
~~~~~
 q_{cc}^\prime+q_{ss}^\prime=q_{cc}+q_{ss}
 \label{eq:rotation}
\ee
\begin{eqnarray*}
\left(\!\!
\begin{array}{c}
Q_{cc}^\prime\minus \frac{1}{2}\\ Q_{cs}^\prime
\end{array}\!\!\right)
&=&\left(\!\!\begin{array}{cc} \cos(2\psi) & -\sin(2\psi) \\
\sin(2\psi) & \cos(2\psi)\end{array}\!\! \right)
\left(\!\!\begin{array}{c} Q_{cc}\minus \frac{1}{2}\\ Q_{cs}
\end{array}\!\!
\right)
 \\[1mm]
  \left(\!\!
 \begin{array}{c}
 \frac{1}{2}(q_{cc}^\prime\minus q_{ss}^\prime)\\ q_{cs}^\prime
 \end{array}\!\!\right)
 &=&\left(\!\!\begin{array}{cc} \cos(2\psi) & -\sin(2\psi) \\
 \sin(2\psi) & \cos(2\psi)\end{array}\!\! \right) \left(\!\!
 \begin{array}{c} \frac{1}{2}(q_{cc}\minus q_{ss})\\ q_{cs}
 \end{array}\!\!\right)
\end{eqnarray*}
Invariance under all global rotations implies
$Q_{cc}\!=\!\frac{1}{2}$, $Q_{cs}\!=\!q_{cs}\!=\!0$,
$q_{cc}\!=\!q_{ss}\!=\!q$ (rotation-invariance implies
reflection-invariance), leaving just one order parameter in
rotation invariant states. The invariant manifolds in order
parameter space are
\begin{eqnarray}
(Q_{cc}-\frac{1}{2})^2+Q_{cs}^2&=& \epsilon_1 \label{eq:manif1}
\\
\frac{1}{4}(q_{cc}-q_{ss})^2+q_{cs}^2&=&\epsilon_2
\label{eq:manif2}
\end{eqnarray} with the invariant state corresponding to
$\epsilon_1=\epsilon_2=0$. We will also find instances of
rotation-invariant states without the pinning field distribution
having rotational symmetry. Insertion of
$\{Q_{cc}=\frac{1}{2},~Q_{cs}=q_{cs}=0,~q_{cc}=q_{ss}=q\}$ as an
ansatz into our RS order parameter equations reveals that (for
nonzero $h$) the following condition is necessary and sufficient
for the existence of a rotation-invariant solution:
\be
\bra \cos^2(\phi)\ket_\phi=\bra \sin^2(\phi)\ket_\phi
\label{eq:rot_inv_condition} \ee Hence, unless explicitly stated
we will in this section not assume $p(\phi)$ to be uniform, but
rely only on the two properties $p(\phi)=p(-\phi)$ (assumed to
hold throughout this paper) and $\bra\cos(2\phi)\ket_\phi=0$ (to
guarantee the existence of a rotation-invariant state).

\subsection{Implications for free energy and order parameters}

For rotation-symmetric states the  measures
(\ref{eq:symRSMeasureI},\ref{eq:symRSMeasureII}) become identical:
\be
 M(\theta|x,y,\phi)= e^{\beta
 h\cos(\theta\minus\phi)+2\beta K \sqrt{q}[x\cos(\theta)+ y\sin(\theta) ]}
\label{eq:rot_symmetry}
 \ee
For any value of $\phi$ one can carry out a suitable rotation of
the Gaussian variables $(x,y)$ to eliminate $\phi$ from the
measure (\ref{eq:rot_symmetry}), leading to the following general
identity for any set of functions $\{k_\ell\}$ (with
$\ell=0,1,2,\ldots$):
\be
\bra \bras ~k_0(\phi)\prod_{\ell>0}\bra
k_\ell(\theta)\ket_\star~\kets\ket_{\phi}= \bra \bras
~k_0(\phi)\prod_{\ell>0}\bra
k_\ell(\theta\plus\phi)\ket_\circ~\kets\ket_{\phi}
\label{eq:move_angles}
 \ee
where $\bra\ldots\ket_\circ$  refers to averages calculated with
the $\phi$-independent measure $M(\theta|x,y)=M(\theta|x,y,0)$. As
a consequence  our subsequent calculations for rotation-invariant
states will  repeatedly involve various derivatives of the
following generating function:
\begin{eqnarray}
Z[x,y]&=& \log\int\!d\theta~e^{\beta
 h\cos(\theta)+2\beta K \sqrt{q}[x\cos(\theta)+ y\sin(\theta) ]}
\nonumber\\ &=& \log(2\pi)+\log I_0[\Xi] \label{eq:Zxy}
\end{eqnarray}
with the short-hand
 $\Xi= \beta
\sqrt{(h+2Kx\sqrt{q})^2+(2Ky\sqrt{q})^2}$, and in which $I_n[z]$
denotes the $n$-th modified Bessel function \cite{AbramStegun}.
For instance:
\begin{eqnarray}
\hspace*{-10mm}
\frac{1}{2\beta K\sqrt{q}}\frac{\partial}{\partial x}Z[x,y]&=&
\bra\cos(\theta)\ket_\circ \label{eq:dZ1}
\\
\hspace*{-10mm} \frac{1}{2\beta K\sqrt{q}}
\frac{\partial}{\partial y}Z[x,y]&=& \bra\sin(\theta)\ket_\circ
\label{eq:dZ2}
\\
\hspace*{-10mm} \frac{1}{(2\beta
K\sqrt{q})^2}\frac{\partial^2}{\partial x^2}Z[x,y]&=&
\bra\cos^2(\theta)\ket_\circ\minus \bra\cos(\theta)\ket_\circ^2
\label{eq:dZ3}
\\
\hspace*{-10mm} \frac{1}{(2\beta K\sqrt{q})^2}
\frac{\partial^2}{\partial y^2}Z[x,y]&=&
\bra\sin^2(\theta)\ket_\circ\minus \bra\sin(\theta)\ket_\circ^2
\label{eq:dZ4}
\\
\hspace*{-10mm} \frac{1}{(2\beta K\sqrt{q})^2}
\frac{\partial^2}{\partial x\partial y}Z[x,y]&=&
\bra\sin(\theta)\cos(\theta)\ket_\circ\minus
\bra\sin(\theta)\ket_\circ\bra\cos(\theta)\ket_\circ
\label{eq:dZ5}
\\
\hspace*{-10mm} \frac{1}{(2\beta K\sqrt{q})^3}
\frac{\partial^3}{\partial x^2\partial y}Z[x,y]&=&
\bra\sin(\theta)\cos^2(\theta)\ket_\circ\minus
\bra\sin(\theta)\ket_\circ\bra\cos^2(\theta)\ket_\circ \nonumber\\
&&\minus
\bra\sin(2\theta)\ket_\circ\bra\cos(\theta)\ket_\circ\plus
2\bra\sin(\theta)\ket_\circ\bra\cos(\theta)\ket_\circ^2
\label{eq:dZ6}
\\
\hspace*{-10mm} \frac{1}{(2\beta K\sqrt{q})^3}
\frac{\partial^3}{\partial x\partial y^2}Z[x,y]&=&
\bra\sin^2(\theta)\cos(\theta)\ket_\circ\minus
\bra\cos(\theta)\ket_\circ\bra\sin^2(\theta)\ket_\circ \nonumber\\
&&\minus
\bra\sin(2\theta)\ket_\circ\bra\sin(\theta)\ket_\circ\plus
2\bra\cos(\theta)\ket_\circ\bra\sin(\theta)\ket_\circ^2
\label{eq:dZ7}
\end{eqnarray}
 In order to work out the remaining order parameter equation for
$q$  we apply the identity (\ref{eq:move_angles}) to
$k_1(\theta)=k_2(\theta)=\cos(\theta)$ and to
$k_1(\theta)=k_2(\theta)=\sin(\theta)$,  giving
\begin{eqnarray}
\hspace*{-15mm}
 \bra \bras \bra \cos(\theta)\ket^2_\star
\kets\ket_\phi
 &=&
 \bra\cos^2(\phi)\ket_\phi \bras~ \bra\cos(\theta)\ket_\circ^2\kets
 \plus
 \bra\sin^2(\phi)\ket_\phi \bras \bra
 \sin(\theta)\ket_\circ^2  ~\kets
 \label{eq:rot_intermediate1}
\\
\hspace*{-15mm}\room
 \bra \bras \bra \sin(\theta)\ket^2_\star \kets\ket_\phi
 &=&
 \bra\sin^2(\phi)\ket_\phi \bras~ \bra\cos(\theta)\ket_\circ^2\kets
 \plus
 \bra\cos^2(\phi)\ket_\phi \bras \bra
 \sin(\theta)\ket_\circ^2  ~\kets
 \label{eq:rot_intermediate2}
\end{eqnarray}
Thus the remaining order parameter $q$ is, for both models,  to be
solved from
 \begin{eqnarray}
 q&=& \frac{1}{2} \int\!DxDy~
 \bra~ \bra\cos(\theta)\ket_\star^2 + \bra\sin(\theta)\ket_\star^2
 ~\ket_{\phi}\nonumber
 \\
&=& \frac{1}{2} \int\!DxDy~
 \left[ \bra\cos(\theta)\ket_\circ^2 + \bra\sin(\theta)\ket_\circ^2\right]
 \label{eq:q_intermediate}
\end{eqnarray}
Using properties of modified Bessel functions, such
 as $\frac{d}{dz}I_n[z]=\frac{1}{2}(I_{n+1}[z]+I_{n-1}[z])$ and
 $I_n[z]=(z/2n)(I_{n-1}[z]-I_{n+1}[z])$,
the two relevant expressions (\ref{eq:dZ1},\ref{eq:dZ2}) give
\begin{eqnarray}
\bra\cos(\theta)\ket_\circ&=&
\frac{\beta(h+2Kx\sqrt{q})}{\Xi}\frac{I_1[\Xi]}{I_0[\Xi]}
\label{eq:rot_cos_av}
\\
\bra\sin(\theta)\ket_\circ&=&
\frac{\beta(2Ky\sqrt{q})}{\Xi}\frac{I_1[\Xi]}{I_0[\Xi]}
\label{eq:rot_sin_av} \end{eqnarray}
 After writing  $(x,y)$ in polar
 coordinates,
 one then  finds (\ref{eq:q_intermediate}) reducing to
 \begin{eqnarray}
 q
&=& \int_0^\pi\frac{d\psi}{2\pi}\int_0^\infty\!
dr~re^{-\frac{1}{2}r^2}\left\{\frac{I_1[\Xi(r,\psi)]}{I_0[\Xi(r,\psi)]}\right\}^2
\label{eq:rot_q_eqn}
\end{eqnarray}
with
\be
\Xi(r,\psi)=\beta \sqrt{4qK^2 r^2+h^2+4Khr\sqrt{q}\cos(\psi)}
\label{eq:Xi} \ee

Inserting $Q_{cs}=q_{cs}=0$, $Q_{cc}=\frac{1}{2}$, and
$q_{cc}=q_{ss}=q$ into expressions
(\ref{eq:fRS},\ref{eq:UI},\ref{eq:UII}) shows that for
rotation-symmetric states and for both models I and II the
disorder-averaged free energy per oscillator equals
$\overline{f}=\extr_{q}f[q]-\frac{1}{\beta}\log(2\pi)$, with
\be
 f[q]=- 2\beta K^2(q\minus \frac{1}{2})^2 -\frac{1}{\beta}
\int_{0}^{\pi}\!\! \frac{d\psi}{\pi}\int_{0}^\infty\!\!\!dr~r
e^{-\frac{1}{2}r^2}
 \log I_0[\Xi(r,\psi)]
 \label{eq:rot_inv_f}
 \ee
 According to (\ref{eq:extremumI},\ref{eq:extremumII})
the function $f[q]$  is in both cases to be maximized, so the
replica symmetric rotation-invariant ground state would be
$q=\frac{1}{2}$.

\subsection{Implications for covariances, AT lines, and RS susceptibility}

 Various terms involving the  covariances
$\gamma_{\star\star}$ (\ref{eq:gammas1},\ref{eq:gammas2}) can be
simplified  via (\ref{eq:move_angles}). We define
$\overline{\gamma}_{\star\star}=\gamma_{\star\star}|_{\phi=0}$
(i.e. as calculated with the $\phi=0$ averages
$\bra\ldots\ket_\circ$). For instance, focusing on the terms
occurring in (\ref{eq:pinning_AT_II},\ref{eq:pinning_AT_II}) and
using (\ref{eq:rot_inv_condition}):
\begin{eqnarray} \hspace*{-15mm} \bra\bras
\gamma^2_{cc}\kets\ket_\phi  &=& \bras
\frac{1}{4}(\overline{\gamma}_{cc}\plus
\overline{\gamma}_{ss})^2\plus \overline{\gamma}^2_{cs}\kets
+\bra\cos^2(2\phi)\ket_\phi\bras\frac{1}{4}(\overline{\gamma}_{cc}\minus\overline{\gamma}_{ss})^2
 \minus \overline{\gamma}^2_{cs}\kets
 \label{eq:general_gammas1}
\\
 \hspace*{-15mm} \bra\bras\gamma_{ss}^2\kets\ket_\phi &=&
\bras\frac{1}{4}(\overline{\gamma}_{cc}\plus\overline{\gamma}_{ss})^2\plus
\overline{\gamma}_{cs}^2\kets
+\bra\cos^2(2\phi)\ket_\phi\bras\frac{1}{4}(\overline{\gamma}_{cc}\minus\overline{\gamma}_{ss})^2
 \minus \overline{\gamma}_{cs}^2\kets
 \label{eq:general_gammas2}
\\
 \hspace*{-15mm}
\bra\bras\gamma_{cs}^2\kets\ket_\phi &=&
\bra\sin^2(2\phi)\ket_\phi\bra\bras
\frac{1}{4}(\overline{\gamma}_{cc}\minus\overline{\gamma}_{ss})^2\kets
+
 \bra\cos^2(2\phi)\ket_\phi\bras
\overline{\gamma}^2_{cs}\kets
  \label{eq:general_gammas3}
\end{eqnarray}
\be
 \hspace*{-15mm}\frac{1}{4}\bra\bras (\gamma_{cc}\minus
\gamma_{ss})^2\kets\ket_\phi =
\bra\sin^2(2\phi)\ket_\phi\bras\overline{\gamma}_{cs}^2\kets+
\bra\cos^2(2\phi)\ket_\phi\bras\frac{1}{4}(\overline{\gamma}_{cc}\minus\overline{\gamma}_{ss})^2
 \kets
 \label{eq:general_gammas4}
\ee From (\ref{eq:dZ3},\ref{eq:dZ4},\ref{eq:dZ5}), in turn, one
obtains explicit expressions for the required
$\{\overline{\gamma}_{\star\star}\}$:
\begin{eqnarray}
\overline{\gamma}_{cc}&=& \frac{1}{2}-\frac{I_2[\Xi]}{2I_0[\Xi]}
+\frac{\beta^2(h+2Kx\sqrt{q})^2}{\Xi^2}\left\{\frac{I_2[\Xi]}{I_0[\Xi]}-\frac{I^2_1[\Xi]}{I_0^2[\Xi]}\right\}
\label{eq:overline_gammacc}
\\
\overline{\gamma}_{ss}&=& \frac{1}{2}-\frac{I_2[\Xi]}{2I_0[\Xi]}
+\frac{\beta^2(2Ky\sqrt{q})^2}{\Xi^2}\left\{\frac{I_2[\Xi]}{I_0[\Xi]}-\frac{I^2_1[\Xi]}{I_0^2[\Xi]}\right\}
\label{eq:overline_gammass}
\\
\overline{\gamma}_{cs}&=&
\frac{\beta^2(h+2Kx\sqrt{q})(2Ky\sqrt{q})}{\Xi^2}\left\{\frac{I_2[\Xi]}{I_0[\Xi]}-\frac{I^2_1[\Xi]}{I_0^2[\Xi]}\right\}
\label{eq:overline_gammacs}
\end{eqnarray}
Combination of the above results with
(\ref{eq:pinning_AT_I},\ref{eq:pinning_AT_II}) gives the equations
for the AT lines. We note that for rotation invariant states the
two  expressions (\ref{eq:pinning_AT_I},\ref{eq:pinning_AT_II})
become identical:
\begin{eqnarray*}
\left[\frac{T}{K}\right]^2 &=&
4\bras\frac{1}{2}\overline{\gamma}^2_{cc}\plus \frac{1}{2}
\overline{\gamma}^2_{ss}\plus \overline{\gamma}^2_{cs}\kets
\nonumber
\\
&=& \int\!DxDy \left\{
\left[\frac{I_2[\Xi]}{I_0[\Xi]}-\frac{I_1^2[\Xi]}{I_0^2[\Xi]}\right]^2
+\left[1-\frac{I_1^2[\Xi]}{I_0^2[\Xi]}\right]^2\right\}
\end{eqnarray*}
After transformation of the Gaussian variables to polar
coordinates, this gives for the AT lines of our two models the
appealing result \bd \hspace*{-22mm}
 \left[\frac{T}{K}\right]^2 =
\int_0^\pi\!\frac{d\psi}{\pi}\int_0^\infty\!dr~r
e^{-\frac{1}{2}r^2}\left\{
\left[\frac{I_2[\Xi(r,\psi)]}{I_0[\Xi(r,\psi)]}-\frac{I_1^2[\Xi(r,\psi)]}{I_0^2[\Xi(r,\psi)]}\right]^2
\! +\left[1\!-
\frac{I_1^2[\Xi(r,\psi)]}{I_0^2[\Xi(r,\psi)]}\right]^2\right\} \ed
\be
 \label{eq:ATrot1}
\ee

 In
a similar manner we work out expression (\ref{eq:pinning_chi}) for
rotation-invariant states:
\begin{eqnarray}
\frac{\chi_{\rm RS}}{\beta}&=&\frac{1}{2}(1\minus 2q) -\bra\!\bras
\frac{1}{2}\cos(2\phi)(\gamma_{cc}\minus\gamma_{ss})
\plus\sin(2\phi)\gamma_{cs} \kets\!\ket_{\phi} \nonumber \\
&=&\frac{1}{2}(1\minus 2q) -\frac{1}{2}\int\!DxDy~(
\overline{\gamma}_{cc}\minus \overline{\gamma}_{ss}) \nonumber\\
 &=&\frac{1}{2}- q -\int_0^\pi \!\frac{d\psi}{2\pi}
 \int_0^\infty\!dr~r e^{-\frac{1}{2}r^2}
 \left[\frac{I_2[\Xi(r,\psi)]}{I_0[\Xi(r,\psi)]}-\frac{I^2_1[\Xi(r,\psi)]}{I_0^2[\Xi(r,\psi)]}\right]
 \nonumber\\
 &&\hspace*{10mm}\times\left[
\frac{h^2 \plus 4K^2r^2q\cos(2\psi) \plus
4hKr\sqrt{q}\cos(\psi)}{h^2 \plus 4K^2r^2q \plus
4hKr\sqrt{q}\cos(\psi)}\right]
 \label{eq:Chirot1}
\end{eqnarray}
with the abbreviation (\ref{eq:Xi}). Note that $\lim_{h\to
0}\Xi(r,\psi)=2\beta Kr\sqrt{q}$ (i.e. independent of $\psi$), so
that for rotation invariant states one  has
\be
\lim_{h\to 0}\chi_{\rm RS}= \beta(\frac{1}{2}- q)
\label{eq:chi_nofields} \ee

\subsection{Rotation symmetry breaking transitions}

  According to
(\ref{eq:manif1},\ref{eq:manif2}) we can study rotation
symmetry-breaking bifurcations most conveniently after the
following transformation of the order parameters: \bd
\hspace*{-0mm}
 Q_{cc}= \frac{1}{2}+\epsilon_1 \cos(\omega_1),~~~~~~
 Q_{cs}=\epsilon_1 \sin(\omega_1)
\ed \bd \hspace*{-0mm}
 q_{cc}=q+\epsilon_2 \cos(\omega_2), ~~~~~~~~
 q_{ss}=q-\epsilon_2 \cos(\omega_2),~~~~~~~~
 q_{cs}=\epsilon_2 \sin(\omega_2)
\ed Reflection symmetry breaking transitions are already marked by
(\ref{eq:reflec_trans_I},\ref{eq:reflec_trans_II}), so we restrict
ourselves to rotation symmetry breaking transitions which preserve
reflection symmetry, i.e. $Q_{cs}=q_{cs}=0$. Thus we put
$\omega_1=\omega_2=0$  and focus on \bd \hspace*{-0mm}
 Q_{cc}= \frac{1}{2}+\epsilon_1,~~~~~~
 Q_{cs}=q_{cs}=0,~~~~~~ q_{cc}=q+\epsilon_2, ~~~~~~
 q_{ss}=q-\epsilon_2
\ed Our interest is in continuous bifurcations of $\epsilon_1\neq
0$ and/or $\epsilon_2\neq 0$. We expand the  free energy per
oscillator $f[\ldots]$ to be extremized (\ref{eq:fRS}) around the
rotation invariant solution, using the second order derivatives in
\ref{app:fderivatives}. Due to reflection symmetry, only those
second derivatives with an even total number of `s' subscripts in
the two corresponding order parameters can be nonzero. This gives
for $0\ll \epsilon_1,\epsilon_2\ll 1$ (with $f_{\rm RI}[\ldots]$
denoting the free energy of the rotation invariant state):
\begin{eqnarray*}
 f[\ldots]-f_{\rm RI}[\ldots]
 &=& \frac{1}{2}\epsilon_1^2 \frac{\partial^2
 f}{\partial Q_{cc}^2}
 +\frac{1}{2}\epsilon_2^2
 \left\{
 \frac{\partial^2
 f}{\partial q_{cc}^2}+\frac{\partial^2
 f}{\partial q_{ss}^2}-2\frac{\partial^2 f}{\partial
 q_{cc}\partial q_{ss}}
\right\}
\\
&&
 +\epsilon_1\epsilon_2\left\{
\frac{\partial^2 f}{\partial Q_{cc}\partial
 q_{cc}}-
\frac{\partial^2 f}{\partial Q_{cc}\partial q_{ss}} \right\}
+\order(\epsilon^3)
\end{eqnarray*}
Hence the continuous rotation symmetry breaking transitions are
marked by \be \hspace*{-25mm} {\rm Det}\left|\begin{array}{cc}
\room
 \partial^2\! f/\partial Q_{cc}^2  &
\partial^2\! f/\partial Q_{cc}\partial
 q_{cc}\minus
\partial^2\! f/\partial Q_{cc}\partial q_{ss}
\\
 \room
\partial^2\! f/\partial Q_{cc}\partial
 q_{cc}\minus
\partial^2\! f/\partial Q_{cc}\partial q_{ss} &
 \partial^2\!
 f/\partial q_{cc}^2\plus \partial^2\!
 f/\partial q_{ss}^2\minus 2\partial^2\! f/\partial
 q_{cc}\partial q_{ss}
 \end{array}\right|=0
 \label{eq:rotation_trans}
\ee We work out the second derivatives for rotation invariant
states (where the effective measures of the two models are
identical), using (\ref{eq:move_angles}). Given our requirements
$p(\phi)=p(-\phi)$ and $\bra \cos(2\phi)\ket_\phi$, these are
found to depend on the choice made for the pinning field
distribution only through $\bra \cos(4\phi)\ket_\phi$. Again we
deal with both models simultaneously upon defining
$\tau\in\{-1,1\}$, where $\tau=1$ for model I and $\tau=-1$ for
model II. This allows us to write:
\begin{eqnarray}
 \frac{1}{\beta K^2}\frac{\partial^2 f}{\partial Q_{cc}^2} &=&
4\tau-8(\beta K)^2\rho_1
\\
 \frac{1}{\beta K^2}\left\{ \frac{\partial^2 f}{\partial q_{cc}^2}
 \plus \frac{\partial^2 f}{\partial q_{ss}^2}
\minus 2\frac{\partial^2 f}{\partial
 q_{cc}\partial q_{ss}}
 \right\}
&=& -4\tau + 8(\beta K)^2\rho_2
\\
 \frac{1}{\beta K^2}\left\{ \frac{\partial^2 f}{\partial Q_{cc}\partial
 q_{cc}}
 \minus \frac{\partial^2 f}{\partial Q_{cc} q_{ss}}\right\}
&=&
 4(\beta K)^2\rho_3
\end{eqnarray}
where straightforward but tedious bookkeeping shows the quantities
$\{\rho_1,\rho_2,\rho_3\}$ to be
 \bd
\hspace*{-25mm} \rho_1=\frac{1}{4} \bras
 1-\bra\cos(2\theta\ket_\circ^2\!
 -\bra\sin(2\theta\ket_\circ^2\kets
 +\frac{1}{4}\bra\cos(4\phi)\ket_\phi \bras
 \bra\cos(4\theta)\ket_\circ\!-\bra\cos(2\theta\ket_\circ^2\!
 +\bra\sin(2\theta\ket_\circ^2\kets
 \ed
 \be
 \label{eq:rho_1}
\ee \bd \hspace*{-25mm}
 \rho_2= \bras~
 [ \bra\cos^2(\theta)\ket_\circ\!-\bra\cos(\theta)\ket_\circ^2]
  [ \bra\cos^2(\theta)\ket_\circ\!-3\bra\cos(\theta)\ket_\circ^2]
\ed \bd \hspace*{15mm}
 + [ \bra\sin^2(\theta)\ket_\circ\!-\bra\sin(\theta)\ket_\circ^2]
  [
  \bra\sin^2(\theta)\ket_\circ\!-3\bra\sin(\theta)\ket_\circ^2]
\ed \bd \hspace*{-15mm}
-\frac{1}{2}[\bra\cos(2\theta)\ket_\circ\!-\bra\cos(\theta)\ket_\circ^2\!+\bra\sin(\theta)\ket_\circ^2]
[\bra\cos(2\theta)\ket_\circ\!-3\bra\cos(\theta)\ket_\circ^2\!+3\bra\sin(\theta)\ket_\circ^2]
  ~\kets
\ed \bd \hspace*{-25mm}
  +  \bra\cos(4\phi)\ket_\phi\bras~
\frac{1}{2}[\bra\cos(2\theta)\ket_\circ\!-\bra\cos(\theta)\ket_\circ^2\!+\bra\sin(\theta)\ket_\circ^2]
[\bra\cos(2\theta)\ket_\circ\!-3\bra\cos(\theta)\ket_\circ^2\!+3\bra\sin(\theta)\ket_\circ^2]
\ed \bd \hspace*{-15mm} -2
[\bra\sin(\theta)\cos(\theta)\ket_\circ\!-\bra\sin(\theta)\ket_\circ\bra\cos(\theta)\ket_\circ]
[\bra\sin(\theta)\cos(\theta)\ket_\circ\!-3\bra\sin(\theta)\ket_\circ\bra\cos(\theta)\ket_\circ]
~\kets \ed
\be
\label{eq:rho_2}
 \ee \bd \hspace*{-25mm}
 \rho_3=
 2\bras
 \bra\sin^2(\theta)\ket_\circ \bra\cos(\theta)\ket_\circ^2\!+
 \bra\cos^2(\theta)\ket_\circ
 \bra\sin(\theta)\ket_\circ^2\!-\bra\sin(2\theta)\ket_\circ\bra\sin(\theta)\ket_\circ
 \bra\cos(\theta)\ket_\circ\kets
\ed \bd \hspace*{-20mm} +2\bra\cos(4\phi)\ket_\phi\bras~
\bra\sin^2(\theta)\ket_\circ \bra\cos(\theta)\ket_\circ^2\!+
\bra\cos^2(\theta)\ket_\circ \bra\sin(\theta)\ket_\circ^2\!
+\bra\sin(2\theta)\ket_\circ\bra\sin(\theta)\ket_\circ
\bra\cos(\theta)\ket_\circ\! \ed \be \hspace*{-10mm}
-2\bra\sin^2(\theta)\cos(\theta)\ket_\circ
\bra\cos(\theta)\ket_\circ
-2\bra\cos^2(\theta)\sin(\theta)\ket_\circ
 \bra\sin(\theta)\ket_\circ \kets
 \label{eq:rho_3}
\ee Insertion of the second derivatives into
(\ref{eq:rotation_trans}), using
(\ref{eq:rho_1},\ref{eq:rho_2},\ref{eq:rho_3}), shows that
rotation symmetry breaking transitions which preserve reflection
symmetry are marked by the highest temperature for which the
following functions become negative:
\begin{eqnarray}\hspace*{-10mm}
{\rm Model~I}:&~~~~& \Sigma^{\rm rot}_{\rm I}(T)=
(T/K)^2\!-\sqrt{(\rho_1\minus \rho_2)^2\minus
\rho_3^2}-\rho_1-\rho_2
 \label{eq:symtrans_I}
 \\
 \hspace*{-10mm}
{\rm Model~II}:&~~~~& \Sigma^{\rm rot}_{\rm II}(T)=
(T/K)^2\!-\sqrt{(\rho_1\minus
\rho_2)^2\minus\rho_3^2}+\rho_1+\rho_2
 \label{eq:symtrans_II}
\end{eqnarray}
What remains is to work out
(\ref{eq:rho_1},\ref{eq:rho_2},\ref{eq:rho_3}). For most terms we
can use
(\ref{eq:Zxy},\ref{eq:dZ1}-\ref{eq:dZ7},\ref{eq:overline_gammacc}-\ref{eq:overline_gammacs}).
One further object in $\rho_1$, $\bras\bra
\cos(4\theta)\ket_\circ\kets$, can be calculated directly as
follows
\begin{eqnarray*}
\bras\bra\cos(4\theta)\ket_\circ\kets&=&\bras\frac{\int\!d\theta~\cos(4\theta+4~{\rm
atan}[\frac{2Ky\sqrt{q}}{h+2Kx\sqrt{q}}])
e^{\Xi\cos(\theta)}}{2\pi I_0(\Xi)}\kets
\\
&=&\bras\cos(4~{\rm
atan}[\frac{2Ky\sqrt{q}}{h+2Kx\sqrt{q}}])\frac{I_4[\Xi]}{I_0[\Xi]}\kets
\\
&=&\bras\left[1-\frac{8\beta^4(2Ky\sqrt{q})^2(h+2Kx\sqrt{q})^2}{\Xi^4}\right]\frac{I_4[\Xi]}{I_0[\Xi]}\kets
\end{eqnarray*}
 The
final result of the exercise is, after transformation of $(x,y)$
to polar coordinates:  \bd
 \hspace*{-25mm}
 \rho_1=\frac{1}{4}\int_0^\infty\!\!dr~r
e^{-\frac{1}{2}r^2}\!\int_0^\pi\!\frac{d\psi}{\pi}\left\{1\minus
\frac{I_2^2[\Xi(r,\psi)]}{I_0^2[\Xi(r,\psi)]} \right\} \ed \bd
 \hspace*{-15mm}
 +\frac{1}{4}\bra\cos(4\phi)\ket_\phi \int_0^\infty\!\!dr~r
e^{-\frac{1}{2}r^2}\!\int_0^\pi\!\frac{d\psi}{\pi}
\left\{\frac{I_4[\Xi(r,\psi)]}{I_0[\Xi(r,\psi)]}\minus\frac{I_2^2[\Xi(r,\psi)]}{I_0^2[\Xi(r,\psi)]}
\right\} \ed
\be
\hspace*{15mm} \times \left\{1\minus
\frac{8\beta^4[2Kr\sqrt{q}\sin(\psi)]^2[h\plus
2Kr\sqrt{q}\cos(\psi)]^2}{\Xi^4(r,\psi)}\right\}
\label{eq:final_rot_rho1} \ee \bd
 \hspace*{-25mm}
 \rho_2=
\frac{1}{2}\int_0^\infty\!\!dr~r
e^{-\frac{1}{2}r^2}\!\int_0^\pi\!\frac{d\psi}{\pi} \left\{1\minus
\frac{I_1^2[\Xi(r,\psi)]}{I_0^2[\Xi(r,\psi)]}\right\}
\left\{1\minus
\frac{3I_1^2[\Xi(r,\psi)]}{I_0^2[\Xi(r,\psi)]}\right\} \ed \bd
 \hspace*{-25mm} +\frac{1}{2}\bra\cos(4\phi)\ket_\phi \int_0^\infty\!\!dr~r
e^{-\frac{1}{2}r^2}\!\int_0^\pi\!\frac{d\psi}{\pi}
\left\{\frac{I_2[\Xi(r,\psi)]}{I_0[\Xi(r,\psi)]}\minus
\frac{I_1^2[\Xi(r,\psi)]}{I_0^2[\Xi(r,\psi)]} \right\}
\left\{\frac{I_2[\Xi(r,\psi)]}{I_0[\Xi(r,\psi)]}\minus
\frac{3I_1^2[\Xi(r,\psi)]}{I_0^2[\Xi(r,\psi)]} \right\} \ed
\be
\hspace*{15mm} \times \left\{1\minus
\frac{8\beta^4[2Kr\sqrt{q}\sin(\psi)]^2[h\plus
2Kr\sqrt{q}\cos(\psi)]^2}{\Xi^4(r,\psi)}\right\}
\label{eq:final_rot_rho2} \ee \bd
 \hspace*{-25mm}
 \rho_3=
\int_0^\infty\!\!dr~r
e^{-\frac{1}{2}r^2}\!\int_0^\pi\!\frac{d\psi}{\pi} \left\{1\minus
\frac{I_2[\Xi(r,\psi)]}{I_0[\Xi(r,\psi)]}\right\}
\frac{I_1^2[\Xi(r,\psi)]}{I_0^2[\Xi(r,\psi)]}
 \ed \bd
 \hspace*{-20mm} +\bra\cos(4\phi)\ket_\phi \int_0^\infty\!\!dr~r
e^{-\frac{1}{2}r^2}\!\int_0^\pi\!\frac{d\psi}{\pi} \left\{1\minus
\frac{I_2[\Xi(r,\psi)]}{I_0[\Xi(r,\psi)]} \right\}
\left\{\frac{I_1^2[\Xi(r,\psi)]}{I^2_0[\Xi(r,\psi)]}\minus
\frac{2I_2[\Xi(r,\psi)]}{I_0[\Xi(r,\psi)]} \right\} \ed
\be
\hspace*{15mm} \times \left\{1\minus
\frac{8\beta^4[2Kr\sqrt{q}\sin(\psi)]^2[h\plus
2Kr\sqrt{q}\cos(\psi)]^2}{\Xi^4(r,\psi)}\right\}
\label{eq:final_rot_rho3} \ee The general identity
$I_{n+1}[z]=I_{n-1}[z]+nI_n[z](I_1[z]\minus I_0[z])/I_1[z]$ allows
us to express any $I_{n>2}[z]$ in terms of the trio
$\{I_0[z],I_1[z],I_2[z]\}$. Since the $\{\rho_i\}$ are  bounded,
expressions
 (\ref{eq:symtrans_I}, \ref{eq:symtrans_II}) confirm that, if
 rotation invariant states exist (i.e. if $\bra\cos(2\phi)\ket_\phi=0$),
rotation symmetry will be stable for sufficiently large
temperatures. \vsp

To round off our discussion we return to reflection symmetry
breaking transitions. We work out
(\ref{eq:lambda_1},\ref{eq:lambda_2},\ref{eq:lambda_3}) for
rotation invariant states, and find (after a lengthy but
straightforward calculation) that
 \bd
 \hspace*{-25mm}
 \lambda_1=\frac{1}{4}\int_0^\infty\!\!dr~r
e^{-\frac{1}{2}r^2}\!\int_0^\pi\!\frac{d\psi}{\pi}\left\{1\minus
\frac{I_2^2[\Xi(r,\psi)]}{I_0^2[\Xi(r,\psi)]} \right\} \ed \bd
 \hspace*{-15mm}
 -\frac{1}{4}\bra\cos(4\phi)\ket_\phi \int_0^\infty\!\!dr~r
e^{-\frac{1}{2}r^2}\!\int_0^\pi\!\frac{d\psi}{\pi}
\left\{\frac{I_4[\Xi(r,\psi)]}{I_0[\Xi(r,\psi)]}\minus\frac{I_2^2[\Xi(r,\psi)]}{I_0^2[\Xi(r,\psi)]}
\right\} \ed
\be
\hspace*{15mm} \times \left\{1\minus
\frac{8\beta^4[2Kr\sqrt{q}\sin(\psi)]^2[h\plus
2Kr\sqrt{q}\cos(\psi)]^2}{\Xi^4(r,\psi)}\right\}
\label{eq:final_rot_lambda1} \ee \bd
 \hspace*{-25mm}
 \lambda_2=
\frac{1}{2}\int_0^\infty\!\!dr~r
e^{-\frac{1}{2}r^2}\!\int_0^\pi\!\frac{d\psi}{\pi} \left\{1\minus
\frac{I_1^2[\Xi(r,\psi)]}{I_0^2[\Xi(r,\psi)]}\right\}
\left\{1\minus
\frac{3I_1^2[\Xi(r,\psi)]}{I_0^2[\Xi(r,\psi)]}\right\} \ed \bd
 \hspace*{-25mm} -\frac{1}{2}\bra\cos(4\phi)\ket_\phi \int_0^\infty\!\!dr~r
e^{-\frac{1}{2}r^2}\!\int_0^\pi\!\frac{d\psi}{\pi}
\left\{\frac{I_2[\Xi(r,\psi)]}{I_0[\Xi(r,\psi)]}\minus
\frac{I_1^2[\Xi(r,\psi)]}{I_0^2[\Xi(r,\psi)]} \right\}
\left\{\frac{I_2[\Xi(r,\psi)]}{I_0[\Xi(r,\psi)]}\minus
\frac{3I_1^2[\Xi(r,\psi)]}{I_0^2[\Xi(r,\psi)]} \right\} \ed
\be
\hspace*{15mm} \times \left\{1\minus
\frac{8\beta^4[2Kr\sqrt{q}\sin(\psi)]^2[h\plus
2Kr\sqrt{q}\cos(\psi)]^2}{\Xi^4(r,\psi)}\right\}
\label{eq:final_rot_lambda2} \ee \bd
 \hspace*{-25mm}
 \lambda_3=
\int_0^\infty\!\!dr~r
e^{-\frac{1}{2}r^2}\!\int_0^\pi\!\frac{d\psi}{\pi} \left\{1\minus
\frac{I_2[\Xi(r,\psi)]}{I_0[\Xi(r,\psi)]}\right\}
\frac{I_1^2[\Xi(r,\psi)]}{I_0^2[\Xi(r,\psi)]}
 \ed \bd
 \hspace*{-20mm} -\bra\cos(4\phi)\ket_\phi \int_0^\infty\!\!dr~r
e^{-\frac{1}{2}r^2}\!\int_0^\pi\!\frac{d\psi}{\pi} \left\{1\minus
\frac{I_2[\Xi(r,\psi)]}{I_0[\Xi(r,\psi)]} \right\}
\left\{\frac{I_1^2[\Xi(r,\psi)]}{I^2_0[\Xi(r,\psi)]}\minus
\frac{2I_2[\Xi(r,\psi)]}{I_0[\Xi(r,\psi)]} \right\} \ed
\be
\hspace*{15mm} \times \left\{1\minus
\frac{8\beta^4[2Kr\sqrt{q}\sin(\psi)]^2[h\plus
2Kr\sqrt{q}\cos(\psi)]^2}{\Xi^4(r,\psi)}\right\}
\label{eq:final_rot_lambda3} \ee Apparently, for
rotation-invariant states we obtain the trio
$\{\lambda_1,\lambda_2,\lambda_3\}$ from
$\{\rho_1,\rho_2,\rho_3\}$ by making in the latter the replacement
$\bra\cos(4\phi)\ket_\phi\to -\bra\cos(4\phi)\ket_\phi$:
\be
\lambda_i(\bra\cos(4\phi)\ket_\phi)=\rho_i(-\bra\cos(4\phi)\ket_\phi)~~~~~~i=1,2,3
\label{eq:lambda_vs_rho} \ee The same is true for the expressions
(\ref{eq:reflec_trans_I},\ref{eq:reflec_trans_II}) for the
reflection symmetry breaking transition lines, which can be
obtained by making the substitution $\bra\cos(4\phi)\ket_\phi\to
-\bra\cos(4\phi)\ket_\phi$ in the two expressions
(\ref{eq:symtrans_I},\ref{eq:symtrans_II}) for those rotation
symmetry breaking transitions which leave reflection symmetry
intact.

\section{Solution for specific choices for the pinning field statistics}

We now work out our results for specific choices of the pinning
field distribution $p(\phi)$ with decreasing levels of symmetry,
restricting ourselves (as throughout this paper) to those with
$p(\phi)=p(-\phi)$. We start with the benchmark case $h=0$, i.e.
absent pinning fields. Wherever possible we have tested our
theoretical predictions against extensive numerical simulations.
Here the Langevin dynamics defined by equation (\ref{eq:forces})
was iterated with a stochastic Euler method with time step $\Delta
t = 10^{-3}$, for systems of size either $N=400$ (with the
advantage of better equilibration within experimentally accessible
time-scales) or $N=800$ (with the advantage of reduced finite size
effects). All simulation results shown are averages over 10
experiments.

\subsection{Absent pinning fields}

For $h=0$ the natural solution is the one with full rotational
symmetry. Given the RS ansatz  we are left with a single order
parameter, $q$, and the two measures
(\ref{eq:RSMeasureI},\ref{eq:RSMeasureII}) become identical.
Insertion of $h=0$ into (\ref{eq:rot_q_eqn}) and
 (\ref{eq:rot_inv_f})
immediately leads us to
\be
\hspace*{-15mm}
q=\frac{1}{2}\int_0^\infty\!dr~r~e^{-\frac{1}{2}r^2}
\frac{I_1^2[\Xi(r)]}{I^2_0[\Xi(r)]} \label{eq:SPnopinning} \ee
\be
\hspace*{-15mm}
 \overline{f}=\max_{q}\left\{ - 2\beta
K^2(q\minus \frac{1}{2})^2 -\frac{1}{\beta}
\int_{0}^\infty\!\!\!dr~r e^{-\frac{1}{2}r^2}
 \log I_0[\Xi(r)]-\frac{1}{\beta}\log(2\pi)\right\}
 \label{eq:fnopinning}
 \ee
 with $\Xi(r)= 2 \beta K \sqrt{q}r$.
Expanding the right-hand side of (\ref{eq:SPnopinning}) gives
${\rm RHS}(q)=(\beta K)^2 q-4(\beta K)^4 q^2 +\order(q^3)$. There
is a second order transition at $T_c=K$ from a paramagnetic state
($q=0$) to an ordered state ($q>0$), and no evidence for
first-order transitions. As the temperature is lowered further,
$q$ increases monotonically towards its maximum value at $T=0$, as
$q=\frac{1}{2}-T\sqrt{\pi}/4K+\order(T^2)$. Close to the critical
point we can expand $q$ in powers of $\tau=1-T/T_c$ and find
$q=\frac{1}{2}\tau+\order(\tau^2)$ $(\tau\downarrow 0)$. All this
is in perfect agreement with the results obtained earlier for
Gaussian interactions \cite{SK}.

For $h=0$ the RS susceptibility reduces to $\chi_{\rm
RS}=\beta(\frac{1}{2}-q)$, according to (\ref{eq:chi_nofields}),
and thus obeys $\lim_{T\to 0}\chi_{\rm RS}=\sqrt{\pi}/4K$ and
$\chi_{\rm RS}=1/2T$ for $T\geq K$. Close to the critical point,
expansion  in $\tau=1-T/T_c$ reveals that $\chi_{\rm
RS}=1/2K+\order(\tau^2)$. $\chi_{\rm RS}$ thus increases from the
value $\sqrt{\pi}/4K$ at  $T=0$ to a cusp with value $1/2K$ at
$T=T_c$, followed by a monotonic $1/2T$ decay to zero in the
regime $T>T_c$ (see also figure \ref{fig:homopinning} ).

 Expression (\ref{eq:ATrot1}) for the AT instability of
rotation invariant states  simplifies similarly upon putting $h=0$
to \be \hspace*{-15mm}
 \left[\frac{T}{K}\right]^2 =
\int_0^\infty\!dr~r e^{-\frac{1}{2}r^2}\left\{
\left[\frac{I_2[\Xi(r)]}{I_0[\Xi(r)]}-\frac{I_1^2[\Xi(r)]}{I_0^2[\Xi(r)]}\right]^2
\! +\left[1\!-
\frac{I_1^2[\Xi(r)]}{I_0^2[\Xi(r)]}\right]^2\right\}
 \label{eq:ATnopinning}
\ee (with stability if the left-hand side is larger than the
right-hand side). From this, in combination with the properties
$I_{n>0}(0)=0$, we find upon inserting $q=0$ that the AT
instability occurs at $T_{\rm AT}=K$, and thus coincides with the
second order transition from a paramagnetic state to a $q>0$ state
found earlier. In both models replica symmetry breaks as soon as
we leave the paramagnetic region at $T=T_c=T_{\rm AT}=K$. Using
the properties $I_n(z)/I_0(z)=1\minus n^2/2z\plus \order(z^{-2})$
$(|z|\to\infty)$ of the modified Bessel functions
\cite{AbramStegun} we can also study the behaviour of both sides
of  (\ref{eq:ATnopinning}) for $T\to 0$. This reveals that the
degree of RS instability diverges near $T=0$: \bd \lim_{T\to
0}\frac{\rm RHS}{\rm LHS}=\int_0^\infty\!\frac{dr}{r}
e^{-\frac{1}{2}r^2}=\infty \ed

 Finally we work out  the functions
 (\ref{eq:symtrans_I},\ref{eq:symtrans_II})
who's zeros mark continuous transitions away from the
rotation-invariant state.
 For $h=0$ it follows
 from  (\ref{eq:lambda_vs_rho}) that the two types of
symmetry breaking coincide, i.e. $\Sigma_{\rm I,II}^{\rm
rot}=\Sigma_{\rm I,II}^{\rm ref}=\Sigma_{\rm I,II}$. For $h\to 0$
all integrals over $\psi$ become trivial, and the quantities
(\ref{eq:final_rot_rho1},\ref{eq:final_rot_rho2},\ref{eq:final_rot_rho3})
reduce to
 \be
 \rho_1=\frac{1}{4}\int_0^\infty\!\!dr~r
e^{-\frac{1}{2}r^2}\left\{1\minus
\frac{I_2^2[\Xi(r)]}{I_0^2[\Xi(r)]} \right\}
 \ee \be
 \rho_2=
\frac{1}{2}\int_0^\infty\!\!dr~r e^{-\frac{1}{2}r^2}
\left\{1\minus \frac{I_1^2[\Xi(r)]}{I_0^2[\Xi(r)]}\right\}
\left\{1\minus \frac{3I_1^2[\Xi(r)]}{I_0^2[\Xi(r)]}\right\}
 \ee \be
 \rho_3=
\int_0^\infty\!\!dr~r e^{-\frac{1}{2}r^2}\left\{1\minus
\frac{I_2[\Xi(r)]}{I_0[\Xi(r)]}\right\}
\frac{I_1^2[\Xi(r)]}{I_0^2[\Xi(r)]}
 \ee
For $T\geq T_c=K$, where $\Xi(r)=0$, one simply obtains
$(\rho_1,\rho_2,\rho_3)=(\frac{1}{4}, \frac{1}{2},0)$. Hence \bd
T\geq K:~~~~~~~~~~~~~
 \Sigma_{\rm I}(T)= (T/K)^2\!-1
 ~~~~~~~~~
 \Sigma_{\rm II}(T)= (T/K)^2
\!+\frac{1}{2} \ed
 Close to the critical
point, expansion  in $\tau=1-T/K$ reveals that
$(\rho_1,\rho_2,\rho_3)=(\frac{1}{4},\frac{1}{2}-2\tau,\tau)+\order(\tau^{3/2})$,
so
 \bd \tau=1\minus  T/K:~~~~~~~~
 \Sigma_{\rm I}(T)= 2\tau\plus \order(\tau^{\frac{3}{2}})
 ~~~~~~~~
 \Sigma_{\rm II}(T)=\frac{3}{2}\minus 2\tau\plus \order(\tau^{\frac{3}{2}}) \ed
Thus for $T>K$ the rotation-invariant RS solution is stable, with
in the case of model I this stability becoming marginal at
$T=T_c=K$, followed by restored stability for $T<T_c$. Close to
$T=0$ one finds
$(\rho_1,\rho_2,\rho_3)=(T\sqrt{\pi}/2K)(1,-1,2)+\order(T^2)$, so
either $\Sigma_{\rm I,II}(T)$ does not exist, or $\Sigma_{\rm
I,II}(T)=\order(T^2)$.
 \vsp

 In figure \ref{fig:homopinning} we
show our theoretical results, together with those of homogeneously
distributed pinning angles (to be studied next), by plotting the
RS order parameter $q$ and the susceptibility $\chi_{\rm RS}$ as
functions of temperature, and testing them against numerical
simulations. We find reasonable agreement, given the CPU
limitations on system size and equilibration times. In spite of
the differences between models I and II at the microscopic level
(notably in terms of frustration properties of spin loops), in the
absence of pinning fields there is no macroscopic distinction
between their physical behaviour in the replica-symmetric state,
probably due to the overruling amount of frustration. The models
have identical $q\neq 0$ and RSB transition lines and identical
values of the replica-symmetric physical observables, with a
paramagnetic state for $T>K$, and a spin-glass state  for $T<K$.
The only difference is the degree of stability of the rotation
invariant state against non-rotationally-invariant fluctuations
(since $\Sigma_{\rm I}(T)\neq \Sigma_{\rm II}(T)$), which cannot
be measured directly.

\subsection{Homogeneously distributed pinning angles}

\begin{figure}[t]
\vspace*{-5mm} \setlength{\unitlength}{1.4mm}
\begin{picture}(100,55)
\put(2,10){\epsfysize=40\unitlength\epsfbox{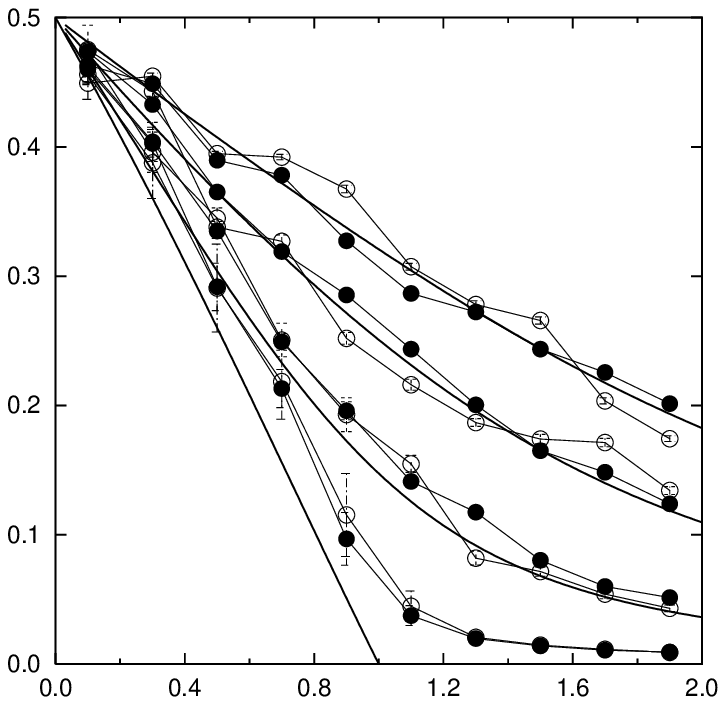}}
\put(50,10){\epsfysize=40\unitlength\epsfbox{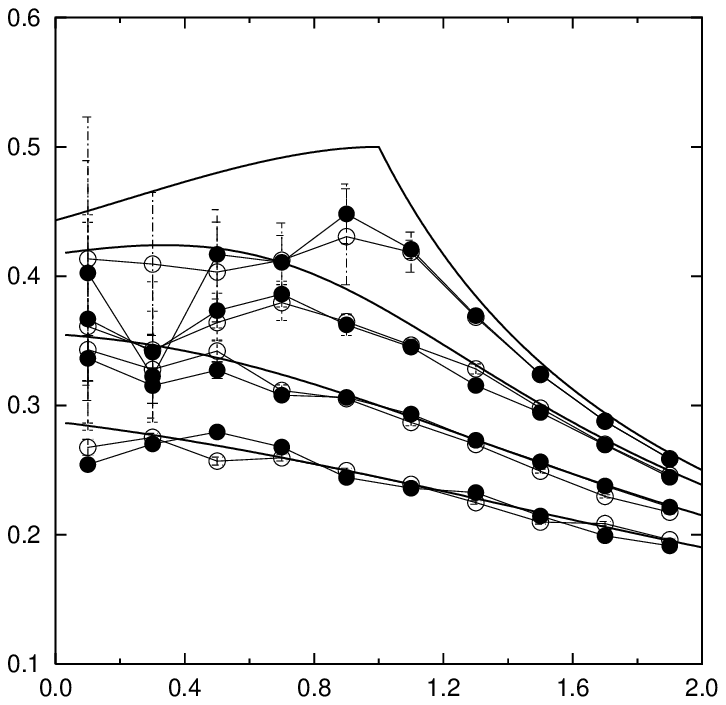}}
\put(-1,30){\large $q$} \put( 22,5){ $T/K$} \put(47,30){\large
$\chi_{\rm RS}$} \put( 70,5){ $T/K$}
\end{picture}
\vspace*{-8mm} \caption{Theoretical predictions for the RS
 order parameter $q$ (left) and the susceptibility
$\chi_{\rm RS}$ (right) as functions of temperature, for models I
and II and with $K=1$, in the case of homogeneously distributed
pinning angles. Different curves correspond to different values of
the pinning field strength $h$, taken from $\{0,1,2,3\}$ (lower to
upper in left picture, upper to lower in right picture). Connected
markers: results of numerical simulations with $N=400$ (model I:
$\bullet$, model II: o).} \label{fig:homopinning}
\end{figure}

\begin{figure}[t]
\vspace*{-5mm} \hspace*{30mm} \setlength{\unitlength}{1.4mm}
\begin{picture}(100,55)
\put(2,10){\epsfysize=40\unitlength\epsfbox{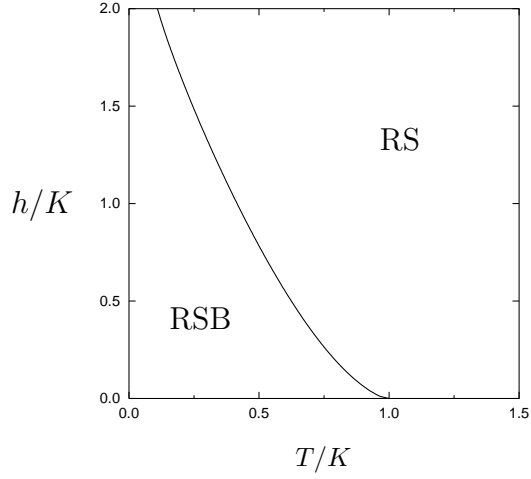}}
\put(-5,29){\large $h/K$} \put( 22,5){$T/K$} \put(10,18){\large
RSB} \put(30,35){\large RS}
\end{picture}
\vspace*{-8mm}
 \caption{The phase diagram for models I and II, in
the case of homogeneously distributed pinning angles. Solid: the
AT instability, signaling a second order transition from a replica
symmetric paramagnetic state (RS)  to a locally ordered state
without replica symmetry (RSB), and possibly to states without
global spherical symmetry. The transition field strength diverges
as $T/K\to 0$.} \label{fig:homopinning_phases}
\end{figure}

Our second choice is the homogeneous pinning angle distribution
$p(\phi)=(2\pi)^{-1}$. The Hamiltonian (\ref{eq:Hamiltonian}) is
no longer invariant under simultaneous rotation of all spins, but
we still expect the system globally to have full rotation
symmetry. We cannot simplify equations
(\ref{eq:rot_q_eqn},\ref{eq:rot_inv_f},\ref{eq:Chirot1})  further:
 \begin{eqnarray}
 q
&=& \int_0^\pi\frac{d\psi}{2\pi}\int_0^\infty\!
dr~re^{-\frac{1}{2}r^2}\left\{\frac{I_1[\Xi(r,\psi)]}{I_0[\Xi(r,\psi)]}\right\}^2
\label{eq:rot_q_eqn_again}
\end{eqnarray}
\be
 f[q]=- 2\beta K^2(q\minus \frac{1}{2})^2 -\frac{1}{\beta}
\int_{0}^{\pi}\!\! \frac{d\psi}{\pi}\int_{0}^\infty\!\!\!dr~r
e^{-\frac{1}{2}r^2}
 \log I_0[\Xi(r,\psi)]
 \label{eq:rot_inv_f_again}
 \ee
\begin{eqnarray}
\frac{\chi_{\rm RS}}{\beta}
 &=&\frac{1}{2}- q -\int_0^\pi \!\frac{d\psi}{2\pi}
 \int_0^\infty\!dr~r e^{-\frac{1}{2}r^2}
 \left[\frac{I_2[\Xi(r,\psi)]}{I_0[\Xi(r,\psi)]}-\frac{I^2_1[\Xi(r,\psi)]}{I_0^2[\Xi(r,\psi)]}\right]
 \nonumber\\
 &&\hspace*{10mm}\times\left[
\frac{h^2 \plus 4K^2r^2q\cos(2\psi) \plus
4hKr\sqrt{q}\cos(\psi)}{h^2 \plus 4K^2r^2q \plus
4hKr\sqrt{q}\cos(\psi)}\right]
 \label{eq:Chirot1_again}
\end{eqnarray}
 with
$\Xi(r,\psi)=\beta \sqrt{4qK^2 r^2+h^2+4Khr\sqrt{q}\cos(\psi)}$.
For $h\neq 0$ one no longer expects to find a phase transition
from a $q=0$ to a $q> 0$ state; this is clear upon expanding
(\ref{eq:rot_q_eqn_again}) for $\beta\to 0$, giving
$q=\frac{1}{8}(\beta h)^2+\order(\beta^3)$. For weak fields one
has $q=h^2/8(T^2-K^2)+\order(h^3)$, in the regime $T>K$. Expansion
for strong fields, $h\to \infty$, gives the leading orders \bd
q\to \frac{1}{2},~~~~~~\overline{f}\to -h,~~~~~~\chi_{\rm RS}\to
0~~~~~~(h\to \infty) \ed For $T\to 0$, on the other hand, one
finds
\be
\hspace*{-10mm}
q=\frac{1}{2}-T\int_0^{\pi}\!\frac{d\psi}{2\pi}\int_0^\infty\!\!
\frac{dr~r e^{-\frac{1}{2}r^2}}{\sqrt{h^2+2K^2
r^2+2Khr\sqrt{2}\cos(\psi)}}+\order(T^2) \ee \be \hspace*{-10mm}
\lim_{T\to 0}\chi_{\rm RS}=
\int_0^{\pi}\!\frac{d\psi}{\pi}\int_0^\infty\! dr~r
e^{-\frac{1}{2}r^2}\frac{[h+\sqrt{2}Kr\cos(\psi)]^2}{[ h^2+2K^2
r^2+2Khr\sqrt{2}\cos(\psi)]^{3/2}} \ee All expression reduce for
$h\to 0$ to those obtained in the previous sub-section (as they
should). Away from $T,h\in \{0,\infty\}$ we must resort mainly to
numerical evaluation of our equations. In figure
\ref{fig:homopinning} we compare the result of this exercise with
numerical simulations, carried out for $N=400$ and
$h\in\{0,1,2,3\}$. The agreement is satisfactory, apart from the
$h=0$ and low temperature results, where finite size effects and
equilibration problems are most prominent. We note, in comparison
with the phase diagram of figure \ref{fig:homopinning_phases},
that the the most serious deviations, observed in the
susceptibility curves, occur in the RSB region where $\chi_{\rm
RS}$ is indeed not expected to be correct.

Next we turn to our expression  (\ref{eq:ATrot1again}) for the AT
line:
 \bd \hspace*{-22mm}
 \left[\frac{T}{K}\right]^2 =
\int_0^\pi\!\frac{d\psi}{\pi}\int_0^\infty\!dr~r
e^{-\frac{1}{2}r^2}\left\{
\left[\frac{I_2[\Xi(r,\psi)]}{I_0[\Xi(r,\psi)]}-\frac{I_1^2[\Xi(r,\psi)]}{I_0^2[\Xi(r,\psi)]}\right]^2
\! +\left[1\!-
\frac{I_1^2[\Xi(r,\psi)]}{I_0^2[\Xi(r,\psi)]}\right]^2\right\} \ed
\be
 \label{eq:ATrot1again}
\ee
 For $h\to 0$ we recover
(\ref{eq:ATnopinning}) with replica instability for $T<K$.
Expanding (\ref{eq:ATrot1again}) for weak fields in the regime
$T>K$ gives $(T/K)^2=1-h^2/2(T^2\minus K^2)+\order(h\sqrt{h})$,
which has no solution; hence weak fields strengthen replica
stability for $T>K$ (as expected). For $h\to\infty$ we find
$\Xi(r,\psi)\to\beta h$, as a result of which we get, using
$I_n(z)/I_0(z)=1\minus n^2/2z\plus \order(z^{-2})$
$(|z|\to\infty)$, for the right-hand side of
(\ref{eq:ATrot1again}) the asymptotic form
$RHS=T^2/2h^2\plus\order(h^{-3})$ $(h\to\infty)$. Hence for every
nonzero temperature there is a pinning field strength above which
replica symmetry holds. Investigation of the limit $T\to 0$ in
(\ref{eq:ATrot1again}), upon again using the asymptotic forms of
the modified Bessel functions inside integrals, we now end up with
the divergent expression
\begin{eqnarray}
\hspace*{-10mm} \lim_{T\to 0}\frac{\rm RHS}{(T/K)^2} &=&
\int_{0}^\pi\!\!\frac{d\psi}{\pi} \int_{0}^\infty\!\!\!\frac{ dr~r
e^{-\frac{1}{2}r^2}}{r^2\plus \frac{1}{2}h^2/K^2\plus \sqrt{2}r
h\cos\psi/K} \nonumber\\ &=&
\int\!\frac{DxDy}{(x+h/K\sqrt{2})^2+y^2}
=e^{-\frac{1}{4}h^2/K^2}\int\!\frac{DxDy}{x^2+y^2}~e^{xh/K\sqrt{2}}\nonumber\\
 &=& e^{-\frac{1}{4}h^2/K^2}\!\int_0^\infty
\!\!\frac{dr}{r} e^{-\frac{1}{2}r^2} I_0[hr/K\sqrt{2}]=\infty
\nonumber
\end{eqnarray}
Thus the field strength $h_{\rm AT}(T)$ for which the AT
instability occurs diverges as $T/K\to 0$. For general fields and
temperatures the integrals in (\ref{eq:ATrot1again}) will have to
be evaluated numerically. This gives rise to the phase diagram
shown in figure \ref{fig:homopinning_phases}.


For homogeneously distributed pinning angles, where
$\bra\cos(4\phi)\ket_\phi=0$,
  it again follows from (\ref{eq:lambda_vs_rho}) that the two types of
rotation symmetry breaking  coincide: for all $T$ one has
$\Sigma_{\rm I,II}^{\rm rot}(T)=\Sigma_{\rm I,II}^{\rm
ref}(T)=\Sigma_{\rm I,II}(T)$. The constituents
$\{\rho_1,\rho_2,\rho_3\}$ of
(\ref{eq:symtrans_I},\ref{eq:symtrans_II}) become
 \be
 \hspace*{-10mm}
 \rho_1=\frac{1}{4}\int_0^\infty\!\!dr~r
e^{-\frac{1}{2}r^2}\!\int_0^\pi\!\frac{d\psi}{\pi}\left\{1\minus
\frac{I_2^2[\Xi(r,\psi)]}{I_0^2[\Xi(r,\psi)]} \right\}
\label{eq:homo_rho1} \ee
\be
 \hspace*{-10mm}
 \rho_2=
\frac{1}{2}\int_0^\infty\!\!dr~r
e^{-\frac{1}{2}r^2}\!\int_0^\pi\!\frac{d\psi}{\pi} \left\{1\minus
\frac{I_1^2[\Xi(r,\psi)]}{I_0^2[\Xi(r,\psi)]}\right\}
\left\{1\minus
\frac{3I_1^2[\Xi(r,\psi)]}{I_0^2[\Xi(r,\psi)]}\right\}
\label{eq:homo_rho2}  \ee \be
 \hspace*{-10mm}
 \rho_3=
\int_0^\infty\!\!dr~r
e^{-\frac{1}{2}r^2}\!\int_0^\pi\!\frac{d\psi}{\pi} \left\{1\minus
\frac{I_2[\Xi(r,\psi)]}{I_0[\Xi(r,\psi)]}\right\}
\frac{I_1^2[\Xi(r,\psi)]}{I_0^2[\Xi(r,\psi)]} \label{eq:homo_rho3}
\ee Since $|I_n[z]/I_0[z]|\leq 1$ it is  clear that
$\rho_1\in[0,\frac{1}{4}]$, $\rho_2\in[-1,\frac{1}{2}]$, and
$\rho_3\in[0,1]$. Hence, given existence of the square root, one
immediately obtains from
 (\ref{eq:symtrans_I},\ref{eq:symtrans_II})
the bounds $\Sigma_{\rm I}(T)\geq (T/K)^2-1$ and $\Sigma_{\rm
II}(T)\geq (T/K)^2-\frac{3}{4}$. Thus the rotation invariant state
is guaranteed to be  stable for $T\geq K$ (model I) and
$T>\frac{1}{2}\sqrt{3}$ (model II). For weak fields we may again,
in the regime $T>K$, expand in powers of $h$. This gives
\begin{eqnarray}
\Sigma_{\rm
I}(T)&=&\left[\frac{T}{K}\right]^2\!-1+\frac{h^2}{T^2\minus
K^2}+\order(h^3) \label{eq:weak_fields_sigmaI}
\\  \Sigma_{\rm
II}(T)&=&\left[\frac{T}{K}\right]^2\!+\frac{1}{2}+\order(h^3)
\label{eq:weak_fields_sigmaII}
\end{eqnarray}
In leading order, weak pinning fields increase the stability of
the rotation-invariant state for model I, but do not effect the
stability for model II. For strong pinning fields we expand our
equations in powers of $h^{-1}$, which gives $\Sigma_{\rm
I,II}(T)=(T/K)^2+\order(h^{-2})$, so that rotation symmetry is now
stable for any temperature.

For arbitrary pinning field strengths and temperatures the
integrals in
(\ref{eq:homo_rho1},\ref{eq:homo_rho2},\ref{eq:homo_rho3}) must be
evaluated numerically. This reveals that for all
 non-zero temperatures and all field strengths
$\Sigma_{\rm I,II}(T)\geq 0$, implying (in combination with
lacking evidence of first order transitions, and within the RS
ansatz) the prediction that the system is always in a
rotation-invariant state. Thus there are no further RS
transitions, and the phase diagram is given by figure
\ref{fig:homopinning_phases}. As with absent pinning angles,
models I and II differ only in the degree of stability of the
rotation invariant state against non-rotationally-invariant
fluctuations.

\subsection{Inhomogeneous distributions:
$p(\phi)=\frac{1}{2}\delta[\phi\minus\alpha]+\frac{1}{2}\delta[\phi\plus\alpha]$}

Finally we turn to pinning angle distributions of the form
$p(\phi)=\frac{1}{2}\delta[\phi\minus\alpha]+\frac{1}{2}\delta[\phi\plus\alpha]$,
with $\alpha\in[0,\pi]$. By varying the parameter $\alpha$ we can
control this distribution to be either uni-modal or bi-modal. One
now has $\bra\cos(2\phi)\ket_\phi=\cos(2\alpha)$ and
$\bra\cos(4\phi)\ket_\phi=\cos(4\alpha)$. Hence, according to
(\ref{eq:rot_inv_condition})  we can now have rotation-invariant
solutions only when $\alpha\in\{\pi/4,3\pi/4\}$; in the latter two
cases one has $\bra\cos(4\phi)\ket_\phi=-1$, so as far as
rotation-invariant solutions and their stability properties are
concerned the  $\alpha\in\{\pi/4,3\pi/4\}$ models behave
identically. For $\alpha\notin\{\pi/4,3\pi/4\}$ there is generally
only the overall reflection symmetry $\phi_i\to -\phi_i$ to be
exploited, and we therefore must resort mainly to evaluating our
equations for reflection symmetry states (where $q_{cs}=Q_{cs}=0$)
(\ref{eq:symRSMeasureI},\ref{eq:symRSMeasureII},\ref{eq:symm_saddle}),
 numerically.

 In view of the above  we concentrated on the cases
$\alpha\in\{0,\pi/4,\pi/2\}$, which cover bi-modal
($\alpha=\pi/4$, $\pi/2$) and uni-modal ($\alpha=0$)
distributions, and models with rotation invariant solutions
($\alpha=\pi/4$) as well as cases without ($\alpha=0$, $\pi/2$).
We tested our theory by comparison with numerical simulations. The
general effect of random pinning fields is to break the symmetry
between models I and II, and the values of otherwise identical
order parameter pairs such as $\{q_{cc},q_{ss}\}$ and
$\{Q_{cc},Q_{ss}\}$. The agreement between theory and simulations
is generally satisfactory, except for the susceptibility when
measured at low values of $T$ and $h$, where one faces
difficulties associated with equilibration and replica symmetry
breaking.

\subsection*{A: $~~p(\phi)=\frac{1}{2}\delta[\phi\minus\frac{1}{4}\pi]+\frac{1}{2}\delta[\phi\plus\frac{1}{4}\pi]$}

\begin{figure}[t]
\vspace*{-5mm} \hspace*{30mm} \setlength{\unitlength}{1.4mm}
\begin{picture}(100,55)
\put(2,10){\epsfysize=40\unitlength\epsfbox{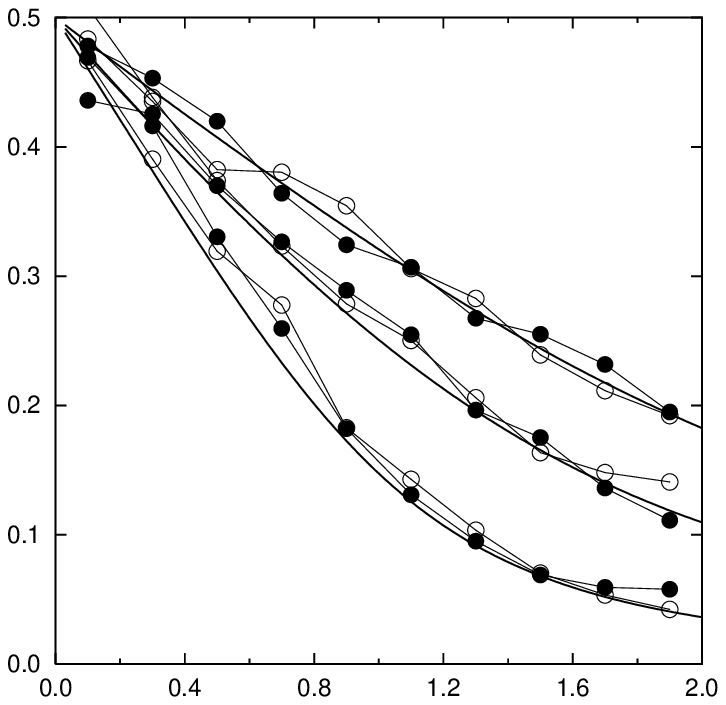}}
\put(-7,29){\large $q_{cc},q_{ss}$} \put( 22,5){$T/K$}
\end{picture}
\vspace*{-8mm} \caption{ Theoretical predictions for the RS order
parameters $q_{cc}$ and $q_{ss}$ as functions of temperature, for
model I with $K=1$, in the case of pinning angle distribution
$p(\phi)=\frac{1}{2}\delta[\phi-\frac{1}{4}\pi]+\frac{1}{2}\delta[\phi+\frac{1}{4}\pi]$
(where the theory predicts that $q_{cc}=q_{ss}$). Different curves
correspond to different values of the pinning field strength $h$,
taken from $\{1,2,3\}$ (lower to upper). Markers: results of
numerical simulations with $N=400$ ($q_{cc}$: $\bullet$, $q_{ss}$:
o). Error bars in the measurements are of the order of $0.02$; the
deviations observed are mainly finite size effects. Note that the
property $q_{cc}=q_{ss}$ is the fingerprint of rotation-invariant
states.} \label{fig:pm_pi_over_4}
\end{figure}

For $\alpha=\pi/4$ one still has rotation-invariant solutions to
our RS order parameter equations. In rotation invariant states all
results obtained earlier for $p(\phi)=(2\pi)^{-1}$ relating to the
order parameter $q$ (\ref{eq:rot_q_eqn_again}), the free energy
per oscillator (\ref{eq:rot_inv_f_again}), the RS susceptibility
$\chi_{\rm RS}$ (\ref{eq:Chirot1_again}) and the AT line
(\ref{eq:ATrot1again}) can be taken over without alteration. The
difference with $p(\phi)=(2\pi)^{-1}$ is in the stability
properties against fluctuations which violate rotational symmetry.
Since here $\bra\cos(4\pi)\ket_\phi=-1$, we now obtain the
symmetry breaking transitions as the zeros of the following
functions (using (\ref{eq:lambda_vs_rho})): \begin{eqnarray}
\hspace*{-10mm} \Sigma^{\rm rot}_{\rm I}(T)&=&
(T/K)^2\!-\sqrt{(\rho_1(\minus 1) \minus\rho_2(\minus
1))^2\minus\rho_3^2(\minus 1)}-\rho_1(\minus 1) -\rho_2(\minus 1)
\label{eq:break_rot_I} \\
 \hspace*{-10mm}
\Sigma^{\rm ref}_{\rm I}(T)&=& (T/K)^2\!-\sqrt{(\rho_1(\plus 1)
\minus\rho_2(\plus 1))^2\minus\rho_3^2(\plus 1)}-\rho_1(\plus 1)
-\rho_2(\plus 1) \label{eq:break_ref_I}
 \\
 \hspace*{-10mm}
\Sigma^{\rm rot}_{\rm II}(T)&=& (T/K)^2\!-\sqrt{(\rho_1(\minus
1)\minus\rho_2(\minus 1))^2\minus\rho_3^2(\minus 1)}+\rho_1(\minus
1)+\rho_2(\minus 1) \label{eq:break_rot_II} \\ \hspace*{-10mm}
\Sigma^{\rm ref}_{\rm II}(T)&=& (T/K)^2\!-\sqrt{(\rho_1(\plus
1)\minus\rho_2(\plus 1))^2\minus\rho_3^2(\plus 1)}+\rho_1(\plus
1)+\rho_2(\plus 1)
 \label{eq:break_ref_II}
\end{eqnarray}
with
 \bd
 \hspace*{-25mm}
 \rho_1(\kappa)=\frac{1}{4}\int_0^\infty\!\!dr~r
e^{-\frac{1}{2}r^2}\!\int_0^\pi\!\frac{d\psi}{\pi}\left\{1\minus
\frac{I_2^2[\Xi(r,\psi)]}{I_0^2[\Xi(r,\psi)]} \right\} \ed \bd
 \hspace*{-15mm}
 +\frac{1}{4}\kappa \int_0^\infty\!\!dr~r
e^{-\frac{1}{2}r^2}\!\int_0^\pi\!\frac{d\psi}{\pi}
\left\{\frac{I_4[\Xi(r,\psi)]}{I_0[\Xi(r,\psi)]}\minus\frac{I_2^2[\Xi(r,\psi)]}{I_0^2[\Xi(r,\psi)]}
\right\} \ed
\be
\hspace*{15mm} \times \left\{1\minus
\frac{8\beta^4[2Kr\sqrt{q}\sin(\psi)]^2[h\plus
2Kr\sqrt{q}\cos(\psi)]^2}{\Xi^4(r,\psi)}\right\}
\label{eq:piover4_rot_rho1} \ee \bd
 \hspace*{-25mm}
 \rho_2(\kappa)=
\frac{1}{2}\int_0^\infty\!\!dr~r
e^{-\frac{1}{2}r^2}\!\int_0^\pi\!\frac{d\psi}{\pi} \left\{1\minus
\frac{I_1^2[\Xi(r,\psi)]}{I_0^2[\Xi(r,\psi)]}\right\}
\left\{1\minus
\frac{3I_1^2[\Xi(r,\psi)]}{I_0^2[\Xi(r,\psi)]}\right\} \ed \bd
 \hspace*{-25mm} +\frac{1}{2}\kappa \int_0^\infty\!\!dr~r
e^{-\frac{1}{2}r^2}\!\int_0^\pi\!\frac{d\psi}{\pi}
\left\{\frac{I_2[\Xi(r,\psi)]}{I_0[\Xi(r,\psi)]}\minus
\frac{I_1^2[\Xi(r,\psi)]}{I_0^2[\Xi(r,\psi)]} \right\}
\left\{\frac{I_2[\Xi(r,\psi)]}{I_0[\Xi(r,\psi)]}\minus
\frac{3I_1^2[\Xi(r,\psi)]}{I_0^2[\Xi(r,\psi)]} \right\} \ed
\be
\hspace*{15mm} \times \left\{1\minus
\frac{8\beta^4[2Kr\sqrt{q}\sin(\psi)]^2[h\plus
2Kr\sqrt{q}\cos(\psi)]^2}{\Xi^4(r,\psi)}\right\}
\label{eq:piover4_rot_rho2} \ee \bd
 \hspace*{-25mm}
 \rho_3(\kappa)=
\int_0^\infty\!\!dr~r
e^{-\frac{1}{2}r^2}\!\int_0^\pi\!\frac{d\psi}{\pi} \left\{1\minus
\frac{I_2[\Xi(r,\psi)]}{I_0[\Xi(r,\psi)]}\right\}
\frac{I_1^2[\Xi(r,\psi)]}{I_0^2[\Xi(r,\psi)]}
 \ed \bd
 \hspace*{-20mm} +\kappa \int_0^\infty\!\!dr~r
e^{-\frac{1}{2}r^2}\!\int_0^\pi\!\frac{d\psi}{\pi} \left\{1\minus
\frac{I_2[\Xi(r,\psi)]}{I_0[\Xi(r,\psi)]} \right\}
\left\{\frac{I_1^2[\Xi(r,\psi)]}{I^2_0[\Xi(r,\psi)]}\minus
\frac{2I_2[\Xi(r,\psi)]}{I_0[\Xi(r,\psi)]} \right\} \ed
\be
\hspace*{15mm} \times \left\{1\minus
\frac{8\beta^4[2Kr\sqrt{q}\sin(\psi)]^2[h\plus
2Kr\sqrt{q}\cos(\psi)]^2}{\Xi^4(r,\psi)}\right\}
\label{eq:piover4_rot_rho3} \ee The inequality
$|I_n[z]/I_0[z]|\leq 1$ allows one to obtain the crude general
bound $T/K\leq 3$ for the zeros of
(\ref{eq:break_rot_I},\ref{eq:break_ref_I},\ref{eq:break_rot_II},\ref{eq:break_ref_II}).
For strong pinning fields one has \bd
(\rho_1,\rho_2,\rho_3)=\frac{1\minus \kappa}{\beta
h}(1,-1,2)+\order(h^{-2})~~~~~~~~~~(h\to \infty) \ed and thus
$\Sigma_{\rm I,II}^{\rm rot,ref}(T)=(T/K)^2+\order(h^{-2})$,
confirming stability of the rotation invariant state  for
sufficiently strong fields.
 For weak fields we may
again in the regime $T>K$ expand in powers of $h$. This reveals
that those terms in $\{\rho_1,\rho_2,\rho_3\}$ which are
proportional to $\kappa$ are of sub-leading order $\order(h^{-3})$
as $h\to 0$. From this it follows that we revert back to
expressions
(\ref{eq:weak_fields_sigmaI},\ref{eq:weak_fields_sigmaII}).
Apparently, the leading order effect of switching on the pinning
fields is again to stabilize the rotation invariant state.

For intermediate values of temperatures and fields we have
evaluated
(\ref{eq:break_rot_I},\ref{eq:break_ref_I},\ref{eq:break_rot_II},\ref{eq:break_ref_II})
numerically, which showed that for all
 non-zero temperatures and all field strengths
one has $\Sigma^{\rm rot,ref}_{\rm I,II}(T)\geq 0$. This, in turn,
implies (in combination with lacking evidence of first order
transitions, and within the RS ansatz) the non-trivial prediction
that for
$p(\phi)=\frac{1}{2}\delta[\phi-\frac{1}{4}\pi]+\frac{1}{2}\delta[\phi+\frac{1}{4}\pi]$,
which is a pinning angle distribution {\em without} rotation
invariance, the system is still {\em always} in a
rotation-invariant state. This statement and its quantitative
consequences are confirmed convincingly for model I by the
numerical simulation data shown in \ref{fig:pm_pi_over_4}; similar
results can be shown for model II. We conclude that for
$p(\phi)=\frac{1}{2}\delta[\phi-\frac{1}{4}\pi]+\frac{1}{2}\delta[\phi+\frac{1}{4}\pi]$
the phase diagram in the $(T/K,h/K)$ plane for both models I and
II is simply identical to that of homogeneously distributed
pinning angles, as shown in figure \ref{fig:homopinning_phases}.

\subsection*{B: $~~p(\phi)=\delta[\phi]$}

For $p(\phi)=\delta[\phi]$ we no longer have rotation invariant
solutions, and the analysis becomes more complicated. Given the
assumption of overall reflection symmetry ($Q_{cs}=q_{cs}=0$,
whose stability we will calculate below) we are left with three RS
order parameters, $\{Q_{cc},q_{cc},q_{ss}\}$, and with the
effective measures
(\ref{eq:symRSMeasureI},\ref{eq:symRSMeasureII}) which now become
\begin{eqnarray}
\hspace*{-15mm}
 M_{\rm I}(\theta|x,y)&=& e^{\beta
h\cos(\theta) +(\beta K)^2[2Q_{cc}-1+q_{ss}-q_{cc}]\cos(2\theta)
+2\beta K[\cos(\theta)x \sqrt{q_{cc}} +\sin(\theta)y
\sqrt{q_{ss}}]} \nonumber
\\
\label{eq:MI_phi0}
\\
 \hspace*{-15mm} M_{\rm I\!I}(\theta|x,y)&=&
 e^{\beta h\cos(\theta)
-(\beta K)^2[2Q_{cc}-1+q_{ss}-q_{cc}]\cos(2\theta) +2\beta
K[\cos(\theta)x\sqrt{q_{ss}} +\sin(\theta)y\sqrt{q_{cc}}]}
\nonumber\\ \label{eq:MII_phi0}
\end{eqnarray}
The remaining order parameters are to be solved from  the coupled
equations
 \be \hspace*{-10mm} Q_{cc}=
\bra\bras \bra \cos^2(\theta)\ket_\star \kets\ket_{\phi}~~~~~~
 q_{cc}=
\bra\bras\bra\cos(\theta)\ket_\star^2\kets\ket_{\phi}~~~~~~
q_{ss}= \bra\bras\bra\sin(\theta)\ket_\star^2\kets\ket_{\phi}
\label{eq:RSeqns_phi0} \ee At the RS ground state  we know that
$q_{cc}\plus q_{ss}=1$ and $Q_{cc}=\frac{1}{2}[1\plus q_{cc}\minus
q_{ss}]$. Due to $p(\phi)=\delta[\phi]$ the RS susceptibility
(\ref{eq:pinning_chi}) simplifies immediately to:
\begin{eqnarray}
\chi_{\rm RS}&=&\beta(1 -Q_{cc}- q_{ss})
\end{eqnarray}
For strong pinning fields one finds $\lim_{h\to
\infty}Q_{cc}=\lim_{h\to \infty}q_{cc}=1$ and $\lim_{h\to
\infty}q_{ss}=0$. Hence $\lim_{h\to\infty}\chi_{\rm RS}=0$.
 In the high temperature
regime an expansion in powers of $\beta$ shows the solution of
(\ref{eq:RSeqns_phi0}) to behave as
\be
Q_{cc}=\frac{1}{2}+\order(\beta^2),~~~~~~q_{cc}=\frac{1}{4}\beta^2h^2\!+\order(\beta^3),~~~~~~
q_{ss}=\order(\beta^3) \label{eq:largeT_phi0} \ee Hence $\chi_{\rm
RS}=1/2T+\order(T^{-3})$ as $T\to \infty$.

The expressions for the AT line(s)
(\ref{eq:pinning_AT_I},\ref{eq:pinning_AT_II}) and reflection
symmetry transitions
(\ref{eq:reflec_trans_I},\ref{eq:reflec_trans_II}) cannot be
simplified further, except in special limits and in the special
case described below. For $h\to\infty$ one trivially extracts from
these equations that both replica symmetry and reflection symmetry
are stable for all finite temperatures; the same is true for high
temperatures and arbitrary field strengths (as expected). \vsp

\begin{figure}[t]
\vspace*{-5mm} \setlength{\unitlength}{1.4mm}
\begin{picture}(100,55)
\put(2,10){\epsfysize=40\unitlength\epsfbox{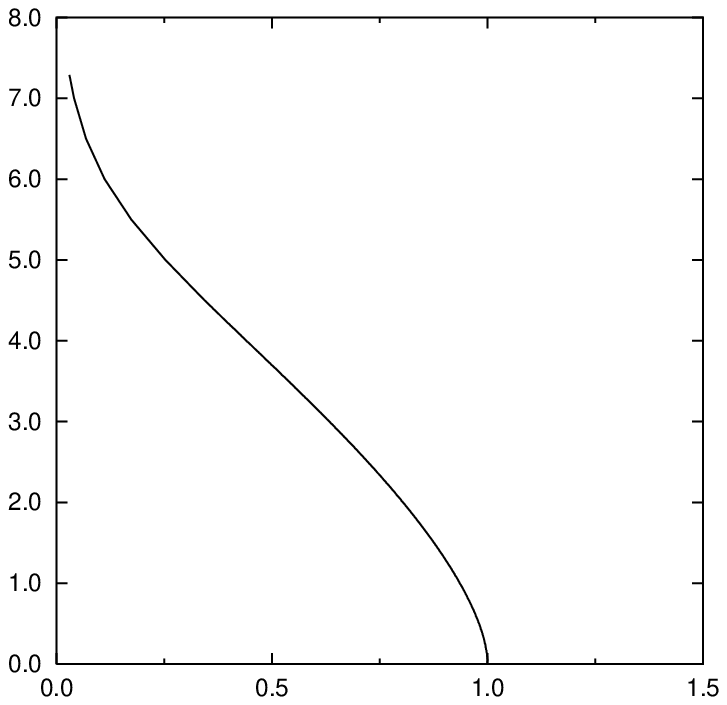}}
\put(50,10){\epsfysize=40\unitlength\epsfbox{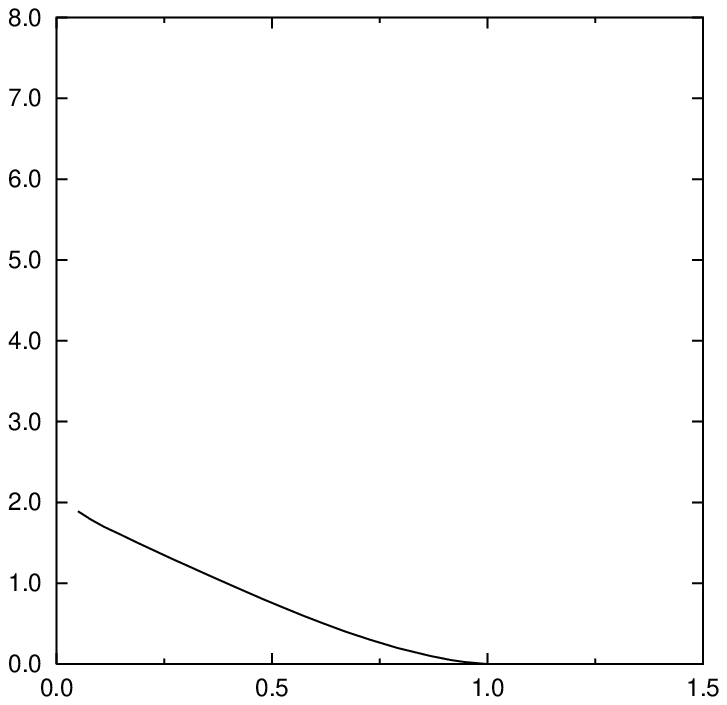}}
\put(-3,30){ $h/K$} \put( 22,5){$T/K$} \put(45,30){ $h/K$} \put(
70,5){$T/K$} \put(22,35){\large RS, $q_{ss}\!=\!0$}
\put(76,35){\large RS} \put(9,15){\large RSB, $q_{ss}\!\neq\!
0$}\put(57,14){\large RSB}
\end{picture}
\vspace*{-8mm} \caption{ The phase diagrams for models I (left)
and II (right),  in the case of $p(\phi)=\delta[\phi]$.
 Solid: the AT instability, signaling a second order transition
from a replica symmetric state (RS) to a state without replica
symmetry (RSB). For model I the AT line coincides with line
marking a continuous bifurcation of $q_{ss}\neq 0$ solutions of
the RS order parameter equations. } \label{fig:phi0diagrams}
\end{figure}

However, analytical progress can be made for model I. Here one
observes, in line with (\ref{eq:largeT_phi0}), that the order
parameter equations allow for solutions with $q_{ss}=0$. Given the
identification $q_{ss}=\lim_{N\to \infty}N^{-1}\sum_i
\overline{\bra\sin(\theta_i)\ket^2}$ such solutions imply that
$\bra\sin(\theta_i)\ket=0$ for each oscillator. This can be
understood as a result of the action of the pinning fields, which
for $p(\phi)=\delta(\phi)$ tend to drive the oscillator phases
towards $\phi_i=0$. Insertion of $q_{ss}=0$ as an ansatz
 into (\ref{eq:MI_phi0})  gives $\bra
 \sin(\theta)\ket_\star=\bra\sin(\theta)\cos(\theta)\ket_\star=0$,
 confirming $q_{ss}=0$ self-consistently (this is not possible for
 model II). This results in considerable analytical simplifications, such as
 $\gamma_{cs}=0$ and $\gamma_{ss}=\bra\sin^2(\theta)\ket_\star$,
 and also reduces the computational effort in the various numerical integrations.
One now has only two RS order parameters, $Q_{cc}$ and $q_{cc}$,
to be solved from
\be
Q_{cc}=\int\!Dx~ \bra \cos^2(\theta)\ket_\star,~~~~~~~~
q_{cc}=\int\!Dx~\bra \cos(\theta)\ket_\star^2 \ee with the
simplified effective measure
 \be
 M_{\rm I}(\theta|x)= e^{\beta\cos(\theta)[
h+2 K x \sqrt{q_{cc}}]
 +(\beta K)^2[2Q_{cc}-1-q_{cc}]\cos(2\theta)  } \ee
The ground state has $Q_{cc}=q_{cc}=1$. The equation for the AT
line of model I becomes
 \be
 \hspace*{-15mm}
  (T/2K)^2={\rm max}\left\{
\int\!\!Dx~\left[\bra\cos^2(\theta)\ket_\star\minus
\bra\cos(\theta)\ket_\star^2\right]^2,\int\!\!Dx~\bra\sin^2(\theta)\ket_\star^2\right\}
\label{eq:qss_ATline}
 \ee
Similarly we can for $q_{ss}=0$ simplify the constituents
$\{\lambda_i\}$ of (\ref{eq:reflec_trans_I}) to
 \begin{eqnarray}
 \lambda_1
 &=&
2\int\!Dx~
 \bra
 \sin^2(\theta)\cos^2(\theta)
 \ket_\star
\\
\lambda_2&=& 2\int\!Dx~ \bra\sin^2(\theta)\ket_\star
 [\bra\cos^2(\theta)\ket_\star\minus 2\bra\cos(\theta)\ket_\star^2]
 \\
\lambda_3&=& 4\int\!Dx~
   \bra\sin^2(\theta)\cos(\theta)\ket_\star\bra \cos(\theta)\ket_\star
\end{eqnarray}
These are to be inserted into (\ref{eq:reflec_trans_I}), whose
zeros mark  reflection symmetry breaking. Finally we now have a
third type of transition: de-stabilization of $q_{ss}= 0$. The
condition for this, $\partial^2\overline{f}_{I}[\ldots]/\partial
q_{ss}^2=0$, can with help of \ref{app:fderivatives} be written as
\be
(T/2K)^2 =\int\!Dx~\bra\sin^2(\theta)\ket_\star^2
\label{eq:qss_bifline} \ee
 Clearly, for $h\to 0$ (absent pinning
fields, as studied in the first part of this section) the
$q_{ss}\neq 0$ bifurcation occurs at $T=K$, and coincides with a
bifurcation of $q_{cc}\neq 0$ and with the AT line. Comparison
with (\ref{eq:qss_ATline}) shows that the critical temperature
$T_c$ defined by (\ref{eq:qss_bifline}) obeys $T_c\leq T_{\rm
AT}$. Hence the AT instability does occur for $q_{ss}=0$ and is
thus given by (\ref{eq:qss_ATline}). Numerical analysis reveals
that along the line (\ref{eq:qss_ATline}) one always has
$\int\!Dx~\left[\bra\cos^2(\theta)\ket_\star\minus
\bra\cos(\theta)\ket_\star^2\right]^2\leq
\int\!Dx~\bra\sin^2(\theta)\ket_\star^2$, and hence $T_{AT}=T_c$
(the two transition lines coincide, in agreement with
\cite{Cragg,Elder83}).

\begin{figure}[t]
\vspace*{-5mm} \setlength{\unitlength}{1.4mm}
\begin{picture}(100,55)
\put(2,10){\epsfysize=40\unitlength\epsfbox{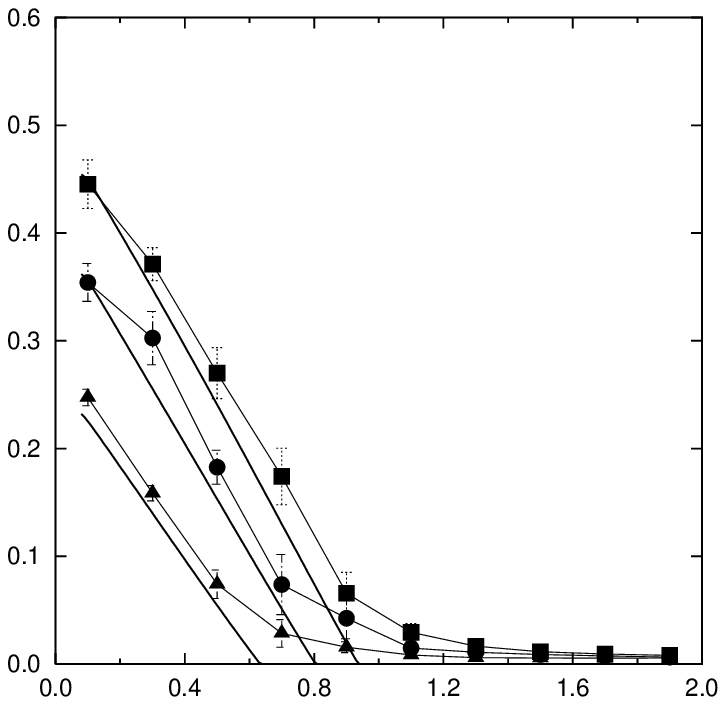}}
\put(50,10){\epsfysize=40\unitlength\epsfbox{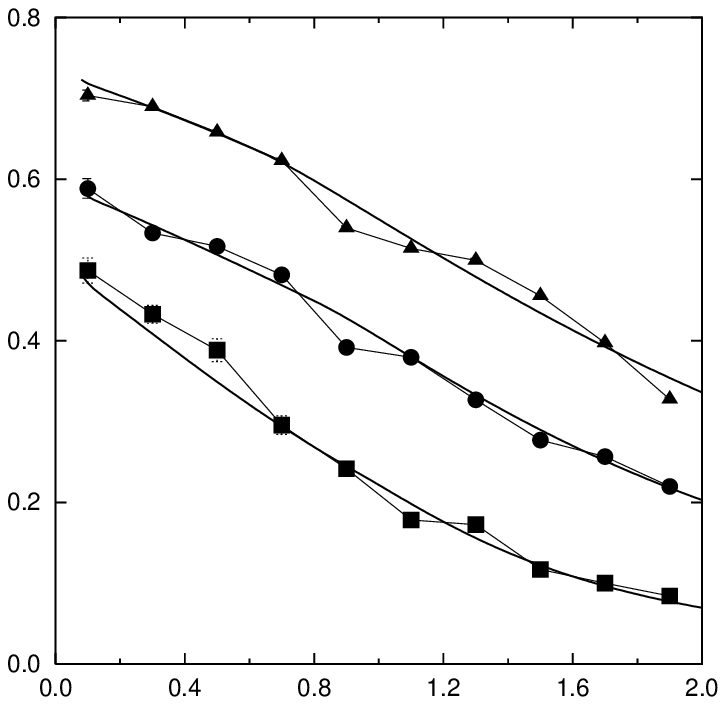}}
\put(-1,30){\large $q_{ss}$} \put( 22,5){$T/K$} \put(47,30){\large
$q_{cc}$} \put( 70,5){$T/K$}
\end{picture}
\vspace*{-15mm}

\setlength{\unitlength}{1.4mm}
\begin{picture}(100,55)
\put(2,10){\epsfysize=40\unitlength\epsfbox{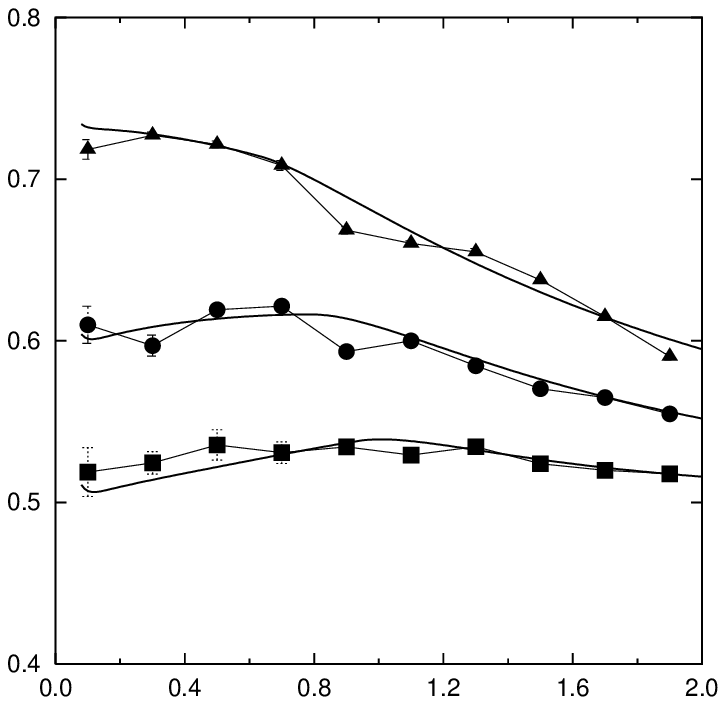}}
\put(50,10){\epsfysize=40\unitlength\epsfbox{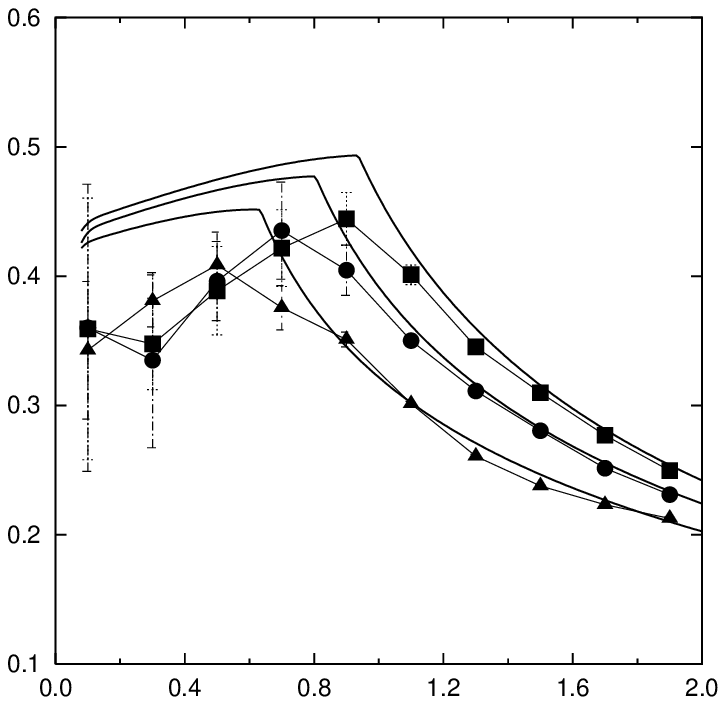}}
\put(-2,30){\large $Q_{cc}$} \put( 22,5){$T/K$} \put(46,30){\large
$\chi_{\rm RS}$} \put( 70,5){$T/K$}
\end{picture}
 \vspace*{-8mm} \caption{ The theoretical predictions for the RS
order parameters $\{q_{ss},q_{cc},Q_{cc}\}$ and the RS
susceptibility $\chi_{\rm RS}$ as functions of temperature, for
model I with $K=1$, in the case of $p(\phi)=\delta[\phi]$. Thick
solid lines: theoretical predictions. Markers: results of
numerical simulations with $N=800$. Different curves correspond to
different values of the pinning field strength $h$, taken from
$\{1,2,3\}$ (indicated by squares, circles and triangles,
respectively).} \label{fig:phi0_modelI}
\end{figure}

\begin{figure}[t]
\vspace*{-5mm} \setlength{\unitlength}{1.4mm}
\begin{picture}(100,55)
\put(2,10){\epsfysize=40\unitlength\epsfbox{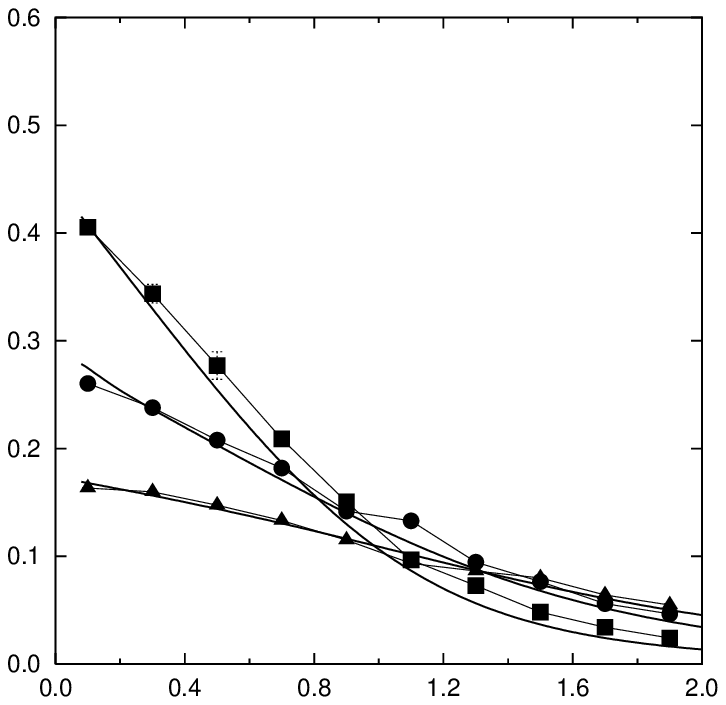}}
\put(50,10){\epsfysize=40\unitlength\epsfbox{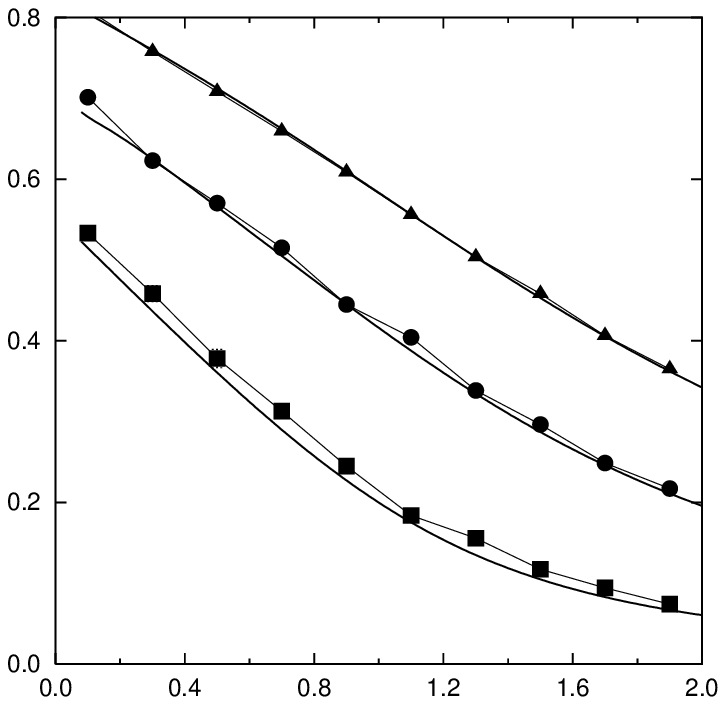}}
\put(-1,30){\large $q_{ss}$} \put( 22,5){$T/K$} \put(47,30){\large
$q_{cc}$} \put( 70,5){$T/K$}
\end{picture}
\vspace*{-15mm}

\setlength{\unitlength}{1.4mm}
\begin{picture}(100,55)
\put(2,10){\epsfysize=40\unitlength\epsfbox{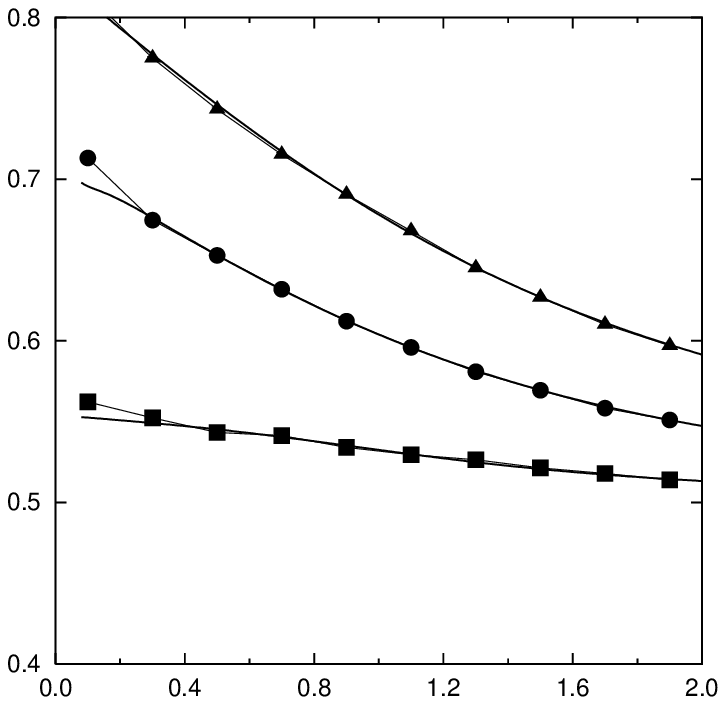}}
\put(50,10){\epsfysize=40\unitlength\epsfbox{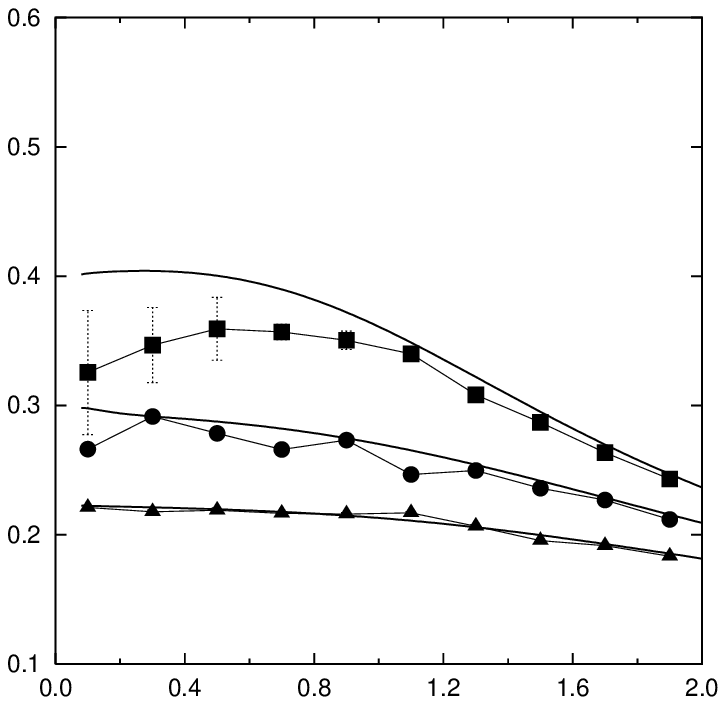}}
\put(-2,30){\large $Q_{cc}$} \put( 22,5){$T/K$} \put(46,30){\large
$\chi_{\rm RS}$} \put( 70,5){$T/K$}
\end{picture}
 \vspace*{-8mm} \caption{ The theoretical predictions for the RS
order parameters $\{q_{ss},q_{cc},Q_{cc}\}$ and the RS
susceptibility $\chi_{\rm RS}$ as functions of temperature, for
model II with $K=1$, in the case of $p(\phi)=\delta[\phi]$. Thick
solid lines: theoretical predictions. Markers: results of
numerical simulations with $N=800$. Different curves correspond to
different values of the pinning field strength $h$, taken from
$\{1,2,3\}$ (indicated by squares, circles and triangles,
respectively).} \label{fig:phi0_modelII}
\end{figure}

Numerical solution of equations
(\ref{eq:pinning_AT_I},\ref{eq:pinning_AT_II}) and
(\ref{eq:qss_bifline}) lead us to the phase diagrams shown in
figure \ref{fig:phi0diagrams}. The $q_{ss}\neq 0$ transition for
model I coincides with the AT line. We observe that the RS
stabilizing effect of the pinning fields is much greater for model
II than for model I, as reflected in a significantly lower RSB
transition temperature for any nonzero field strength. Numerical
evaluation of the conditions
(\ref{eq:reflec_trans_I},\ref{eq:reflec_trans_II}) marking the
continuous breaking of reflection symmetry shows for both models
that this does not happen at any finite temperature; reflection
symmetry remains locally stable. \vsp

The results of comparing the solutions of our RS order parameter
equations (obtained numerically) with simulations are shown in
figures \ref{fig:phi0_modelI} (model I) and \ref{fig:phi0_modelII}
(model II). It is clear from the figures \ref{fig:phi0_modelI} and
\ref{fig:phi0_modelII} that, due to the breaking of global
rotation symmetry by the external pinning fields, there is a now a
non-trivial difference between the macroscopic behaviour of our
two models. The energetic conflicts in the two models are resolved
differently. With the exception of the susceptibility, the
agreement between theory and numerical experiment is good, taking
into account finite size effects which lead as usual to a
smoothening of the second order phase transition marking
$q_{ss}\neq 0$ bifurcations. Comparison with the locations of the
AT lines in figure \ref{fig:phi0diagrams} suggests again that the
serious deviations in the susceptibility are due to replica
symmetry breaking; $\chi_{\rm RS}$ cannot be expected to be
correct in the RSB region.

\subsection*{C: $~~p(\phi)=\frac{1}{2}\delta[\phi\minus\frac{1}{2}\pi]
+\frac{1}{2}\delta[\phi\plus\frac{1}{2}\pi]$}

Our final example is a bi-modal pinning angle distribution which,
due to $\bra\cos(2\theta)\ket_\phi=-1\neq 0$, again does not allow
for rotation invariant states. According to transformation
(\ref{eq:gauge}) (with $\eta=\lambda=0$) this choice is related to
the case $p(\phi)=\delta[\phi]$ by a simple gauge transformation,
and our models must have identical free energies and phase
diagrams. As a consistency test we will try to extract this
property from the RS saddle-point equations. Given reflection
symmetry, the two effective measures
(\ref{eq:symRSMeasureI},\ref{eq:symRSMeasureII}) now become
\begin{eqnarray}
\hspace*{-25mm}
 M_{\rm I}(\theta|x,y,\pm\frac{\pi}{2})&=& e^{\pm\beta
h\sin(\theta) +(\beta K)^2[2Q_{cc}-1+q_{ss}-q_{cc}]\cos(2\theta)
+2\beta K[\cos(\theta)x \sqrt{q_{cc}} +\sin(\theta)y
\sqrt{q_{ss}}]}\nonumber \\
\label{eq:finalex_MI}
\\
\hspace*{-25mm} M_{\rm I\!I}(\theta|x,y,\pm\frac{\pi}{2})&=&
 e^{\pm\beta h\sin(\theta)
-(\beta K)^2[2Q_{cc}-1+q_{ss}-q_{cc}]\cos(2\theta) +2\beta
K[\cos(\theta)x\sqrt{q_{ss}} +\sin(\theta)y\sqrt{q_{cc}}]}
\nonumber \\ \label{eq:finalex_MII}
\end{eqnarray}
We observe that our models can indeed be mapped onto those
obtained for $p(\phi)=\delta[\phi]$. We introduce the
transformation $\theta=\pm (\frac{1}{2}\pi-\theta^\prime)$
(permutation and reflection of axes), which, via the order
parameter equations (\ref{eq:RSeqns_phi0}), induces a
corresponding transformation of the order parameters:
\be
Q_{cc}=1-Q_{cc}^\prime,~~~~~~q_{cc}=q_{ss}^\prime,~~~~~~q_{ss}=q_{cc}^\prime
\label{eq:mapping_models} \ee The effect on the measures
(\ref{eq:finalex_MI},\ref{eq:finalex_MII}) is that the latter can
now be expressed in terms of the measures
(\ref{eq:MI_phi0},\ref{eq:MII_phi0}) describing the distribution
$p(\phi)=\delta[\phi]$ studied previously: \bd \hspace*{-5mm}
M_{\rm I}(\theta|x,y,\pm\frac{\pi}{2}) = M_{\rm
I}(\theta^\prime|\pm y,x) ~~~~~~~~ M_{\rm
II}(\theta|x,y,\pm\frac{\pi}{2}) = M_{\rm II}(\theta^\prime|\pm
y,x) \ed Since operations such as $(x,y)\to (\pm y,x)$ have no
physical consequences, and since all observables and transition
lines are (within RS) constructed from the measures
(\ref{eq:finalex_MI},\ref{eq:finalex_MII}), we conclude that the
macroscopic physics of the cases $p(\phi)=\delta[\phi]$ and
$p(\phi)=\frac{1}{2}\delta[\phi\minus\frac{1}{2}\pi]
+\frac{1}{2}\delta[\phi\plus\frac{1}{2}\pi]$ are indeed identical.
The relations (\ref{eq:mapping_models}) map the order parameters
$\{Q_{cc},q_{cc},q_{ss}\}$ of the present bi-modal distribution to
the parameters $\{Q_{cc}^\prime,q_{cc}^\prime,q_{ss}^\prime\}$ of
the  case $p(\phi)=\delta[\phi]$, and both cases have identical
free energies for all values of  $\{T/K,h/K\}$ and identical phase
diagrams. In particular, the $q_{ss}=0$ solution of model I for
$p(\phi)=\delta[\phi]$ corresponds here to a solution with
$q_{cc}=0$ (again for model I only),  and  the RS susceptibility
now becomes $\chi_{\rm RS}=\beta(Q_{cc}- q_{cc})$. The equivalence
of the $p(\phi)=\delta[\phi]$ and
$p(\phi)=\frac{1}{2}\delta[\phi\minus\frac{1}{2}\pi]
+\frac{1}{2}\delta[\phi\plus\frac{1}{2}\pi]$ models is also
immediately obvious in numerical simulations (which we do not show
for brevity) .

\section{Discussion}

We have studied models of frustrated coupled Kuramoto oscillators
in the presence of random pinning fields. To simplify our problem
we have assumed that the natural frequency of all oscillators are
the same and that the pinning fields distribution is
reflection-symmetric, i.e. $p(\phi)=p(-\phi)$ (a simple gauge
transform allows one to map the present models also to those with
larger families of pinning field distributions). We have
calculated the disorder-averaged free energy using the replica
method for two types of random pair interactions (model I and
model II, the first we standard ferro- and anti-ferromagnetic
interactions, the second with chiral interactions), which differ
in the level of frustration. In terms of e.g. oscillator triplets,
model I is partially bond-frustrated, whereas model II is fully
bond-frustrated. For small system sizes the two models have
significantly different ground state properties. Our main interest
here is to understand the difference(s) in physical behaviour
between the two models in equilibrium, in the thermodynamic limit.
At the mathematical level our two models differ in the details of
the effective (replicated) single-spin measure, but they have the
same types of order parameters. Within the replica symmetric
ansatz, one finds different expressions for the AT instabilities;
more unexpectedly, the two models are also found to differ in the
nature of the extremum of the free energy.

We inspected the effects of different types of symmetries which
our models could inherit from the pinning field distribution.
Global reflection symmetry leads to a simplified measure and a
reduced number of order parameters. Reflection symmetry breaking
transitions are found to be possible at most in the RSB regime. In
the case of global rotation symmetry, only one order parameter
remains (within RS), and the effective measure, the free enrgy,
and the AT lines are identical for both models. We analyzed our
order parameter equations in detail for four specific choices for
the pinning fields, and tested our theoretical predictions against
computer simulations. Without pinning fields, our two models both
behave identically to a standard long-range XY model with
non-chiral Gaussian interactions \cite{SK}. The same is found to
be true for
 uniformly distributed pinning angles.
 The agreement between theory and simulation is quite
 satisfactory, modulo finite size effects,
except for the susceptibility at low temperatures, where RSB
effects in combination with  equilibration difficulties play a
major role. We then studied inhomogeneous pinning angle
distributions, of the form
$p(\phi)=\frac{1}{2}\delta[\phi\minus\alpha]+\frac{1}{2}\delta[\phi\plus\alpha]$.
For the special value $\alpha=\frac{\pi}{4}$ one finds,
remarkably, that our models are again always in a rotation
invariant state (in spite of the fact that the pinning field
distribution does not have rotation invariance); this is confirmed
by simulations. For $\alpha=0$ the symmetry between our two models
is broken. They now differ significantly in terms of the location
of the AT line, which for model I in addition coincides with a
further order parameter  bifurcation, as in \cite{Cragg,Elder83}.
The $\alpha0{\pi}{2}$ case can be mapped onto the $\alpha=0$ one,
by a suitable gauge transformation. Again, there is a good
agreement between theory and simulations.

As a next stage it would be interesting to explore the behavior of
these models for oscillators with random distributions of natural
frequencies, as well as in the presence of more general chiral
interactions. In both extensions/generalizations the standard
equilibrium replica formalism can no longer be used; in the first
case  one has to rely on dynamical formalisms
\cite{Bonil1,Bonil2}, whereas in the second case one has to call
upon renormalization tricks \cite{Raeetal}.

\subsubsection*{Acknowledgements}

The authors are grateful for support from the British Council (UK)
and the Mcyt (grant BSM2000-0696) (Spain), and from the European
Science Foundation (under the SPHINX programme).

\clearpage

\appendix

\section{Derivatives of the RS free energy}
\label{app:fderivatives}

 The (first and second order) derivatives of $f[\ldots]$ in (\ref{eq:fRS}) with
respect to the RS order parameters are calculated by working out
the following general relations (where $\gamma$ and
$\gamma^\prime$ denote any two order parameters from the set
$\{Q_{\star\star},q_{\star\star}\}$):
\begin{eqnarray*}
\hspace*{-15mm} \frac{1}{\beta K^2}
 \frac{\partial
 f}{\partial \gamma}&=& \frac{\partial
 U}{\partial\gamma} -\frac{1}{\beta^2 K^2 }\bigbra\bigbras \bra
 \frac{\partial\log M(\ldots)}{\partial \gamma}\ket_\star
 \bigkets\bigket_{\!\!\phi}
\\
\hspace*{-15mm} \frac{1}{\beta K^2}
 \frac{\partial^2
 f}{\partial \gamma \partial\gamma^\prime}&=&
 \frac{\partial^2
 U}{\partial\gamma\partial\gamma^\prime} -\frac{1}{\beta^2 K^2 }\bigbra\bigbras
 \bra
 \frac{\partial^2\log M(\ldots)}{\partial \gamma\partial\gamma^\prime}\ket_\star
 \right.\right.\right.
  \\
  &&\left.\left.\left.
  \hspace*{-10mm}
 +
 \bra
 \frac{\partial\log M(\ldots)}{\partial \gamma}\frac{\partial\log M(\ldots)}{\partial \gamma^\prime}
 \ket_\star
 -
\bra
 \frac{\partial\log M(\ldots)}{\partial \gamma}\ket_\star\bra \frac{\partial\log M(\ldots)}{\partial \gamma^\prime}]
 \ket_\star
 \bigkets\bigket_{\!\!\phi}
\end{eqnarray*}
The Gaussian disorder variables $\{x,y,u,v\}$ generated by the
differentiations are eliminated via integration by parts. Below we
give the final results of the calculations, which, although not
fundamentally complicated, can be lengthy.

\subsection{First order derivatives}

For model I one finds the following first order derivatives:
\begin{eqnarray*}
\hspace*{-15mm}
  \frac{1}{\beta K^2}\frac{\partial
 f_{\rm I}}{\partial Q_{cc}}&=& 4\left\{
 Q_{cc}\minus \bra\bras \bra
 \cos^2(\theta)\ket_\star
 \kets\ket_{\phi}
 \right\}
\\
 \hspace*{-15mm}
 \frac{1}{\beta K^2}
 \frac{\partial
 f_{\rm I}}{\partial Q_{cs}}&=& 4\left\{
 Q_{cs} \minus \bra\bras \bra
 \sin(\theta)\cos(\theta)\ket_\star
 \kets\ket_{\phi}
 \right\}
\\
 \hspace*{-15mm}
  \frac{1}{\beta K^2}\frac{\partial
 f_{\rm I}}{\partial q_{cc}}&=&
 2\left\{
\bra\bras \bra\cos(\theta)\ket_\star^2
 \kets\ket_{\phi} -q_{cc}
 \right\}
\\
 \hspace*{-15mm}
 \frac{1}{\beta K^2} \frac{\partial
 f_{\rm I}}{\partial q_{ss}}&=&
 2\left\{
  \bra\bras \bra
\sin(\theta)\ket^2_\star
 \kets\ket_{\phi}-q_{ss}
 \right\}
\\
 \hspace*{-15mm}
  \frac{1}{\beta K^2}
 \frac{\partial
 f_{\rm I}}{\partial q_{cs}}&=&
 4\left\{
  \bra\bras
\bra\cos(\theta)\ket_\star\bra \sin(\theta)\ket_\star
 \kets\ket_{\phi} \!-q_{cs}
 \right\}
\end{eqnarray*}
 For model II one finds:
 \begin{eqnarray*}
 \hspace*{-15mm}
 \frac{1}{\beta K^2} \frac{\partial
 f_{\rm I\!I}}{\partial Q_{cc}}&=& \minus 4\left\{
 Q_{cc}\minus \bra\bras \bra
 \cos^2(\theta)\ket_\star
 \kets\ket_{\phi}
 \right\}
\\
\hspace*{-15mm}
 \frac{1}{\beta K^2}  \frac{\partial
 f_{\rm I\!I}}{\partial Q_{cs}}&=& \minus 4\left\{
 Q_{cs} \minus \bra\bras \bra
 \sin(\theta)\cos(\theta)\ket_\star
 \kets\ket_{\phi}
 \right\}
\\
\hspace*{-15mm}
 \frac{1}{\beta K^2} \frac{\partial
 f_{\rm I\!I}}{\partial q_{cc}}&=& 2\left\{
 \bra\bras \bra\sin(\theta)\ket_\star^2
 \kets\ket_{\phi}
 -q_{ss}
 \right\}
\\
\hspace*{-15mm}
 \frac{1}{\beta K^2}
 \frac{\partial
 f_{\rm I\!I}}{\partial q_{ss}}&=& 2\left\{
 \bra\bras \bra \cos(\theta) \ket^2_\star
 \kets\ket_{\phi}
-q_{cc} \right\}
\\
\hspace*{-15mm}
 \frac{1}{\beta K^2}
 \frac{\partial
 f_{\rm I\!I}}{\partial q_{cs}}&=&
\minus 4\left\{
 \bra\bras
\bra\cos(\theta)\ket_\star\bra \sin(\theta)\ket_\star
 \kets\ket_{\phi}
  \!- q_{cs}
 \right\}
\end{eqnarray*}
 The above results re-confirm our replica-symmetric
saddle-point equations.

\subsection{Second order derivatives for model I}

\begin{eqnarray*}
\hspace*{-25mm}
 \frac{1}{\beta K^2}\frac{\partial^2 f_{\rm I}}{\partial Q_{cc}^2} &=&
 4-4(\beta K)^2\bra\bras
 \bra\cos^2(2\theta)\ket_\star-\bra\cos(2\theta)\ket^2_\star\kets\ket_\phi
\\
\hspace*{-25mm}
 \frac{1}{\beta K^2}\frac{\partial^2 f_{\rm I}}{\partial Q_{cs}^2} &=&
 4-4(\beta K)^2\bra\bras
 \bra\sin^2(2\theta)\ket_\star-\bra\sin(2\theta)\ket^2_\star\kets\ket_\phi
\\
\hspace*{-25mm}
 \frac{1}{\beta K^2}\frac{\partial^2 f_{\rm I}}{\partial Q_{cc}\partial Q_{cs}}&=&
 -4(\beta K)^2\bra\bras
 \bra\sin(2\theta)\cos(2\theta)\ket_\star-\bra\sin(2\theta)\ket_\star
 \bra\cos(2\theta)\ket_\star\kets\ket_\phi
 \\[3mm]
 \hspace*{-25mm}
 \frac{1}{\beta K^2}\frac{\partial^2 f_{\rm I}}{\partial Q_{cc}\partial q_{cc}} &=&
 8(\beta K)^2\bra\bras
 \bra\cos(2\theta)\cos(\theta)\ket_\star\bra\cos(\theta)\ket_\star
   -\bra\cos(2\theta)\ket_\star\bra\cos(\theta)\ket_\star^2\kets\ket_\phi
\\
 \hspace*{-25mm}
 \frac{1}{\beta K^2}\frac{\partial^2 f_{\rm I}}{\partial Q_{cc}\partial q_{ss}} &=&
 8(\beta K)^2\bra\bras
 \bra\cos(2\theta)\sin(\theta)\ket_\star\bra\sin(\theta)\ket_\star
   -\bra\cos(2\theta)\ket_\star\bra\sin(\theta)\ket_\star^2\kets\ket_\phi
\\
\hspace*{-25mm}
 \frac{1}{\beta K^2}\frac{\partial^2 f_{\rm I}}{\partial Q_{cc}\partial q_{cs}}&=&
 8(\beta K)^2\bra\bras
 \bra\cos(2\theta)\cos(\theta)\ket_\star\bra\sin(\theta)\ket_\star
    + \bra\cos(2\theta)\sin(\theta)\ket_\star\bra\cos(\theta)\ket_\star
    \\ && \hspace*{25mm}
    -2\bra\cos(2\theta)\ket_\star\bra\sin(\theta)\ket_\star \bra\cos(\theta)\ket_\star\kets\ket_\phi
 \\
 \hspace*{-25mm}
 \frac{1}{\beta K^2}\frac{\partial^2 f_{\rm I}}{\partial Q_{cs}\partial q_{cc}} &=&
 8(\beta K)^2\bra\bras
 \bra\sin(2\theta)\cos(\theta)\ket_\star\bra\cos(\theta)\ket_\star
  -\bra\sin(2\theta)\ket_\star\bra\cos(\theta)\ket_\star^2\kets\ket_\phi
\\
 \hspace*{-25mm}
 \frac{1}{\beta K^2}\frac{\partial^2 f_{\rm I}}{\partial Q_{cs}\partial q_{ss}} &=&
 8(\beta K)^2\bra\bras
 \bra\sin(2\theta)\sin(\theta)\ket_\star\bra\sin(\theta)\ket_\star
  -\bra\sin(2\theta)\ket_\star\bra\sin(\theta)\ket_\star^2\kets\ket_\phi
\\
 \hspace*{-25mm}
 \frac{1}{\beta K^2}\frac{\partial^2 f_{\rm I}}{\partial Q_{cs}\partial q_{cs}}&=&
 8(\beta K)^2\bra\bras
 \bra\sin(2\theta)\cos(\theta)\ket_\star\bra\sin(\theta)\ket_\star
   +\bra\sin(2\theta)\sin(\theta)\ket_\star\bra \cos(\theta)\ket_\star
   \\&&\hspace*{25mm}
   -2\bra\sin(2\theta)\ket_\star
 \bra\sin(\theta)\ket_\star\bra\cos(\theta)\ket_\star\kets\ket_\phi
 \\[3mm]
 \hspace*{-25mm}
 \frac{1}{\beta K^2}\frac{\partial^2 f_{\rm I}}{\partial q^2_{cc}} &=&
 -2+ 8(\beta K)^2\bra\bras  [\bra\cos^2(\theta)\ket_\star-\bra\cos(\theta)\ket_\star^2]
 [\bra\cos^2(\theta)\ket_\star-3\bra\cos(\theta)\ket_\star^2]
 \kets\ket_\phi
\\
 \hspace*{-25mm}
 \frac{1}{\beta K^2}\frac{\partial^2 f_{\rm I}}{\partial q^2_{ss}} &=&
 -2+ 8(\beta K)^2\bra\bras  [\bra\sin^2(\theta)\ket_\star-\bra\sin(\theta)\ket_\star^2]
 [\bra\sin^2(\theta)\ket_\star-3\bra\sin(\theta)\ket_\star^2]
 \kets\ket_\phi
\\
 \hspace*{-25mm}
 \frac{1}{\beta K^2}\frac{\partial^2 f_{\rm I}}{\partial q^2_{cs}} &=&
 -4+ 8(\beta K)^2\bra\bras
[\bra\sin^2(\theta)\ket_\star\minus \bra\sin(\theta)\ket_\star^2]
 [\bra\cos^2(\theta)\ket_\star\minus 3\bra\cos(\theta)\ket_\star^2]
\\[-1mm]&&
 \hspace*{21mm} + [\bra\cos^2(\theta)\ket_\star\minus \bra\cos(\theta)\ket_\star^2]
 [\bra\sin^2(\theta)\ket_\star\minus 3\bra\sin(\theta)\ket_\star^2]
\\[1mm]&&
 \hspace*{-15mm}
 +2[\bra\sin(\theta)\cos(\theta)\ket_\star\minus \bra\sin(\theta)\ket_\star\bra\cos(\theta)\ket_\star]
 [\bra\sin(\theta)\cos(\theta)\ket_\star\minus 3\bra\sin(\theta)\ket_\star\bra\cos(\theta)\ket_\star]
 \kets\ket_\phi
\\
\hspace*{-25mm}
 \frac{1}{\beta K^2}\frac{\partial^2 f_{\rm I}}{\partial q_{cc}\partial q_{ss}}&=&
 8(\beta K)^2\bra\bras
[\bra\sin(\theta)\cos(\theta)\ket_\star\minus
\bra\sin(\theta)\ket_\star\bra\cos(\theta)\ket_\star]
\\&&\hspace*{10mm}\times
 [\bra\sin(\theta)\cos(\theta)\ket_\star\minus 3\bra\sin(\theta)\ket_\star\bra\cos(\theta)\ket_\star]
\kets\ket_\phi
 \\
 \hspace*{-25mm}
 \frac{1}{\beta K^2}\frac{\partial^2 f_{\rm I}}{\partial q_{cc}\partial q_{cs}} &=&
 8(\beta K)^2\bra\bras
 [\bra\sin(\theta)\cos(\theta)\ket_\star\minus\bra\sin(\theta)\ket_\star\bra\cos(\theta)\ket_\star]
 [\bra\cos^2(\theta)\ket_\star\minus 3\bra\cos(\theta)\ket_\star^2]
\\&&\hspace*{10mm}
+[\bra\sin(\theta)\cos(\theta)\ket_\star\minus 3
\bra\sin(\theta)\ket_\star\bra\cos(\theta)\ket_\star]
 [\bra\cos^2(\theta)\ket_\star\minus \bra\cos(\theta)\ket_\star^2]
 \kets\ket_\phi
\\
 \hspace*{-25mm}
 \frac{1}{\beta K^2}\frac{\partial^2 f_{\rm I}}{\partial q_{ss}\partial q_{cs}} &=&
 8(\beta K)^2\bra\bras
 [\bra\sin(\theta)\cos(\theta)\ket_\star\minus\bra\sin(\theta)\ket_\star\bra\cos(\theta)\ket_\star]
 [\bra\sin^2(\theta)\ket_\star\minus 3\bra\sin(\theta)\ket_\star^2]
\\&&\hspace*{10mm}
+[\bra\sin(\theta)\cos(\theta)\ket_\star\minus 3
\bra\sin(\theta)\ket_\star\bra\cos(\theta)\ket_\star]
 [\bra\sin^2(\theta)\ket_\star\minus \bra\sin(\theta)\ket_\star^2]
\kets\ket_\phi
\end{eqnarray*}

\clearpage
\subsection{Second order derivatives for model II}

\begin{eqnarray*}
\hspace*{-25mm}
 \frac{1}{\beta K^2}\frac{\partial^2 f_{\rm I\!I}}{\partial Q_{cc}^2} &=&
 -4-4(\beta K)^2\bra\bras
 \bra\cos^2(2\theta)\ket_\star-\bra\cos(2\theta)\ket^2_\star\kets\ket_\phi
\\
\hspace*{-25mm}
 \frac{1}{\beta K^2}\frac{\partial^2 f_{\rm I\!I}}{\partial Q_{cs}^2} &=&
 -4-4(\beta K)^2\bra\bras
 \bra\sin^2(2\theta)\ket_\star-\bra\sin(2\theta)\ket^2_\star\kets\ket_\phi
\\
\hspace*{-25mm}
 \frac{1}{\beta K^2}\frac{\partial^2 f_{\rm I\!I}}{\partial Q_{cc}\partial Q_{cs}}&=&
 -4(\beta K)^2\bra\bras
 \bra\sin(2\theta)\cos(2\theta)\ket_\star-\bra\sin(2\theta)\ket_\star
 \bra\cos(2\theta)\ket_\star\kets\ket_\phi
 \\[3mm]
 \hspace*{-25mm}
 \frac{1}{\beta K^2}\frac{\partial^2 f_{\rm I\!I}}{\partial Q_{cc}\partial q_{cc}} &=&
 -8(\beta K)^2\bra\bras
 \bra\cos(2\theta)\sin(\theta)\ket_\star\bra\sin(\theta)\ket_\star
   -\bra\cos(2\theta)\ket_\star\bra\sin(\theta)\ket_\star^2\kets\ket_\phi
\\
 \hspace*{-25mm}
 \frac{1}{\beta K^2}\frac{\partial^2 f_{\rm I\!I}}{\partial Q_{cc}\partial q_{ss}} &=&
 -8(\beta K)^2\bra\bras
 \bra\cos(2\theta)\cos(\theta)\ket_\star\bra\cos(\theta)\ket_\star
   -\bra\cos(2\theta)\ket_\star\bra\cos(\theta)\ket_\star^2\kets\ket_\phi
\\
\hspace*{-25mm}
 \frac{1}{\beta K^2}\frac{\partial^2 f_{\rm I\!I}}{\partial Q_{cc}\partial q_{cs}}&=&
 8(\beta K)^2\bra\bras
 \bra\cos(2\theta)\cos(\theta)\ket_\star\bra\sin(\theta)\ket_\star
    + \bra\cos(2\theta)\sin(\theta)\ket_\star\bra\cos(\theta)\ket_\star
    \\ && \hspace*{25mm}
    -2\bra\cos(2\theta)\ket_\star\bra\sin(\theta)\ket_\star \bra\cos(\theta)\ket_\star\kets\ket_\phi
 \\
 \hspace*{-25mm}
 \frac{1}{\beta K^2}\frac{\partial^2 f_{\rm I\!I}}{\partial Q_{cs}\partial q_{cc}} &=&
- 8(\beta K)^2\bra\bras
 \bra\sin(2\theta)\sin(\theta)\ket_\star\bra\sin(\theta)\ket_\star
  -\bra\sin(2\theta)\ket_\star\bra\sin(\theta)\ket_\star^2\kets\ket_\phi
\\
 \hspace*{-25mm}
 \frac{1}{\beta K^2}\frac{\partial^2 f_{\rm I\!I}}{\partial Q_{cs}\partial q_{ss}} &=&
 -8(\beta K)^2\bra\bras
 \bra\sin(2\theta)\cos(\theta)\ket_\star\bra\cos(\theta)\ket_\star
  -\bra\sin(2\theta)\ket_\star\bra\cos(\theta)\ket_\star^2\kets\ket_\phi
\\
 \hspace*{-25mm}
 \frac{1}{\beta K^2}\frac{\partial^2 f_{\rm I\!I}}{\partial Q_{cs}\partial q_{cs}}&=&
 8(\beta K)^2\bra\bras
 \bra\sin(2\theta)\cos(\theta)\ket_\star\bra\sin(\theta)\ket_\star
   +\bra\sin(2\theta)\sin(\theta)\ket_\star\bra \cos(\theta)\ket_\star
   \\&&\hspace*{25mm}
   -2\bra\sin(2\theta)\ket_\star
 \bra\sin(\theta)\ket_\star\bra\cos(\theta)\ket_\star\kets\ket_\phi
 \\[3mm]
 \hspace*{-25mm}
 \frac{1}{\beta K^2}\frac{\partial^2 f_{\rm I\!I}}{\partial q^2_{cc}} &=&
  8(\beta K)^2\bra\bras  [\bra\sin^2(\theta)\ket_\star-\bra\sin(\theta)\ket_\star^2]
 [\bra\sin^2(\theta)\ket_\star-3\bra\sin(\theta)\ket_\star^2]
 \kets\ket_\phi
\\
 \hspace*{-25mm}
 \frac{1}{\beta K^2}\frac{\partial^2 f_{\rm I\!I}}{\partial q^2_{ss}} &=&
 8(\beta K)^2\bra\bras  [\bra\cos^2(\theta)\ket_\star-\bra\cos(\theta)\ket_\star^2]
 [\bra\cos^2(\theta)\ket_\star-3\bra\cos(\theta)\ket_\star^2]
 \kets\ket_\phi
\\
 \hspace*{-25mm}
 \frac{1}{\beta K^2}\frac{\partial^2 f_{\rm I\!I}}{\partial q^2_{cs}} &=&
 4+ 8(\beta K)^2\bra\bras
[\bra\sin^2(\theta)\ket_\star\minus \bra\sin(\theta)\ket_\star^2]
 [\bra\cos^2(\theta)\ket_\star\minus 3\bra\cos(\theta)\ket_\star^2]
\\[-1mm]&&
 \hspace*{21mm} + [\bra\cos^2(\theta)\ket_\star\minus \bra\cos(\theta)\ket_\star^2]
 [\bra\sin^2(\theta)\ket_\star\minus 3\bra\sin(\theta)\ket_\star^2]
\\[1mm]&&
 \hspace*{-15mm}
 +2[\bra\sin(\theta)\cos(\theta)\ket_\star\minus \bra\sin(\theta)\ket_\star\bra\cos(\theta)\ket_\star]
 [\bra\sin(\theta)\cos(\theta)\ket_\star\minus 3\bra\sin(\theta)\ket_\star\bra\cos(\theta)\ket_\star]
 \kets\ket_\phi
\\
\hspace*{-25mm}
 \frac{1}{\beta K^2}\frac{\partial^2 f_{\rm I\!I}}{\partial q_{cc}\partial
 q_{ss}}&=&-2+ 8(\beta K)^2\bra\bras
[\bra\sin(\theta)\cos(\theta)\ket_\star\minus
\bra\sin(\theta)\ket_\star\bra\cos(\theta)\ket_\star]
\\&&\hspace*{10mm}\times
 [\bra\sin(\theta)\cos(\theta)\ket_\star\minus 3\bra\sin(\theta)\ket_\star\bra\cos(\theta)\ket_\star]
\kets\ket_\phi
 \\
 \hspace*{-25mm}
 \frac{1}{\beta K^2}\frac{\partial^2 f_{\rm I\!I}}{\partial q_{cc}\partial q_{cs}} &=&
 -8(\beta K)^2\bra\bras
 [\bra\sin(\theta)\cos(\theta)\ket_\star\minus\bra\sin(\theta)\ket_\star\bra\cos(\theta)\ket_\star]
 [\bra\sin^2(\theta)\ket_\star\minus 3\bra\sin(\theta)\ket_\star^2]
\\&&\hspace*{10mm}
+[\bra\sin(\theta)\cos(\theta)\ket_\star\minus 3
\bra\sin(\theta)\ket_\star\bra\cos(\theta)\ket_\star]
 [\bra\sin^2(\theta)\ket_\star\minus \bra\sin(\theta)\ket_\star^2]
 \kets\ket_\phi
\\
 \hspace*{-25mm}
 \frac{1}{\beta K^2}\frac{\partial^2 f_{\rm I\!I}}{\partial q_{ss}\partial q_{cs}} &=&
 -8(\beta K)^2\bra\bras
 [\bra\sin(\theta)\cos(\theta)\ket_\star\minus\bra\sin(\theta)\ket_\star\bra\cos(\theta)\ket_\star]
 [\bra\cos^2(\theta)\ket_\star\minus 3\bra\cos(\theta)\ket_\star^2]
\\&&\hspace*{10mm}
+[\bra\sin(\theta)\cos(\theta)\ket_\star\minus 3
\bra\sin(\theta)\ket_\star\bra\cos(\theta)\ket_\star]
 [\bra\cos^2(\theta)\ket_\star\minus \bra\cos(\theta)\ket_\star^2]
\kets\ket_\phi
\end{eqnarray*}

\section{Derivation of the AT Instability}
\label{app:AT}

\subsection{Calculation of the RS Hessian for replicon fluctuations}

We calculate the Hessian of the disorder-averaged free energy per
oscillator  by expanding (\ref{eq:f_rsb}) in powers of the
fluctuations $\{\delta \bq^{\star\star},\delta\hbq^{\star\star}\}$
around the replica-symmetric saddle-point. In this appendix we
will use pairs of Roman indices $(ab)$ and $(de)$ to label the
four combinations $\{cc,ss,cs,sc\}$, and abbreviate the
corresponding functions in the obvious way as
 $a(\theta),b(\theta),c(\theta),d(\theta)\in\{\cos(\theta),\sin(\theta)\}$.
 We put
$\hat{q}_{\alpha\beta}^{\star\star}=2i(\beta K)^2
k_{\alpha\beta}^{\star\star}$ and $\delta\overline{f}[\ldots]=
\overline{f}[\ldots]-\overline{f}_{\rm RS}[\ldots]$, and obtain
(since linear terms are absent):
\begin{eqnarray}
\hspace*{-20mm}
 -\frac{n}{\beta
 K^2} \delta\overline{f}[\ldots]
 &=&
  2\sin^2(A^\star) \sum_{\alpha\beta}\left[\delta
 q^{ss}_{\alpha\beta} \delta q^{cc}_{\alpha\beta} \minus \delta
 q^{sc}_{\alpha\beta}\delta  q^{cs}_{\alpha\beta}\right]
  +\cos^2(A^\star) \sum_{ab}\sum_{\alpha\beta} [ \delta
 q^{ab}_{\alpha\beta}]^2
\nonumber\\&& \hspace*{-25mm}
 -2\sum_{ab}\sum_{\alpha\beta}\delta
 k^{ab}_{\alpha\beta} \delta q^{ab}_{\alpha\beta}
  +
 \frac{1}{(\beta K)^2} \bigbra\!  \log\left\{
 \frac{\int\!d\btheta~ M(\btheta|\{\bk^{\star\star}\!\!\plus\delta\bk^{\star\star}\})}
 {\int\!d\btheta~ M(\btheta|\{\bk^{\star\star}\})}
 \right\}\!
 \bigket_\phi+\ldots
\label{eq:ATfluct}
\end{eqnarray}
with \bd
 M(\btheta|\{\bk^{\star\star}\})=
 e^{\beta h\sum_{\alpha}\cos(\theta^\alpha\minus\phi) +2(\beta
 K)^2\sum_{ab}\sum_{\alpha\beta}
 k^{ab}_{\alpha\beta}a(\theta_\alpha)b(\theta_\beta)}
 \ed
 Working out the fraction in the last term of (\ref{eq:ATfluct})  gives
 \bd
 \hspace*{-20mm}
  \frac{\int\!d\btheta~ M(\btheta|\{\bk^{\star\star}\!\!\plus\delta\bk^{\star\star}\})}
 {\int\!d\btheta~ M(\btheta|\{\bk^{\star\star}\})}
=
 1+ 2(\beta K)^2\!\sum_{\alpha\beta}\sum_{ab} \delta
 k_{\alpha\beta}^{ab}
 \frac{\int\!d\btheta~M(\btheta|\{\bk^{\star\star}\})a(\theta_\alpha)b(\theta_\beta)}
 {\int\!d\btheta~M(\btheta|\{\bk^{\star\star}\})}
\ed \bd \hspace*{-10mm}
 +
 2(\beta K)^4\sum_{\alpha\beta\gamma\delta}\sum_{ab,de}
 \delta k_{\alpha\beta}^{ab}\delta k_{\gamma\delta}^{de}
 \frac{\int\!d\btheta~M(\btheta|\{\bk^{\star\star}\})a(\theta_\alpha)b(\theta_\beta)
 d(\theta_\gamma)e(\theta_\delta)}
 {\int\!d\btheta~M(\btheta|\{\bk^{\star\star}\})}
~ +~\ldots \ed
 Thus, upon expanding $\log(x)=x-\frac{1}{2}x^2+\ldots$, and
 upon introducing the short-hand
$M(\btheta|{\rm RS})=M(\btheta|\{\bk^{\star\star}_{\rm RS}\})$, we
arrive at \begin{eqnarray*} \hspace*{-25mm}
 -\frac{n}{\beta
 K^2} \delta\overline{f}[\ldots]
 &=&
  2\sin^2(A^\star) \sum_{\alpha\beta}\left[\delta
 q^{ss}_{\alpha\beta} \delta q^{cc}_{\alpha\beta} \minus \delta
 q^{sc}_{\alpha\beta}\delta  q^{cs}_{\alpha\beta}\right]
  +\cos^2(A^\star) \sum_{ab}\sum_{\alpha\beta} [ \delta
 q^{ab}_{\alpha\beta}]^2
\nonumber\\&& \hspace*{-30mm}
 -2\sum_{ab}\sum_{\alpha\beta}\delta
 k^{ab}_{\alpha\beta} \delta q^{ab}_{\alpha\beta}
 + 2(\beta K)^2 \sum_{\alpha\beta\gamma\delta}\sum_{ab,de}\delta
 k_{\alpha\beta}^{ab}\delta k_{\gamma\delta}^{de} \bigbra
 \frac{\int\!d\btheta~M(\btheta|{\rm RS})a(\theta_\alpha)b(\theta_\beta)d(\theta_\gamma)e(\theta_\delta)}
 {\int\!d\btheta~M(\btheta|{\rm RS})}
 \right.
 \\&&
  \left.
\hspace*{8mm}
  - \frac{\int\!d\btheta~M(\btheta|{\rm RS})a(\theta_\alpha)b(\theta_\beta)}
 {\int\!d\btheta~M(\btheta|{\rm RS})}
 \frac{\int\!d\btheta~M(\btheta|{\rm RS})d(\theta_\gamma)e(\theta_\delta)}
 {\int\!d\btheta~M(\btheta|{\rm RS})}
 \bigket_{\!\!\phi} +\ldots
\end{eqnarray*}
We next work out the last of the above quadratic terms for
fluctuations around the RS solution (\ref{eq:rs_ansatz}), for
small $n$. We use the short-hand $Dxyuv=DxDyDuDv$, and denote by
$M(\theta)$ either the measure (\ref{eq:RSMeasureI}) (for model I)
or (\ref{eq:RSMeasureII}) (for model II): \bd \hspace*{-20mm}
\bigbra\ldots\bigket_{\!\!\phi}= \bigbra
\frac{\int\!Dxyuv[\prod_\lambda\int\!d\theta M(\theta_\lambda)]
a(\theta_\alpha)b(\theta_\beta)c(\theta_\gamma)d(\theta_\delta)}
{\int\!Dxyuv[\int\!d\theta M(\theta)]^n} \bigket_{\!\!\phi} \ed
\bd \hspace*{-15mm}
  -\bigbra
\frac{\int\!Dxyuv[\prod_\lambda\int\!d\theta_\lambda
M(\theta_\lambda)]a(\theta_\alpha)b(\theta_\beta)}
{\int\!Dxyuv[\int\!d\theta M(\theta)]^n}
\frac{\int\!Dxyuv[\prod_\lambda\int\!d\theta_\lambda
M(\theta_\lambda)]c(\theta_\gamma)d(\theta_\delta)}
{\int\!Dxyuv[\int\!d\theta M(\theta)]^n} \bigket_{\!\!\phi} \ed If
we restrict ourselves to replicon fluctuations, where $\delta
q_{\alpha\alpha}^{\star\star}=\delta
k_{\alpha\alpha}^{\star\star}=0$, $\sum_\alpha \delta
k_{\alpha\beta}=\sum_{\alpha}\delta q_{\alpha\beta}=0$ and
$\sum_\beta \delta k_{\alpha\beta}=\sum_{\beta}\delta
q_{\alpha\beta}=0$, we can proceed by inserting a string of
Kronecker symbols (and complementary symbols
$\overline{\delta}_{\alpha\beta}=1\minus\delta_{\alpha\beta}$) to
streamline the bookkeeping of possibly identical combinations of
replica indices: \bd \hspace*{-10mm}
 \delta_{\alpha\delta}    \delta_{\beta\gamma}
+\delta_{\alpha\gamma} \delta_{\beta\delta} +\delta_{\alpha\delta}
\ndelta_{\beta\gamma} +\delta_{\alpha\gamma} \ndelta_{\beta\delta}
+\ndelta_{\alpha\gamma}\delta_{\beta\delta}
+\ndelta_{\alpha\delta}   \delta_{\beta\gamma}
+\ndelta_{\alpha\delta}\ndelta_{\alpha\gamma}
\ndelta_{\beta\delta}\ndelta_{\beta\gamma} \ed giving, for $n\to
0$, and upon using again the replicon properties which allow us to
drop contributions which will not survive the summation over
$(\alpha,\beta;ab)$ and $(\gamma,\delta;de)$: \bd \hspace*{-18mm}
\bigbra\ldots\bigket_{\!\!\phi}
=
 \delta_{\alpha\delta} \delta_{\beta\gamma}~
 \bra\bras
 \left[
 \bra a(\theta)e(\theta)\ket_\star \minus  \bra
 a(\theta)\ket_\star\bra
 e(\theta)\ket_\star\right]
 \left[
 \bra b(\theta)d(\theta)\ket_\star \minus  \bra
 b(\theta)\ket_\star\bra
 d(\theta)\ket_\star\right]
\kets\ket_\phi
 \ed \bd
 \hspace*{-5mm}
 +~\delta_{\beta\delta} \delta_{\alpha\gamma}~
 \bra\bras
 \left[
 \bra b(\theta)e(\theta)\ket_\star \minus  \bra
 b(\theta)\ket_\star\bra
 e(\theta)\ket_\star\right]
 \left[
 \bra a(\theta)d(\theta)\ket_\star \minus  \bra
 a(\theta)\ket_\star\bra
 d(\theta)\ket_\star\right]
\kets\ket_\phi \ed
 These two remaining terms will give
identical contributions to the fluctuations around the RS free
energy, after the summation over $(\gamma,\delta;de)$ has been
carried out, and the result can be written in the compact form
\begin{eqnarray} \hspace*{-15mm}
 -\frac{n}{\beta
 K^2} \delta\overline{f}[\ldots]
 &=&
 \sum_{\alpha\neq \beta}\sum_{ab,de}\left\{
 \room
 \delta q_{\alpha\beta}^{ab}\delta q_{\alpha\beta}^{de}\left[
  \sin^2(A^\star) C_{ab,de}
  +\cos^2(A^\star) \delta_{ab,de}\right]
  \nonumber
  \right.
  \\
  &&\hspace*{0mm}\left.\room
  -2\delta q_{\alpha\beta}\delta
  k_{\alpha\beta}^{de}~\delta_{ab,de}+4(\beta K)^2
   \delta k_{\alpha\beta}^{ab}\delta k_{\alpha\beta}^{de}
   E_{ab,de}\right\}
+\ldots \label{eq:hessian_halfway}
\end{eqnarray}
with the two $4\times 4$ matrices (note: $ab,de\in \{
cc,ss,cs,sc\}$)
\be
\hspace*{-15mm}
C_{ab,de}=\delta_{ab,ss}\delta_{de,cc}+\delta_{ab,cc}\delta_{de,ss}
-\delta_{ab,cs}\delta_{de,sc}-\delta_{ab,sc}\delta_{de,cs}
\label{eq:fullmatrixC}
 \ee
\be
\hspace*{-15mm}
 E_{ab,de}=
 \bra\bras
 \left[
 \bra b(\theta)e(\theta)\ket_\star \minus  \bra
 b(\theta)\ket_\star\bra
 e(\theta)\ket_\star\right]
 \left[
 \bra a(\theta)d(\theta)\ket_\star \minus  \bra
 a(\theta)\ket_\star\bra
 d(\theta)\ket_\star\right]
\kets\ket_\phi \label{eq:fullmatrixE} \ee Expression
(\ref{eq:hessian_halfway}) shows that the replica indices
$(\alpha,\beta)$ have become irrelevant labels, and that for $n\to
0$ and within the sub-space of replicon fluctuations $\{\delta
k_{\alpha\beta}^{\star\star},\delta
q_{\alpha\beta}^{\star\star}\}$, the spectrum of the Hessian
reduces to that of the following $8\times 8$ matrix: \be {\cal H}
=\left(\!\!
\begin{array}{cc}\cos^2(A^\star)\unity+\sin^2(A^\star)\bC &
-\unity \\ -\unity & 4(\beta K)^2\bE \end{array}\!\!\right) \ee
(apart from an overall multiplicative constant), with the building
blocks  $\bC=\{ C_{ab,de}\}$ and $\bE=\{ E_{ab,de}\}$ as given in
(\ref{eq:fullmatrixC},\ref{eq:fullmatrixE}) and with the $4\times
4$ unit matrix $\unity_{ab,de}=\delta_{ab,de}$.

\subsection{Replicon instabilities}

Requiring the Hessian to have a zero eigenvalue (replicon
instability), using the above results, leads us to the following
condition: \bd \hspace*{-10mm}
 \exists\left(\!\!
\begin{array}{c}\bx
\\\by\end{array}\!\!\right) \neq \left(\!\! \begin{array}{c}\bnul \\
\bnul \end{array}\!\!\right): ~~~~~~ \left(\!\!
\begin{array}{cc}\cos^2(A^\star)\unity+\sin^2(A^\star)\bC &
-\unity \\ -\unity & 4(\beta K)2\bE \end{array}\!\!\right)
\left(\!\! \begin{array}{c}\bx \\\by\end{array}\!\!\right)
=\left(\!\! \begin{array}{c}\bnul \\ \bnul \end{array}\!\!\right)
\ed
 \bd
\hspace*{-10mm}
 \exists\left(\!\!
\begin{array}{c}\bx
\\\by\end{array}\!\!\right) \neq \left(\!\! \begin{array}{c}\bnul \\
\bnul \end{array}\!\!\right): ~~~~~~
 \left\{
\begin{array}{l}
[\cos^2(A^\star)\unity+\sin^2(A^\star)\bC]\bx=\by\\
\bx=4(\beta K)^2 \bE\by
\end{array}
\right.
\ed
 Equivalently: \bd {\rm Det}\left\{4(\beta
K)^2[\cos^2(A^\star)\unity+\sin^2(A^\star)\bC]\bE-\unity\right\}=0
\ed For our two models I (where $A^\star=0$) and II (where
$A^\star=\frac{1}{2}\pi$) this translates into
\begin{eqnarray}
{\rm Model~I:} &~~~& {\rm Det}\left[\bE-(T/2K)^2\unity\right]=0
\label{eq:fullAT_model1}
\\
{\rm Model~II:} &~~~& {\rm Det}\left[\bE-(T/2K)^2\bC\right]=0
\label{eq:fullAT_model2}
\end{eqnarray} where we have used the property $\bC^2=\unity$.

 In the
representation where the various entries of the matrices are
ordered as $\{cc,ss,cs,sc\}$, and with assistance of the
abbreviations  (\ref{eq:gammas1},\ref{eq:gammas1}), we find our
matrices $\bC$ and $\bE$ (which will generally not commute) to
acquire the following form: \be \hspace*{-25mm}
\bC=\left(\!\!\begin{array}{cccc} 0 & 1 & 0 & 0\\ 1 & 0 & 0 & 0\\
0 & 0 & 0 & \minus 1\\ 0 & 0 & \minus 1 & 0
\end{array}
\!\!\right) ~~~~~~ \bE=\bigbra\!\bigbras\!
\left(\!\!\begin{array}{cccc}
 \gamma_{cc}^2   & \gamma_{cs}^2 & \gamma_{cc}\gamma_{cs} & \gamma_{cc}\gamma_{cs} \\
 \gamma_{cs}^2   & \gamma_{ss}^2 & \gamma_{ss}\gamma_{cs} & \gamma_{ss}\gamma_{cs} \\
 \gamma_{cc}\gamma_{cs} & \gamma_{ss}\gamma_{cs} & \gamma_{cc}\gamma_{ss} & \gamma_{cs}^2 \\
 \gamma_{cc}\gamma_{cs} & \gamma_{ss}\gamma_{cs} & \gamma_{cs}^2 &\gamma_{cc}\gamma_{ss}
\end{array}\!\!\right)
\!\bigkets\!\bigket_{\!\!\phi} \label{eq:matrices_on_4basis}
 \ee
In the representation (\ref{eq:matrices_on_4basis}) one observes
that the vector $|0\ket=\frac{1}{2}\sqrt{2}(0,0,1,\minus 1)$ is an
eigenvector of both $\bC$ (with eigenvalue 1) and $\bE$ (with
eigenvalue $\bra\bras \gamma_{cc}\gamma_{ss}\minus
\gamma_{cs}^2\kets\ket_\phi$), and therefore also of the two
matrices in (\ref{eq:fullAT_model1},\ref{eq:fullAT_model2}). This
immediately leads us to the first replicon instability condition:
\be
(T/2K)^2=\bra\bras \gamma_{cc}\gamma_{ss}\minus
\gamma_{cs}^2\kets\ket_\phi \label{eq:first_replicon} \ee (for
both models I and II).

The remaining three eigenvalues of the relevant matrices in
(\ref{eq:fullAT_model1},\ref{eq:fullAT_model2}), viz. $\bE-
(T/2K)^2\unity$ (for model I) and  $\bE- (T/2K)^2\bC$ (for model
II), must all be orthogonal to $|0\ket$, and are thus in the
sub-space spanned by the following three orthogonal eigenvectors
of $\bC$ (with eigenvalues $\{1,-1,-1\}$, respectively):  \bd
\hspace*{-10mm} |1\ket=\frac{1}{2}\sqrt{2}(1,1,0,0),~~~~~
|2\ket=\frac{1}{2}\sqrt{2}(1,\minus 1,0,0),~~~~~
|3\ket=\frac{1}{2}\sqrt{2}(0,0,1,1) \ed After some simple algebra
one finds that  on the basis $\{|1\ket,|2\ket,|3\ket\}$ the
matrices (\ref{eq:matrices_on_4basis}) reduce to the following
$3\times 3$ ones
\be
\hspace*{0mm} \bC=\left(\!\!\begin{array}{ccc} 1 & 0 & 0 \\ 0 &
\minus 1 & 0 \\ 0 & 0 &  \minus 1
\end{array}
\!\!\right) \label{eq:thematrixC}
\ee
\be
\hspace*{-25mm} \bE=\left(\!\begin{array}{ccc} \bra\bras
\frac{1}{2}(\gamma_{cc}^2\plus \gamma_{ss}^2)\plus
\gamma_{cs}^2\kets\ket_\phi &
\bra\bras\frac{1}{2}(\gamma_{cc}^2\minus
\gamma_{ss}^2)\kets\ket_\phi &
\bra\bras\gamma_{cs}(\gamma_{cc}\plus \gamma_{ss})\kets\ket_\phi
\\[1mm]
\bra\bras \frac{1}{2}(\gamma_{cc}^2\minus
\gamma_{ss}^2)\kets\ket_\phi &
\bra\bras\frac{1}{2}(\gamma_{cc}^2\plus \gamma_{ss}^2)\minus
\gamma_{cs}^2\kets\ket_\phi &
\bra\bras\gamma_{cs}(\gamma_{cc}\minus \gamma_{ss})\kets\ket_\phi
\\[1mm]
\bra\bras \gamma_{cs}(\gamma_{cc}\plus\gamma_{ss})\kets\ket_\phi &
\bra\bras \gamma_{cs}(\gamma_{cc}\minus \gamma_{ss})\kets\ket_\phi
& \bra\bras \gamma_{cc}\gamma_{ss}\plus\gamma_{cs}^2\kets\ket_\phi
\end{array}\!
\right) \label{eq:thematrixE} \ee The remaining replicon
instabilities now follow upon inserting
(\ref{eq:thematrixC},\ref{eq:thematrixE}) into the conditions
(\ref{eq:fullAT_model1},\ref{eq:fullAT_model2}).
 The physical RSB transition
associated with the combined replicon instabilities is the one
occurring at the highest temperature.


\section*{References}

\begin{thebibliography}{99}

\bibitem{Winfr}
Winfree AT 1967 1980 {\em The geometry of biological time}
Springer-Verlag, New York.
\bibitem{Kuram1}
Kuramoto Y 1984 {\em Chemical oscillations, waves and turbulence}
Springer-Verlag, Berlin.
\bibitem{Kuram2}
Kuramoto Y 1984 {\em Prog. Theor. Phys. Suppl} {\bf 79} 223
\bibitem{Strog}
Strogatz SH and Mirollo RE 1991 {\em J. Stat. Phys.} {\bf 63} 613
\bibitem{Bonil1}
Bonilla LL, Neu JC and Spigler R 1992 {\em  J. Stat. Phys.} {\bf
67} 313
\bibitem{Bonil2}
Bonilla LL, Perez Vicente CJ and Spigler R 1998 {\em Physica D}
{\bf 113} 79
\bibitem{Aceb}
Acebron JA, Bonilla LL, De Leo S and Spigler R 1998 {\em Phys.
Rev. E} {\bf 57} 5287
\bibitem{Daid92}
Daido H 1988 {\ Phys. Rev. Lett.} {\bf 61} 231
\bibitem{Stro88} Strogatz SH and Mirollo RE 1988 {\em J. Phys. A}
{\bf 21} L699
\bibitem{Boni93}
Bonilla LL, Perez Vicente CJ and Rubi JM 1993 {\em J. Stat. Phys.}
{\bf 70} 921
\bibitem{Arenas94}
Arenas A and Perez Vicente CJ 1994 {\em Europhys. Lett.} {\bf 26}
79
\bibitem{Saka88}
Sakaguchi H 1988 {\em Prog. Theor. Phys.} {\bf 79} 39
\bibitem{Reimann99} Reimann P, Van den Broeck C and Kawai R 1999
{\em Phys. Rev. E} {\bf 60} 6402
\bibitem{Schu}
Schuster HG and Wagner P 1990 {\em Biol. Cybern.} {\bf 64} 77,
{\em Biol. Cybern.} {\bf 64} 83
\bibitem{Abb}
Abbott LF 1990 {\em J. Phys. A} {\bf 23} 3835
\bibitem{Wies}
Wiesenfeld K, Colet P and Strogatz SH 1996 {\em Phys. Rev. Lett.}
{\bf 76} 404; 1998 {\em Phys. Rev E} {\bf 57} 1563
\bibitem{PB222a}
Granato E, Kosterlitz JM and Nightingale MP 1996 {\em Physica B}
{\bf 222} 266
\bibitem{PB222b}
Majhofer A 1996 {\em Physica B} {\bf 222} 273
\bibitem{Sakaku}
Sakaguchi H and Kuramoto Y 1986 {\em Prog. Theor. Phys.} {\bf 76}
576
\bibitem{Raeetal}
Rae HC, Coolen ACC and Sollich P 2002 {\em in preparation}
\bibitem{Daido}
Daido H 1992 {\em Phys Rev Lett} {\bf 68} 1073 ,2000 {\em Phys.
Rev. E} {\bf 61} 2145
\bibitem{Still}
Stiller JC and Radons G 1998 {\em Phys. Rev. E} {\bf 58} 1789
\bibitem{Choi} Choi MY and Doniach S 1985 {\em Phys. Rev. B} {\bf
31} 4516
\bibitem{Arenas95}
Arenas A and Perez Vicente CJ 1994 {\em Phys. Rev. E} {\bf 50} 949
\bibitem{Elder82}
Elderfield DJ and Sherrington D 1982 {\em J. Phys. A} {\bf 15}
L513
\bibitem{Cragg}
Cragg DM, Sherrington D and Gabay M {\em Phys. Rev. Lett.} {\bf
49} 1982
\bibitem{Elder83}
Elderfield D and Sherrington D 1983 {\em J. Phys. C} {\bf 16} 4865
\bibitem{AT}
De Almeida JRL and Thouless DJ 1978 {\em J. Phys. A} {\bf 11} 983
\bibitem{Mezardetal}
M\'{e}zard M, Parisi G and Virasoro MA 1987 {\em Spin-Glass Theory
and Beyond} (Singapore: World Scientific)
\bibitem{AbramStegun}
Abramowitz M and Stegun IA 1972 {\em Handbook of Mathematical
Functions} (New York: Dover)
\bibitem{SK}
Kirkpatrick S and Sherrington D 1978 {\em Phys. Rev. B} {\bf 17}
4384
\end{thebibliography}
\end{document}